\documentclass[aps,superscriptaddress,tightenlines,nofootinbib,floatfix,longbibliography,notitlepage,10pt,pra]{revtex4-1}
\usepackage[left=18mm,right=19mm,top=23mm,bottom=16mm]{geometry}
\usepackage{amsmath,amssymb}
\usepackage{bm,bbm}
\usepackage{comment}
\usepackage{graphicx}
\usepackage{color}
\usepackage[dvipsnames]{xcolor}
\usepackage{slashed}
\usepackage[multidot]{grffile} 
\usepackage[
    colorlinks=true,
    allcolors=blue
]{hyperref}
\usepackage{pgfplots}
\pgfplotsset{compat=1.9}
\usepackage{booktabs} 
\usepackage{placeins}
\usepackage{anyfontsize} 
\usepackage{float} 

\usepackage{siunitx}

\providecommand{\repositoryInformationSetup}{} 
\repositoryInformationSetup

\usepackage{xspace}
\usepackage{bbm}

\newcommand{\Luscher}{L\"{u}scher\xspace}
\newcommand{\nstep}{\ensuremath{{n_s}}\xspace}
\newcommand{\spherical}{\ensuremath{\bigcirc}\xspace}

\newcommand{\counterterm}{\ensuremath{\mathcal{L}}}
\newcommand{\dispersion}{\ensuremath{\boxplus}\xspace}
\newcommand{\PV}{\ensuremath{\mathcal{P}}}
\newcommand{\normalization}{\ensuremath{\mathcal{N}}\xspace}

\newcommand{\F}{\ensuremath{\mathcal{F}}}
\newcommand{\FV}{\ensuremath{\textrm{FV}}}
\newcommand{\xtilde}{\ensuremath{\tilde{x}}\xspace}
\newcommand{\Aoneg}{\ensuremath{A_{1g}}\xspace}
\renewcommand{\l}{\ensuremath{\ell}\xspace}
\newcommand{\Laplacian}{\ensuremath{\mathop{}\!\mathbin\bigtriangleup}}
\newcommand{\BZ}{\text{B.Z.}}

\newcommand{\repoURL}{https://github.com/ckoerber/luescher-nd}

\newcommand{\Secref}[1]{Section~\ref{sec:#1}}

\newcommand{\Appref}[1]{Appendix~\ref{sec:#1}}
\newcommand{\tabref}[1]{Tab.~\ref{tab:#1}\xspace}
\newcommand{\Tabref}[1]{Table~\ref{tab:#1}\xspace}
\newcommand{\figref}[1]{Fig.~\ref{fig:#1}\xspace}
\newcommand{\Figref}[1]{Figure~\ref{fig:#1}\xspace}
\renewcommand{\eqref}[1]{(\ref{eq:#1})\xspace}

\newcommand{\Ref}[1]{Ref.~\cite{#1}}

\newcommand{\Refs}[1]{Refs.~\cite{#1}}

\newcommand{\goesto}{\ensuremath{\rightarrow}}

\newcommand{\one}{\ensuremath{\mathbbm{1}}}
\newcommand{\order}[1]{\ensuremath{\mathcal{O}\left(#1\right)}\xspace}

\newcommand{\inverse}{\ensuremath{^{-1}}}                                       

\newcommand{\abs}[1]{\ensuremath{\left| #1 \right|}\xspace}

\newcommand{\ket}[1]{\ensuremath{\left|\;#1\;\right\rangle}}

\newcommand{\braMket}[3]{\ensuremath{\left\langle\;#1\;\middle|\;#2\;\middle|\;#3\;\right\rangle}}

\let\builtinLaTeX\LaTeX
\def\LaTeX{\builtinLaTeX\xspace}

\renewcommand{\vec}[1]{\boldsymbol{#1}}

\begin{document}

\title{Renormalization of a Contact Interaction on a Lattice}
\newcommand{\ikp}{
    Institut f\"{u}r Kernphysik,
    Forschungszentrum J\"{u}lich, 54245 J\"{u}lich Germany
}

\newcommand{\ias}{
    Institute for Advanced Simulation,
    Forschungszentrum J\"{u}lich, 54245 J\"{u}lich Germany
}

\newcommand{\julich}{
    Institut f\"{u}r Kernphysik and Institute for Advanced Simulation,
    Forschungszentrum J\"{u}lich, 54245 J\"{u}lich Germany
}

\newcommand{\bonn}{
    Helmholtz-Institut f\"{u}r Strahlen- und Kernphysik,
    Rheinische Friedrich-Wilhelms-Universit\"{a}t Bonn, 53012 Bonn Germany
}

\newcommand{\jsc}{
    J\"{u}lich Supercomputing Center,
    Forschungszentrum J\"{u}lich, 54245 J\"{u}lich Germany
}

\newcommand{\berkeley}{
    Department of Physics,
    University of California, Berkeley, CA 94720, USA
}

\newcommand{\lbnl}{
    Nuclear Science Division,
    Lawrence Berkeley National Laboratory, Berkeley, CA 94720, USA
}

\newcommand{\umd}{
    Department of Physics,
    University of Maryland, College Park, MD 20742,
    USA
}

\newcommand{\mcfp}{
    Maryland Center for Fundamental Physics,
    University of Maryland, College Park, MD 20742,
    USA
}

\author{Christopher K\"orber}   \affiliation{\berkeley} \affiliation{\lbnl}
\author{Evan Berkowitz}         \affiliation{\julich} \affiliation{\mcfp}
\author{Thomas Luu}             \affiliation{\julich} \affiliation{\bonn}

\date{\today}

\begin{abstract}
Contact interactions can be used to describe a system of particles at unitarity, contribute to the leading part of nuclear interactions and are numerically non-trivial because they require a proper regularization and renormalization scheme.
We explain how to tune the coefficient of a contact interaction between non-relativistic particles on a discretized space in 1, 2, and 3 spatial dimensions such that we can remove all discretization artifacts.
By taking advantage of a latticized \Luscher zeta function, we can achieve a momentum-independent scattering amplitude at any finite lattice spacing.
\end{abstract}

\maketitle

\section{Introduction}\label{sec:intro}

Many physically interesting systems comprise strongly-interacting fermions.
In three spatial dimensions the scattering of fermions with a short-range interaction can be completely characterized by a scattering length, and when that length diverges the details of the potential are washed out and no dimensionful scales remain.
Such \emph{unitary fermions} exhibit interactions as strong as can be without forming bound states, and provide an interesting guide for understanding other strong interactions because of their universal behavior.
For example, the nuclear interaction in the deuteron channel has an extremely long scattering length, and trapped ultracold atoms can be tuned to unitarity by applying external magnetic fields and leveraging Feshbach resonances.

By tuning a quantum-mechanical two-body contact interaction, one should be able to completely control the scattering length and, absent other interactions, have that scattering length completely describe the scattering.
With such an interaction in hand, a variety of interesting many-body problems are unlocked.
Since all other dimensionful quantities are gone, all observables must be determined by naive dimensional analysis in the density, times some non-perturbative numerical factor, such as the Bertsch parameter\cite{PhysRevC.60.054311} in the case of the energy density.

In fact, a contact interaction can be shown to always produce momentum-independent scattering amplitudes (in three dimensions, for example, a momentum-independent $p \cot \delta$), and it ought to be possible to produce any amplitude, unless otherwise restricted by the Wigner bound\cite{Wigner:1955zz,Phillips:1996ae,Hammer:2010fw}.

Such scale-free results must result from peculiar potentials.
In three dimensions, for example, a delta function potential requires regulation, and to get scale-free dynamics its dimensionful strength must be sent to zero with the removal of the regulator in just such a way as to keep the phase shift at $\pi/2$.
In one dimension the strength of the contact interaction is also dimensionful and a delta function potential needs no regulation, but nevertheless is regulated when space is discretized; in two dimensions the strength of the delta function potential is dimensionless, which entails a more complicated story we discuss in \Secref{2D}.

Numerical computations are often performed in discretized boxes with periodic boundary conditions.
\Luscher's finite-volume formalism\cite{Hamber198399,luscher:1986I,luscher:1986II,wiese1989,Luscher1991,Luscher1991237} is the method by which one can extract infinite-volume real-time scattering data from the finite-volume Euclidean spectrum of a theory, taking advantage of the interplay between the physical scattering and the finite-volume boundary conditions in determining the spectrum.  Recently there has been an investigation of \Luscher's formalism for continuous scattering within a crystal lattice~\cite{Valiente:2015oya}.

The usual understanding of \Luscher's formalism is that one should find the continuum zero-temperature finite-volume energy levels, holding the physical volume fixed, and put that cold, continuum spectrum through \Luscher's formula to extract continuum scattering data.

Understanding the continuum limit of observables is important as it is shown in \Ref{Seki:2005ns} that, in the infinite-volume limit, lattice artifacts induce terms in the scattering data.
In practice, few results of lattice QCD calculations are zero-temperature- or, more seriously, continuum-extrapolated, but are nevertheless put through \Luscher's formula to get an estimate of the continuum scattering data, assuming thermal and discretization effects to be much smaller than the statistical uncertainties.
In particular, to date no continuum-limit study of any baryonic channel exists, even at unphysically heavy pion masses.

While alternatives, including the potential method (\Refs{Ishii:2006ec,Nemura:2008sp,Aoki:2009ji,Murano:2011nz,Aoki:2012bb,Kurth:2013tua,Sugiura:2017vwo,Yamazaki:2019vid,Aoki:2017yru,Yamazaki:2018qut,Iritani:2017rlk,Iritani:2018zbt,Gongyo:2018gou,Akahoshi:2019klc,Namekawa:2019xiy}), the mapping onto harmonic oscillators (\Ref{McElvain:2019ltw}) and the imposition of spherical walls (\Refs{Borasoy:2007vy,Borasoy:2007vi,Lee:2008fa,Epelbaum:2008vj,Epelbaum:2010xt,Lu:2015riz,Elhatisari:2015iga,Elhatisari:2016owd,Elhatisari:2016hby,Klein:2018lqz,Li:2019ldq,Bovermann:2019jbt,Lahde:2019npb}), can be used to translate finite-volume physics to infinite-volume observables, here we focus on the \Luscher finite-volume formalism.
Moreover, to our knowledge, no numerical work leveraging these methods is in the continuum, either.

Here, we construct example Hamiltonians explicitly and diagonalize them exactly, albeit numerically.
This allows us to circumvent all of the issues of statistical uncertainty that accompanies Monte Carlo data, and lets us completely isolate the features of the formalism itself, removing, for example, any finite-temperature effects that should in principle be extrapolated away in any finite-temperature method like Lattice QCD.

We find that it is in practice difficult to reliably extrapolate the spectrum to the continuum limit in a way that reproduces the exact known result, but that taking the continuum limit of the lattice-artifact-contaminated phase shifts sometimes can produce a more reliable result.

Extending the work of \Ref{Seki:2005ns} to finite volume, our main innovation, however, is to explain how to incorporate lattice artifacts into \Luscher's formula, for systems described by a contact interaction, accounting both for the Brillouin zone of the lattice and the lattice-induced dispersion relation.

While not universal, this lattice improvement can be quite useful for a contact interaction.
In pursuit of a lattice formulation of unitary fermions, the authors of \Ref{Endres:2011er} followed the tuning procedure of \Ref{Lee:2007ae}, parametrizing the contact interaction as a sum of a tower of Galilean-invariant operators, tuning their coefficients so as to drive the lowest interacting energy levels to the zeros of the \Luscher finite-volume zeta function.
However, in \Ref{Endres:2012cw} they found that even with a highly-improved construction the states ultimately deviated from a $\pi/2$ phase shift (see, for example, Figure 3).
In \Ref{He:2019ipt} the lattice implementation was smeared to reduce errors due to discretization, however a direct comparison of other methods with theirs was not possible for us since we were not able to identify the discretization parameters for the presented phase shifts (Fig. 7).

We introduce a new continuum-limit prescription for achieving unitarity in lattice simulations by tuning just the simplest, unsmeared contact operator, but taking the discretization effects into account by incorporating the lattice dispersion relation into the finite-volume zeta function, both in the tuning step and in the analysis step.
By re-tuning the interaction at each lattice spacing we can very easily and smoothly take the continuum limit after applying the lattice-aware finite-volume formula.
We demonstrate that this allows us to maintain a constant phase shift deep into the spectrum, covering as many \Aoneg states as exist in the lattice of interest.

This paper is organized as follow.  In \Secref{scattering} we give a brief summary of two particle scattering in $D$ dimensions.
In \Secref{hamiltonian} we give specifics about the latticized contact-interaction Hamiltonians we study numerically.
In \Secref{luescher} we provide a traditional continuum derivation of \Luscher's formula and in \Secref{dispersion} explain how to adapt it to include finite spacing effects by truncating the usual sum to just the momentum modes in the lattice and incorporating the dispersion relation into the appropriate propagators, yielding a lattice-improved generalized \Luscher zeta function.

Then, we leverage our dispersion zeta function, studying concrete examples.
In \Secref{3D} we study the three-dimensional case.
First we compare a continuum-extrapolated energy spectrum fed through the continuum zeta function and the continuum extrapolation of the finite-spacing spectra fed through the continuum zeta.
In \Secref{3D} we tune and analyze the same problem using our lattice-aware dispersion zeta function, and show that the resulting scattering $p\cot\delta$ remains constant deep into the spectrum; when we tune to unitarity the results stay at the expected value as accurately as the initial tuning is made modulo propagated numerical uncertainties.
We then study the one dimensional case in \Secref{1D}, where the absence of a counterterm makes things particularly simple.
In \Secref{2D} we repeat the story for the more intricate two-dimensional case, where here dimensional transmutation and logarithmic singularities require special attention and care.  Such a case was originally considered in~\cite{Fiebig:1994qi}, and subsequently worked out in detail for the s-wave case in~\cite{Beane:2010ny}.
We find that our lattice-aware \Luscher function handles this case with no difficulty.
Further, in all dimensions considered here we provide correction terms that come about when using energies calculated in a discrete space but fed through continuum \Luscher formula, which when applied to three dimensions corrects for the deviation found in \Ref{Endres:2012cw}.
Our corrections are valid only for the case of a contact interaction.
Finally, we recapitulate our findings in \Secref{conclusion} and discuss future directions.
We provide the data used for this publication and the code which generated the data in \Ref{luescher-nd_201}

\section{Two-particle scattering}\label{sec:scattering}

Two non-relativistic particles interacting via a contact interaction of strength $C$ in $D$ dimensions are described by the Hamiltonian
\begin{equation}
    \label{eq:particle hamiltonian}
    \hat H = \frac{\hat p_1^2}{2 m_1} + \frac{\hat p_2^2}{2 m_2} + C \delta^D(\hat x_1 - \hat x_2)
    \,,
\end{equation}
where the subscripts identify the particle of the position and momentum operators.
Moving to center-of-mass and relative coordinates, this Hamiltonian may be rewritten
\begin{equation}
    \label{eq:hamiltonian}
    \hat H = \frac{\hat P^2}{2 M} + \frac{\hat p^2}{2 \mu} + C\delta^D(\hat{x})
\end{equation}
where capital letters represent center-of-mass variables, lower case implies relative coordinates, and $\mu$ is the reduced mass.
Specializing to the center of mass frame by setting $P=0$ we reduce the problem to an effective one-body quantum mechanics in an external delta-function potential.

For a general two-body interaction $V$ in $D$ dimensions we can obtain scattering data by solving the Lippmann-Schwinger equation,
\begin{align}
	T_D(\vec p', \vec p, E)
	&=
	V(\vec p', \vec p) + \lim\limits_{\epsilon \to 0}\int \frac{d \vec k^D}{(2\pi)^D} V(\vec p', \vec k) G(\vec k, E + i \epsilon) T(\vec k, \vec p, E) \, ,
	&
	G(\vec k, E+ i \epsilon) = \frac{1}{E + i \epsilon - \frac{k^2}{2\mu}}
	\, .
\end{align}
where $G$ is the free Green's function.
Projecting onto the set of partial waves in $D$ dimensions labelled by $\l$, the $T$ matrix may be re-expressed in terms of phase shifts.
For a central interaction like the contact interaction, partial waves do not mix and $\l$ labels the orbital angular momentum, which is conserved.
In this case, the phase shifts can be extracted from the scattering or $T$-matrix by
\begin{align}\label{eq:on-shell-T}
	\frac{1}{T_{D\l}(p)}
    \equiv
    \frac{1}{T_{D\l}(p, p, E_p)}
    = \frac{\mu}{2}
    \frac{1}{\mathcal F_{D\l}(p)} \left[\cot (\delta_{D\l}(p)) - i\right] \, ,
\end{align}
where $E_p = p^2 / (2 \mu)$ and $\mathcal F_{l D}(p)$ is a dimension-dependent kinematic function of the on-shell momentum.

At low energy one often considers the expansion of \eqref{on-shell-T} in scattering momentum $p$, called the effective range expansion (ERE), which takes the form \cite{Hammer:2010fw}
\begin{align}
    \label{eq:ere}
    \cot \left(\delta_{D\l}(p)\right)
    &=
    \theta_D \frac{2}{\pi}  \ln \left(p R_{D\l}\right)
    -
    \frac{1}{a_{D\l}} p^{2 - 2 \l - D} +\frac{1}{2} r_{D\l} p^{4 - 2 \l - D} + \order{p^{6 - 2 \l - D}}
    \, , &
    \theta_D &= \begin{cases}
        0 & D \;\text{odd} \\ 1 & D \;\text{even}
    \end{cases}
    \, ,
\end{align}
where $R_{D \l}$ is an arbitrary length scale that enters in even dimensions and $a_{D\l}$, $r_{D\l}$ and subsequent higher-order coefficients describe the properties of the two-particle interaction.
In three spatial dimensions, the S-wave phase shift is described by the \emph{scattering length} $a_{30}$, the \emph{effective range} $r_{30}$ and further shape parameters.

In this paper we refer to $a$ as the scattering length and $r$ the effective range, even when, by simple dimensional analysis, they may not be actual lengths.
Moreover, in this work we will focus on the S-wave or its $D$-dimensional equivalent partial wave for simplicity, and henceforth suppress the $\l$ label
\begin{align}
	\delta_{D} &\equiv \delta_{D0}\, , &
	a_{D} &\equiv a_{D0}\, , &
	r_{D} &\equiv r_{D0}\, , &
	\cdots &
	\,
\end{align}
We work in three, two, and one spatial dimension.

Contact interactions, which are analytically tractable, correspond to a momentum-independent scattering amplitude when properly renormalized (as long as the log dependence is handled carefully in even dimensions).
So, the strength of the contact interaction $C$ may be traded for the scattering length $a$ and all other scattering parameters vanish.
The lattice interactions we will construct, when analyzed appropriately, will exhibit this momentum independence.

\section{Discretized Hamiltonian}\label{sec:hamiltonian}

We consider a cubic finite volume (FV) of linear size $L$ with periodic boundary conditions and lattice spacing $\epsilon$ so that $N=L/\epsilon$ is an even integer that counts the number of sites in one spatial direction.

The contact interaction Hamiltonian \eqref{hamiltonian} is implemented on the lattice as an entirely local operator, vanishing everywhere except at the origin where it is of strength $C$---the interaction is not smeared.
The Hamiltonian is given by
\begin{equation}
    \left\langle \vec{r}' \middle| H \middle| \vec{r} \right\rangle
    \rightarrow
    H_{\vec{r}',\vec{r}}^\dispersion
    =
    \frac{1}{2\mu} K_{\vec{r}',\vec{r}}^\dispersion + \frac{1}{\epsilon^D} C^\dispersion \delta_{\vec{r}',\vec{r}} \delta_{\vec{r},\vec{0}}
\end{equation}
where $K$ is a discretized Laplacian, implementing the momentum squared.
The $\dispersion$ symbol indicates that quantities depend on the lattice spacing $\epsilon$ and the explicit implementation of discretization effects like derivatives.

To ensure we control the discretization effects in generality, we study a variety of kinetic operators $K_{xy}^\dispersion$.
An often-used set of finite-difference kinetic operators are constructed from the one-dimensional finite-difference Laplacian that reaches $\nstep$ nearest neighbors,
\begin{equation}
    \Laplacian^\dispersion_{r'r} = \frac{1}{\epsilon^2}\sum_{s=-\nstep}^{\nstep} \gamma_{\abs{s}}^{(\nstep)} \delta_{r',r+\epsilon s}^{(L)}
\end{equation}
where the $(L)$ index of the Kronecker delta indicates that the spatial indices are understood modulo the periodic boundary conditions of the lattice.
In $D$ dimensions we simply take on-axis finite differences, so that the Laplacian is a $(1+2\nstep D)$-point stencil
\begin{equation}\label{eq:kinetic}
    K_{\vec{r}',\vec{r}}^\dispersion(\epsilon)
    =
    - \sum_{d=1}^D \Laplacian_{r_d'r_d^{}}^\dispersion
    \, .
\end{equation}

In the Fourier transformed space, momentum space, the one-dimensional Laplacian may be written
\begin{equation}
    \label{eq:laplacian}
    -\Laplacian_{r'r^{}}^\dispersion
    \overset{\text{F.T.}}{\longleftrightarrow}
    \Laplacian^\dispersion_{p'p}
    =
    \frac{1}{\epsilon^2}
    \delta_{p'p}
    \sum_{s=0}^{\nstep} \gamma_s^{(\nstep)} \cos(s p \epsilon)
    \, , \qquad p = \frac{2 \pi}{L} n \,
\end{equation}
where $n$ is an integer.
In $D$ dimensions we just sum the same expression over the different components of momentum.
Note that this is a specialization, in the sense that it contains no off-axis differencing (in position space) or products of different components (in momentum space).
However, since the numerical formalism we will describe is valid for every $\nstep$, we believe it holds for every possible kinetic operator.

\begin{figure}
    \input{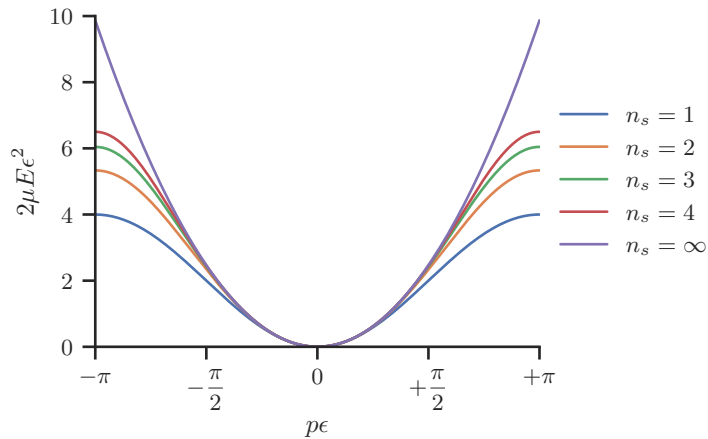}
    \caption{We show the continuum dispersion relation of energy as a function of momentum for different one-dimensional $\nstep$ derivatives.  For a finite number of lattice points $N$, the allowed momenta are evenly-spaced in steps of $2\pi/N$.
    As additional steps are incorporated into the finite difference, the dispersion relation more and more faithfully reproduces the desired $p^2$~behavior of $\nstep=\infty$.
    }
    \label{fig:dispersion relation}
\end{figure}

The coefficients $\gamma_{s}^{(\nstep)}$ are determined by requiring the dispersion relation be as quadratic as possible,
\begin{equation}
    \label{eq:gamma determination}
    \Laplacian^\dispersion_{p'p}
    \overset{!}{=}
    \delta_{p'p} \;
    p^2 \left[
        1 + \order{(\epsilon p)^{2\nstep}}
    \right].
\end{equation}
Additionally, we study a nonlocal operator with $\nstep=\infty$ which, in momentum space, can be implemented by multiplying by $p^2$ directly,
\begin{equation}
    \lim\limits_{n_s \to \infty}
    \Laplacian^\dispersion_{p'p}
    =
    \delta_{p'p} p^2,
\end{equation}
including at the edge of the Brillouin zone, the Laplacian implementation of the ungauged SLAC derivative.
Including the edge of the Brillouin zone does not introduce a discontinuity at the boundary, nor does including the corners pose any problem.
In addition to the $\nstep=\infty$ operator, we also call this kinetic operator the \emph{exact-$p^2$} operator.
The resulting dispersion relations are presented in \Figref{dispersion relation} for a variety of $\nstep$s and in \Appref{coefficients} we collect the required $\gamma$ coefficients.
In \Refs{Endres:2011er,Endres:2012cw} the exact dispersion relation is cut off by a LEGO sphere in momentum space (see equation (6) and the discussion after (9) in those references, respectively).
The formalism we develop here takes into account the implemented dispersion relation and thus is in principle extendable to these cut off operators, though the analytic results are harder to extract and we do not discuss such operators further.

The Hamiltonian in momentum space reads
\begin{equation}
    \label{eq:p space hamiltonian}
    \left\langle \vec{p}' \middle| H \middle| \vec{p} \right\rangle
    \rightarrow
    H_{\vec{p}',\vec{p}}^\dispersion
    =
    \frac{4 \pi^2}{2\mu L^2} \tilde K_{\vec n \vec n}^{N}
    +\frac{1}{L^D}C^\dispersion
\end{equation}
where $\vec{p} = 2\pi \vec{n}/L$ for a $D$-plet of integers $\vec{n} \in (-N/2, +N/2]^D$, and the coefficients $\gamma_{s}^{(\nstep)}$ are determined as described above.
Furthermore, we replaced the lattice-spacing-dependent kinetic Hamiltonian with the $N$-dependent
\begin{equation}\label{eq:normalized-kinetic-hamitlonian}
	\tilde K_{\vec n \vec n}^{N}
	= \frac{L^2}{4\pi^2} K_{\vec p \vec p}^{\dispersion} \bigg|_{\vec p=\frac{2\pi \vec n}{L}}
	= \frac{N^2}{4\pi^2}
    \sum_{i=1}^{D}\sum_{s=0}^{\nstep} \gamma_s^{(\nstep)} \cos\left(\frac{2 \pi s n_i}{N}\right)
\end{equation}
which goes to $n^2$ in the continuum limit $N\goesto\infty$.

Although the non-interacting energy levels are no longer proportional to $n^2$ at generic \nstep, $n^2$ is still a useful classification for states, as long as it is understood simply as the magnitude of the lattice momentum---describing shells---rather than as a proxy for energy.

\subsection{Reduction to  \texorpdfstring{\Aoneg}{A-one-g}}

Because we are interested in contact interactions, infinite-volume arguments suggest that only the s-wave will feel the interaction; such arguments translate to the lattice relatively cleanly.
Since the s-wave is most like \Aoneg we will focus on the spectrum in that irreducible representation of the cubic symmetry group $O_h$ in three dimensions, of the symmetry group of the square $D_{4h}$ in two dimensions, or $Z_2$ in one dimension, where an \Aoneg restriction amounts to focusing on parity-even states.

With a projection operator to the \Aoneg sector $P_{\Aoneg}$ we can raise the energy of all the other states an arbitrary amount $\alpha$ by supplementing the Hamiltonian
\begin{equation}
    H(\alpha) = H + \alpha (\one - P_{\Aoneg}) \, ,
\end{equation}
Because $P_{\Aoneg}$ commutes with $H$, $H$ and $H(\alpha)$ have the same spectrum within the $\Aoneg$ irrep.
If $\alpha$ is much larger than the expected energies of the Hamiltonian, the \Aoneg states remain low-lying and all other states are shifted to much higher energies.
Then, exact diagonalization for low-lying eigenvalues of $H(\alpha)$ provides an easier extraction of \Aoneg eigenenergies.

Because of the simplicity of \Aoneg we can also easily construct the Hamiltonian directly in that sector (a construction for general $O_h$ irreps was recently given in \Ref{Li:2019qvh}).
In momentum space we can label plane wave states by a vector on integers $\vec{n}$.
In the \Aoneg basis we can use one plane wave label and understand that we intend a normalized unweighted average of every plane wave state.
That is,
\begin{equation}
    \ket{\Aoneg\; \vec{n}} = \frac{1}{\sqrt{\normalization}} \sum_{g \in O_h} \ket{ g\vec{n}}
\end{equation}
where $g$ is an element of the group $O_h$, the sum is over all inequivalent states, and $\normalization$ the normalization.
When $\vec{n}$ is large we should be careful not to double-count states that live right on the edge of the Brillouin zone.
The states may be labeled by symmetry-inequivalent vectors with components all as large as $N/2$.
As a simple example, in three dimensions the $N/2(1,1,1)$ plane wave state in one corner of the Brillouin zone is invariant under all the $O_h$ operations modulo periodicity in momentum space, so $\normalization=1$ for that state.

Formulated in this basis, the kinetic energy operator remains diagonal and proportional to $n^2$ when $N\goesto\infty$.
Reading off the momentum-state potential matrix element from \eqref{p space hamiltonian}, the contact interaction is given by
\begin{equation}
    \braMket{\Aoneg\; \vec{n}'}{V}{\Aoneg\; \vec{n}}
    =
    \sum_{g'g \in O_h}
        \frac{1}{\sqrt{\normalization'}\sqrt{\normalization}} \braMket{g'\vec{n}'}{V}{g\vec{n}}
    =
    \frac{C^\dispersion}{L^D} \sqrt{\normalization'\normalization},
\end{equation}
so that every \Aoneg state talks to every other.
So, the Hamiltonian is in this sector is
\begin{equation}
    H_{\vec{n}'\vec{n}}^\dispersion = \frac{4 \pi^2}{2\mu L^2} \tilde K_{\vec n \vec n}^{N} + \frac{C^\dispersion}{L^D} \sqrt{\normalization'\normalization}
\end{equation}
and we divide by $4\pi^2/\mu L^2$ to make everything dimensionless.

We have implemented both this \Aoneg-only Hamiltonian and the general Hamiltonian with an energy penalty for non-\Aoneg states and verified that the spectra match where expected to as much precision as desired.

For a given $N$ multiple momenta inequivalent under the $O_h$ symmetry may have the same $n^2$.
For example, when $N\geq5$ there are two $n^2=9$ shells corresponding to $n=(2,2,1)$ and $n=(3,0,0)$, which lives on the edge of the Brillouin zone for $N=5$.
When $\nstep=\infty$ the corresponding non-interacting eigenstates are degenerate, while with imperfect dispersion relations the degeneracy is, generically, lifted.
For the contact interaction and $\nstep=\infty$, one linear combination of these \Aoneg states overlaps the $S$-wave and has a nontrivial finite-spacing finite-volume energy, and the other overlaps a higher partial wave and has $x^{\dispersion}=2\mu E^{\dispersion}L^2/4\pi^2=9$ to machine precision, sitting right on a pole of the \Luscher zeta function~\eqref{spherical S}.
In contrast, when $N=4$ there is no $n=(3,0,0)$ state, and the $(2,2,1)$ state is itself an eigenstate.
When $N$ is very large sometimes there are multiple eigenstates that have no support for the delta function---$n^2=41, 50, 54\ldots$ have two non-interacting states, while $n^2=81, 89, 101\ldots$ have three non-interacting states, and $n^2=146$ is the first shell with four non-interacting states, for example.
After diagonalizing, we exclude these non-interacting \Aoneg states from our analysis.
We do not discuss these non-interacting states further and omit them from figures without comment.

\section{\Luscher's Formulae}\label{sec:luescher}

In subsequent sections we will extract scattering data from numerical calculations for particular box sizes and discretizations.
We will show that when tuned and analyzed using the traditional \Luscher method, we induce a momentum-dependent scattering amplitude at any finite lattice spacing and explain how to achieve a momentum-independent scattering amplitude, even at finite lattice spacing, by constructing a lattice-aware \Luscher-like method.

For concreteness of our discussion we here provide a derivation of \Luscher's S-wave formula roughly following \Ref{Beane:2003da}, although the technology and sophistication of the finite-volume formalism has grown substantially \cite{Beane:2010ny,Ozaki:2012ce,Hansen:2012tf,Briceno:2013hya,Briceno:2013lba,Li:2014wga,Zhu:2019dho}. What differentiates our derivation from others is our ensuing lattice spacing-corrected procedure.

\subsection{Continuum Procedure}\label{sec:continuum}
\begin{figure}[ht!]
\center
\includegraphics[width=.675\columnwidth]{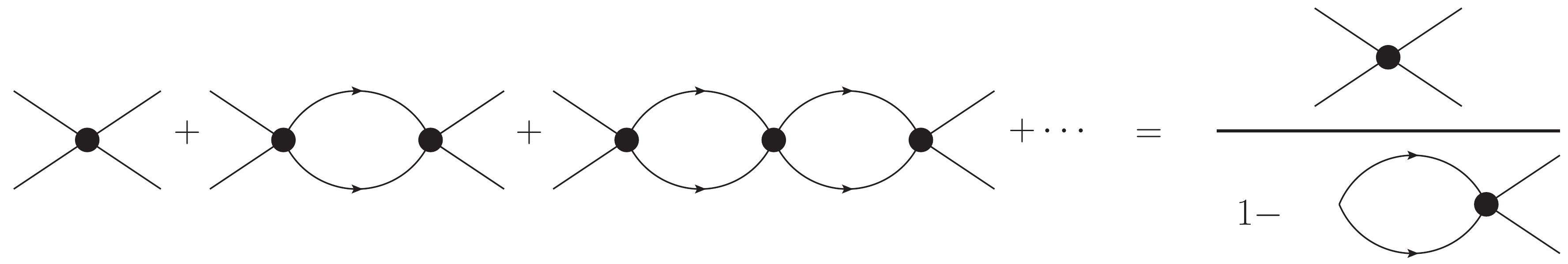}
\hfill
\includegraphics[width=.275\columnwidth]{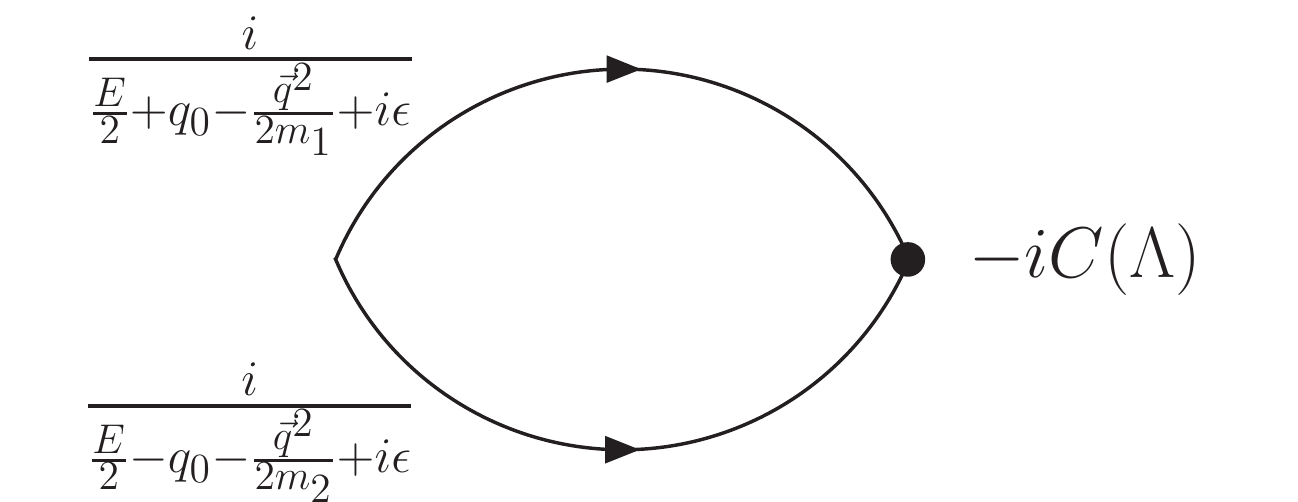}
\caption{(Left) The bubble sum. Each line represents a propagator, each vertex represents $-i C(\Lambda)$, and the bubble is given by $I_D$.
(Right) The single loop diagram needed to calculate $I_D$ in the bubble sum.
\label{fig:bubbleSum}}
\end{figure}

The starting point is a contact interaction\footnote{This derivation generalizes to a tower of contact interactions where $C(\Lambda)$ is replaced by $\sum_n C_{2n}(\Lambda) p^{2n}$ \cite{Kaplan:1998we,Beane:2003da} and dimensional regularization is used to absorb power-law divergencies.} such that the tree amplitude in the center of mass frame is given by
\begin{equation}
    \mathcal A(\Lambda) = + i C(\Lambda)
\end{equation}
where $p$ denotes the relative momentum of incoming nucleons and the interaction strengths $ C(\Lambda)$ depend on the regulator $\Lambda$ and carry dimension-dependent units.
The scattering amplitude is given by the bubble sum depicted in \Figref{bubbleSum}.

This bubble sum is a geometric series and, restricting our attention to the contact interaction causes all other partial wave than the S-wave to vanish.
This restriction gives for the standard on-shell $T$-matrix
\begin{equation}\label{eq:T matrix}
T_{D\l}(p, \Lambda) = \delta_{\l 0} \frac{C(\Lambda)}{1-I_D(p,\Lambda) C(\Lambda)},
\end{equation}
where $p$ is the relative on-shell momentum, $\Lambda$ the regularization scale.
The physical result for the T-matrix is recovered once the parameter $C$ is chosen such that one can remove the regularization scale---in the limit of $\Lambda \to \infty$ for a hard momentum cutoff, for example.

$I_D(p,\Lambda)$ is a $D$-dependent function that arises from integrating the loop shown in the right panel of \Figref{bubbleSum},
\begin{align}
    I_D(p, \Lambda)
    &=-i\int^{\Lambda}
        \frac { \mathrm {d}q_0}{2\pi}\ \frac{\mathrm { d } ^ { D } \vec{ q } } { (2\pi)^ { D } }
        \left( \frac { i } { \frac{E}{2} + q _ { 0 } - \frac{\vec{q}^2}{2m_1} + i \epsilon } \right)
        \left( \frac { i } { \frac{E}{2} - q _ { 0 } - \frac{\vec{q}^2}{2m_2} + i \epsilon } \right)
    \label{eq:I0 in two particle language}\\
    &=\frac{\Omega_D}{(2\pi)^D}\int^{\Lambda}  \mathrm { d } q \ q^{D-1}\left[\PV \left( \frac { 1 } { E - \frac{\vec{q}^2}{2\mu} } \right)
-i\frac{\pi \mu}{q}\delta(q-\sqrt{2 \mu E})\right]
    \label{eq:I0 in relative coordinates}
    \\
    &=\frac{\Omega_D}{(2\pi)^2}\frac{2\mu}{L^{D-2}}\int^{\Lambda L/2\pi}  \mathrm { d } n \ n^{D-1}\left[\PV \left( \frac { 1 } { \left(\frac{pL}{2\pi}\right)^2 - n^2 } \right)
-i\frac{\pi^2}{L n}\delta\left(\frac{2\pi}{L}n -p\right)\right]
    \label{eq:I0}
\end{align}
where $\PV$ refers to Principal (Cauchy) Value, we have used the on-shell condition $2\mu E=p^2$, and the geometric factor
\begin{equation}
\Omega_D=\frac{2\pi^{D/2}}{\Gamma(D/2)}=
    \begin{cases}
    	2       &   (D=1)\\
		2\pi    &   (D=2)\\
        4\pi    &   (D=3)
    \end{cases}\ ,
\end{equation}
accounts for the angular integration in $D$ dimensions.

Because we are focusing on the contact interaction, we can restrict our attention to the $s$-wave, $\l=0$.
Dropping the $\l$ dependence in \eqref{on-shell-T}, the momentum-dependent $T$-matrix is related to the phase shift when
\begin{equation}\label{eq:spherical FD}
    \F_{D}(p)
    \equiv
    \begin{cases}
        p/2     & (D=1)\\
        1       & (D=2)\\
        \pi/p   & (D=3)\\
        \vdots  & \vdots
\end{cases}
\end{equation}
is a dimension-dependent kinematic factor determined by requiring the imaginary parts of the $T$-matrix \eqref{T matrix} from the bubble sum \eqref{I0 in relative coordinates} exactly matches the imaginary part of the amplitude \eqref{on-shell-T}.
This fixes the coefficients $C(\Lambda)$ as a function of the scattering data,
\begin{equation}\label{eq:IV pole}
    \frac{\mu}{2 \F_{D}(p)}\left(\cot \delta_{D}(p) - i\right)
    =
    \lim\limits_{\Lambda \to \infty} \left[ I_D(p, \Lambda) - \frac{1}{C(\Lambda)} \right].
\end{equation}

In a finite volume, the energy eigenstates $E$ appear at poles of the $T$-matrix, so that
\begin{equation}\label{eq:FV pole}
    \frac{1}{2\mu E C_\FV(\Lambda) } - I_{D, \FV}(\sqrt{2\mu E}, \Lambda) = 0
\end{equation}
and the infinite-volume integral $I_D$ has been replaced by the matching finite-volume sum which introduces another scale $L$,
\begin{align}
I_{D,\FV}(\sqrt{2\mu E}, \Lambda)
    &=-i\int \frac { \mathrm {d}q_0}{2\pi} \frac{1}{L^D}\sum_{\vec{q}}^{q < \Lambda} \left( \frac { i } { \frac{E}{2} + q _ { 0 } - \frac{\vec{q}^2}{2m_1} + i \epsilon } \right) \left( \frac { i } { \frac{E}{2} - q _ { 0 } - \frac{\vec{q}^2}{2m_2} + i \epsilon } \right)
    \\
    \label{eq:I0 FV}
    &=\frac{1}{L^D}\sum_{\vec{q}}^{q < \Lambda} \frac { 1 } { E - \frac{\vec{q}^2}{2\mu} }
    =\frac{2\mu}{(2\pi)^2 L^{D-2}} \sum_{\vec{n}}^{n < \frac{\Lambda L}{2\pi}} \frac{1}{x-n^2}
    &
    x &= \frac{2\mu E L^2}{4\pi^2}
    \, .
\end{align}
Combining the infinite-volume and finite-volume relations \eqref{IV pole} and \eqref{FV pole} yields
\begin{equation}\label{eq:spherical zeta}
    \frac{\mu}{2\F_{D}(\sqrt{2\mu E})}(\cot\delta_{D}(\sqrt{2\mu E})-i)
    =
    \lim\limits_{\Lambda \to \infty} \left[ I_D(\sqrt{2\mu E}) - I_{D,\FV}(\sqrt{2\mu E}) \right] \, ,
\end{equation}
the finite-volume quantization condition.
Note that both equations are explicitly evaluated for the same interactions $C_\FV(\Lambda) = C(\Lambda)$ independent of the volume $L$ and using the same regulator.
Furthermore \eqref{spherical zeta} is only valid if evaluated at momenta corresponding to finite-volume eigenenergies $E$.

Plugging our results for the integrals in, one finds
\begin{multline}
    \frac{1}{2\F_{D}(\sqrt{2\mu E})}\left(\cot \delta_{D}(\sqrt{2\mu E}) - i\right)
    =\\
    \frac{2}{(2\pi)^2 L^{D-2}}
    \lim\limits_{\Lambda \to \infty}
    \left[
    	\left(\mathcal{P}\int_{\vec{n}} - \sum_{\vec{n}}\right) \frac{1}{x-n^2} +
		\frac{-i \pi^2\Omega_D}{L} \int \mathrm{d}n\ n^{D-2} \delta\left(\frac{2\pi}{L}n - \sqrt{2\mu E}\right)
	\right]
\end{multline}
where both the sum and integral are cut off by a restriction on the magnitude of $n$, $n^2 < (\Lambda L / 2\pi)^2$,
The principle value integration implicitly carries a factor of $\Omega_D n^{D-1}$ (see \eqref{I0}).
The imaginary part on the left hand side exactly cancels the last term on the right when $E\ge0$.  When $E<0$ the last term on the RHS vanishes and so we have
\begin{multline}\label{eq:general luscher}
   \frac{1}{2\F_{D}(\sqrt{2\mu E})}  \left( \cot \delta_{D}(p)-i\theta(-E)\right)
    =
   \frac{2}{(2\pi)^2 L^{D-2}}
    \lim\limits_{\Lambda \to \infty}\left(\sum_{\vec{n}}-\mathcal{P}\int_{\vec{n}}\right) \frac{1}{n^2-x}\\
    \implies
      \cot \delta_{D}(p)= \frac{\F_{D}(\sqrt{2\mu E})}{\pi^2 L^{D-2}}\left[
    \lim\limits_{\Lambda \to \infty}\left(\sum_{\vec{n}}-\mathcal{P}\int_{\vec{n}}\right) \frac{1}{n^2-x}\right]
    +i\theta(-x)
    \ ,
\end{multline}
with $x$ as in \eqref{I0 FV}, $\theta(x)$ is the heavyside function, and we switched the sign of the sum and integral as well as the sign of the denominator.  In the second line above we moved the term proportional to the $\theta(-E)$ to the RHS.
Because we cut off the sum and the integral in exactly the same way, in dimensions where $I_D$ diverges with $\Lambda$, the divergence cancels against the divergence in the sum.
Let $N=\Lambda L/\pi$.
Then, with a finite cutoff on magnitude $N/2$, we define
\begin{equation}\label{eq:spherical cutoff S}
    S^{\spherical N}_D(x) =
    \left(\sum_{\vec{n}}- \mathcal{P}\int_{\vec{n}}\right) \frac{1}{n^2-x}
    + i \frac{(2\pi)^D}{4 \F_D\left(\sqrt{x}\right)}\theta(-x)\ ,
\end{equation}
where it was used that $\F_D(p) \sim p^{2-D}$ and the $\spherical$ superscript reminds us that we cut off our sum and integral in a spherical way, based on the magnitude of $n<N/2$.   By performing the principal value integral and taking the limit $N\to \infty$, we recover the usual \Luscher zeta functions,
\begin{equation}\label{eq:spherical S}
    S^\spherical_D(x)
    =
    \lim_{N\goesto\infty} S^{\spherical N}_D(x)
    =
    \lim_{N\rightarrow\infty} \sum_{\vec{n}}^{n < N/2}
    \begin{cases}
     \frac{1}{n^2-x} - \counterterm_3^\spherical \frac{N}{2}& (D=3)\\
     \frac{1}{n^2-x} - 2\pi\log\left(\counterterm_2^\spherical\frac{N}{2}x^{-1/2}\right)& (D=2)\\
    \frac{1}{n^2-x} & (D=1)
     \end{cases}
\end{equation}
where the dimension-dependent coefficients $\counterterm_D^\spherical$ of the counterterms come from the principal value integral; we evaluate the spherical-cutoff integrals and extract these coefficients in \Appref{counterterm/spherical}.\footnote{In higher dimensions there will be additional divergences which cancel, for example, in five spatial dimensions there will be a cubic and linear divergence.
}
Finally, we can write the quantization condition~\eqref{general luscher} using the zeta function~\eqref{spherical S},
\begin{equation}\label{eq:spherical quantization}
    \cot \delta_{D}(p) = \frac{\F_{D}(p)}{\pi^2 L^{D-2}} S^\spherical_D(x)
\end{equation}
where we traded the energy dependence for momentum on the left-hand side.  Our result is consistent with those given in~\Ref{Zhu:2019dho}\footnote{In~\Ref{Zhu:2019dho} the zeta functions~\eqref{spherical cutoff S} were defined \emph{without} the term proportional to the heavyside function.  Thus their zeta functions have a different behavior for $x<0$ as ours.  We note that our definition is more common in the literature.}.
This is the \Luscher finite-volume quantization condition, and finite-volume energy levels calculated in the continuum should be fed through it to produce continuum scattering data.
In three dimensions it is common to move the momentum dependence in $\F_{D}$ to the other side, as $p \cot\delta_{D}(p)$ is what appears in the effective range expansion \eqref{ere}.
In two dimensions, it will prove useful to explicitly separate the logarithmic divergence as $N\to\infty$ from the logarithmic singularity as $x\to 0$, and we will rearrange this equation and slightly redefine $S^\spherical_2$ as needed in \Secref{2D}.  Finally, the sum in~\eqref{spherical S} can be analytically done in $D=1$, as we will show in \Secref{1D}.


To approach the continuum limit, the authors of \Ref{Lee:2007ae} proposed tuning the interaction until the ground state, when fed through $S^\spherical$, produced the desired amplitude that corresponds to the desired scattering length.
We will show in \Secref{3D} that this procedure induces a momentum dependence in the scattering amplitude sensitive to discretization.
In the next subsection we give a procedure that produces a momentum-independent amplitude as one approaches the continuum, and discuss the limiting procedure itself.

\subsection{The Dispersion Method}\label{sec:dispersion}

To correctly implement a theory in a finite basis, any observable in this basis must be correctly reproduced in the physical limit.
In case of a lattice theory, one of these limiting procedures is the continuum limit.
One sensible idea for taking the continuum limit is to tune theory parameters such that some lattice observables are held fixed at their continuum value for any lattice spacings.
By construction, these fixed observables recover their continuum value when sending the lattice spacing to zero.
Of course, observables will be infected by lattice artifacts, and so one must readjust the input parameters as one takes the limit.
If this implementation and tuning prescription is well defined, all additional lattice observables will converge in the continuum as well.

For example, in lattice QCD calculations, the continuum limit is sought by finding a line of constant physics where some parts of the single-hadron spectrum are held fixed as the continuum is approached.
Then, at any finite spacing, the hadron-hadron interactions are already determined by the finite-spacing of QCD itself, and the interaction one measures depends on the lattice spacing and approaches the correct interaction in the continuum.
A continuum limit of lattice QCD could, in principle, be taken along a line of constant deuteron-channel scattering length, but practical issues abound, even if simpler scattering channels like $I=2$ $\pi\pi$ scattering are picked instead.

In our setup, non-relativistic nucleons interacting through a contact interaction, the masses are set by hand and only the interaction parameter needs tuning.
Knowing that we must readjust the strength of our contact interaction as a function of lattice spacing raises the question of which observables to tune to.
Such observables can be scattering data, for example, but the interaction itself is not an observable.
One renormalization scheme is to hold one part of the scattering data, such as the scattering length, fixed and independent of lattice spacing.
As mentioned at the end of \Secref{continuum}, in this approach one effectively requires that the lowest energy state matches the desired scattering amplitude, when put through $S^{\spherical}$ (see \Refs{Endres:2011er,Lee:2007ae,Endres:2012cw}).
Tuned this way, one finds induced momentum dependence in the phase shift (see the $N_\mathcal{O}=1$ behavior of the left panel of Figure 2 of \Ref{Endres:2011er}, for example).

In this section we present a procedure for a contact interaction which ensures that computed phase shifts are at their physical value for each finite lattice spacing.
At each spacing we construct a lattice-aware generalized \Luscher zeta function $S^{\dispersion}$ which is used instead of the regular zeta function to tune the lowest energy at that spacing to the desired amplitude.
With that tuning accomplished, other finite-volume energy levels at the same spacing are extracted and analyzed using the spacing-appropriate $S^{\dispersion}$.
We will show that tuning and analysis with $S^{\dispersion}$ yields momentum-independent scattering for the simple lattice contact interaction described in \Secref{hamiltonian}.

To construct such a lattice-aware zeta function we return to the derivation of \Luscher's finite-volume formalism.
By recognizing that we're interested in incorporating lattice artifacts from the start, we replace the continuum dispersion relation with the lattice dispersion relation in the propagators and require that the integrals are cut off consistently---with a momentum cutoff that corresponds to that imposed by the lattice.
We replace $I_{D,FV}$ in the quantization condition \eqref{spherical zeta} with a lattice-aware substitute and match the finite-spacing finite-volume ground state to the continuum infinite volume scattering information using our lattice-aware zeta function.
This replacement result in
\begin{equation}
    \label{eq:finite spacing matching}
    \frac{\mu}{2 \F_D(\sqrt{2\mu E})}\left(\cot\delta_D(\sqrt{2\mu E})-i\right)
    =
    \lim_{\epsilon\goesto0}
    \left[
    	I_D^{\dispersion}(\sqrt{2\mu E}) - I_{D,FV}^{\dispersion}(\sqrt{2\mu E})
	\right]
\end{equation}
where $\F_D$ is the usual continuum kinematic factor \eqref{spherical FD}, $I_D^{\dispersion}$ is the cartesian version of \eqref{I0} term with $\vec{q}^2/2\mu$ replaced by the lattice dispersion relation,
\begin{align}
	\label{eq:dispersion I0}
    I_D^{\dispersion}(\sqrt{2 \mu E})
    &=
    \left(\prod_{i=1}^D
    \int\limits_{-\pi/\epsilon}^{+\pi/\epsilon}
    \frac{\mathrm{d} q_i}{2\pi}
    \right)
        \left[
            \PV \left(
                \frac{1}{
                    E - \frac{1}{2\mu} K_{qq}^\dispersion }
                \right)
            -i \pi \delta\left(E - \frac{1}{2\mu}K_{qq}^\dispersion\right)
        \right]
	\, .
\end{align}
The operator $K_{qq}^\dispersion$ is a momentum-space matrix element of the Laplacian (which, of course, is diagonal in momentum space), and the integral's cutoff $\Lambda$ in \eqref{I0 in relative coordinates} is taken to be $\pi/\epsilon$, matching the lattice's Brillouin zone.
We adopt dispersion $\dispersion \leftrightarrow (L, \epsilon, \nstep)$ superscripts to indicate the quantities are aware of the lattice (and discretization scheme if relevant).
Dispersion quantities need not only the range of momenta in the Brillouin zone (on a square lattice, each momentum component cut off independently), but also the spacing-aware dispersion relation $K$ (from \eqref{kinetic}, for example, though we emphasize other kinetic operators can be used).
The fact that $\F_D$ appears in \eqref{finite spacing matching} is reflected by evaluating the infinite volume $I_D^{\dispersion}$ in the continuum limit, so that the imaginary part of \eqref{dispersion I0} matches the continuum result from \eqref{I0 in relative coordinates}.
It is easy to see that when $\epsilon\goesto0$ the dispersion relation goes to the exact $p^2$ relation and the limits of the integral go to infinity so that we may execute the integral spherically and recover the continuum $\F_D$ in \eqref{spherical FD}.

To match the \Luscher like zeta function we rewrite the quantization condition as
\begin{align}\label{eq:dispersion-counter-integral}
    \cot \delta_D(\sqrt{2 \mu E})-i\theta\left(-E\right)
    &=
    \frac{\F_D(\sqrt{2 \mu E})}{\pi^2 L^{D-2}}
    \lim_{N\goesto\infty}
    \left[
    	\sum_{n\in\BZ} -
		\left(\prod_{i=1}^D
    		\int\limits_{-N/2}^{+N/2}
    		\mathrm{d} n_i
    	\right)\; \PV
	\right]\  \frac{1}{\tilde K_{nn}^{N}-x}
	\, ,
\end{align}
where we rescaled $q\goesto 2\pi n/L$ and replaced the dimension full hamiltonian with the normalized version \eqref{normalized-kinetic-hamitlonian}.
The limits of the integration are understood for each spatial direction independently, and the Brillouin zone (\BZ) runs over all the finite-volume lattice modes.
The $n$-dependent piece of the denominator goes to $n^2$ in the continuum limit (fixed $L$ with $N\goesto\infty$).
But even at finite spacing the denominator only depends on $N$ rather than $L$ and $\epsilon$, which follows from the Laplacians we study \eqref{laplacian} and the elimination of the dimensionful scale (the rescaling from $q$ to $n$).

We can construct, therefore, a \Luscher-like formalism,
\begin{align}
    \label{eq:dispersion quantization}
    \cot \delta_D( \sqrt{2 \mu E} )
    =
	\frac{\F_D(\sqrt{2 \mu E})}{\pi^2 L^{D-2}}
	\lim_{N\goesto\infty}
    \left[
    	\sum_{n\in\BZ}\frac{1}{ \tilde K_{nn}^{N} - x} - \counterterm_D^{\dispersion}\left(\frac{N}{2}\right)^{D-2}
		+ \mathcal O\left(\frac{x}{N}\right)
	\right]\, ,
\end{align}
where the above expression knows about the particular finite-differencing Laplacian or the dispersion relation as well as the discretization of the box into $N$ sites.
In contrast, in the usual finite-volume procedure, no UV details of the box infect the zeta function.
When taking $N$ to infinity, in three dimensions, the sum is divergent and the counter term exactly cancels the this growth; in one dimension there is no divergence to cancel, and we defer the discussion of two dimensions to \Secref{2D}.

There are two ways to view this equation.
First, in the continuum limit both expressions for the zeta function, the continuum-derived \eqref{spherical S} and the lattice-derived \eqref{dispersion quantization} are equivalent.
So, we simply have another way of approaching this limit.
Second, if it was possible to compute the exact error from lattice discretization and reincorporate it into the numerically-computed energy levels, one might leverage this difference to directly compute the physical phase shifts.
That is, numerically compute $x^\dispersion$, corresponding to energy level at finite spacing, and adjust it by a known $\delta x^\dispersion(x^\dispersion)$, so that one exactly lands on the continuum value: $x \equiv x^\dispersion + \delta x^\dispersion(x^\dispersion)$.
Were we to do that, then we would find
\begin{equation}
    \frac{1}{\pi L}S^\spherical_D(x)
    =
    \frac{1}{\pi L}S^\spherical_D\left(x^\dispersion+\delta x^\dispersion(x^\dispersion)\right)
    \equiv
    \frac{1}{\pi L}S^\dispersion_D\left(x^\dispersion\right)
\end{equation}
to be flat when evaluated on those adjusted $x$ values.

The structure of the contact interaction is such that it is also possible to analytically compute these shifts and incorporate them into a dispersion-aware zeta function.
Evaluated at a finite spacing we find
\begin{align}
    \cot \delta_D(\sqrt{2\mu E})
    &=
    \frac{\F_D(\sqrt{2\mu E})}{\pi^2 L^{D-2}} S^{\dispersion }_D\left(\frac{2\mu E^{\dispersion} L^2}{4\pi^2}\right)
    && \text{(No $N\goesto\infty$ limit!)}
    \\
    \label{eq:finite N}
    S^{\dispersion}_D\left(x^\dispersion\right)
    &=
		\sum_{n\in\BZ}\frac{1}{ \tilde K_{nn}^{N} - x^\dispersion} - \counterterm_D^{\dispersion}\left(\frac{N}{2}\right)^{D-2}
	\, ,
\end{align}
which is the zeta in \eqref{dispersion quantization} with the subleading $x$ dependence dropped, at finite $N$.
The sum is over a finite $N$ and the lattice energy levels are used to build $x^\dispersion$.
Unlike the continuum case, there is, strictly speaking, no divergence in the sum in \eqref{finite N}, because we are always interested in a real calculation performed with finite $N$.
Note that the zeta function we define in \eqref{finite N} differs from the expression derived in \eqref{dispersion quantization}, in that it does not include any $x/N$ effects that disappear in the continuum, and that it is valid to feed finite-spacing eigenenergies $x^\dispersion$ through the finite-$N$ formula \eqref{finite N}.
Plugging finite-spacing eigenenergies through the continuum formula induces a momentum dependence arising from the $x/N$ dependence in \eqref{dispersion quantization}---accounting for the seen momentum dependence that was shown to vanish towards the continuum in a variety of prior results.
That dependence is calculable for a contact interaction and is subtracted in our finite-spacing zeta function \eqref{finite N}.
We provide an explicit derivation in three dimensions in section \ref{sec:3D dispersion}.

We want to add further remarks:
\begin{itemize}
\item The quantization condition \eqref{dispersion quantization} can be thought of as \Luscher's zero-center-of-mass-momentum finite-volume formula non-perturbatively improved for discretization effects with our particular interaction.
To arrive at formulas for nonzero center of mass momentum is substantially more complicated, because only at zero center of mass momentum does the change from single-particle coordinates in \eqref{particle hamiltonian} to center-of-mass coordinates in \eqref{hamiltonian} commute with performing the spatial discretization, yielding the same dispersion relation in the effective one-body problem as in the two-body problem.
The ordering matters, as in a realistic many-body calculation (and in physical crystals!), each individual particle sees the lattice discretization.\footnote{
We note that the change to Jacobi coordinates commutes with the discretization of momenta if the dispersion relation is exactly equal to $p^2$ all the way up to the edge of the Brillouin zone ($\nstep=\infty$).}
To construct a lattice-improved finite-volume formula for two particles with finite center-of-mass momentum, one must backtrack even further, earlier than the effective one-body integral \eqref{dispersion I0}, to an equation more like the two-body loop diagram that determines $I_0$ \eqref{I0 in two particle language} before the energy integral is performed, replacing the single-particle dispersion relations there and changing the domain of integration to match the Brillouin zone.
We leave such a construction to future work.

\item In the usual case, the on-shell condition is leveraged to trade $2\mu E$ for the scattering momentum $p$.
However, with a finite lattice spacing the on-shell condition is not so simple to invert.
In fact, there are multiple momenta that all correspond to the same energy, because the lattice dispersion relation begins decreasing once the momentum leaves the lattice's first Brillouin zone, and the energy repeats indefinitely so that there are infinitely many momenta that correspond to that energy.

\item Leaving the dependence on energy alone and not the momentum allows us to account naturally for Umklapp scattering processes and the violation of crystal momentum conservation.
This would be important for capturing physics of physical crystals, were we need to match to an infinite volume lattice instead.
In this context one may define finite-spacing quantization condition through finite-spacing phase shifts according to
\begin{equation}
    \frac{\mu}{2 \F_D^\dispersion(\sqrt{2\mu E})}\left(\cot\delta_D^\dispersion(\sqrt{2\mu E})-i\right)
    =
    I_D^{\dispersion}(\sqrt{2\mu E}) - I_{D,FV}^{\dispersion}(\sqrt{2\mu E})
	\, .
\end{equation}
On the left-hand side of the quantization condition we get the infinite-volume $A_1^+$ phase shift\footnote{For a physical lattice there are UV breaking effects of rotational symmetry, so the irreps still do not carry angular momentum labels.} at scattering energy $E$ while on the right-hand size we need knowledge of the box size $L$, its lattice spacing $\epsilon$, as well as the finite-volume finite-spacing spectrum.
One may calculate a spacing-aware $\F_D^\dispersion$ by considering the imaginary part of the infinite-volume integral \eqref{dispersion I0}.
Unfortunately, achieving a closed-form expression for $\F_D^\dispersion$ is challenging though it is numerically tractable.
Matching to a real physical crystal requires formulating a spacing-dependent kinematic factor $\F_D^\dispersion$ from \eqref{dispersion I0} and keeping $N$ finite in the integral in the dispersion zeta function \eqref{dispersion quantization}, which introduces a whole tower of terms, each down by $N^2$, that vanish because we are matching to the continuum.
\end{itemize}

\section{Three Dimensions}\label{sec:3D}

In this section we describe a two fermion system with a contact interaction, considering both unitarity and, later, a finite scattering length.
We implement the Hamiltonian of this system in \eqref{p space hamiltonian} in a three-dimensional cubic box of linear size $L$ with $N$ sites and lattice spacing $\epsilon=L/N$.
At first, the interaction parameter $C(\Lambda)$ of this system is tuned in the regular way--so that the ground state energy of the system matches the first intersection of the spherical zeta function $S^{\spherical}_3$ (evaluated using software provided by \Refs{Morningstar:2017spu,Morningstar:2hib}) with the physical phase shifts \eqref{spherical quantization}.
After the interaction parameter is tuned to machine precision, the low-lying energy levels for the fixed volume and fixed lattice spacing are extracted using numeric exact diagonalization.

The tuning procedure to intersections of the zeta function with the physical phase shifts ensures that the finite-volume effects are incorporated in the energy levels and thus the contact interaction parameter is independent of the volume length $L$.
However, the interaction strength still depends on the implementation of the kinetic operator and the lattice spacing.
Therefore the strength has to be retuned for each lattice discretization implementation.
This discretization dependence has the consequence that in order to obtain \textit{pure} finite-volume energy levels which can be used to compute physical phase shifts, each lattice energy level (besides the input ground state), has to extrapolated to the continuum first.
Only when using these continuum energy levels in \Luscher's formalism can one expect to extract infinite volume scattering information.

In practice, it is not always possible to compute any energy level in the continuum limit before using it in the finite-volume \Luscher formalism.
We therefore present consequences of the following scenarios; to obtain physical scattering data, we
\begin{enumerate}
	\item perform a continuum limit of the spectrum before inserting it in \Luscher's zeta function,
	\item insert finite-spacing energy levels into \Luscher's zeta function, followed by a continuum limit,
	\item utilize the dispersion zeta function to simultaneously perform a continuum and infinite volume limit,
	\item subtract lattice artifacts from finite-spacing eigenvalues before inserting them in the standard zeta function.
\end{enumerate}
The results for these approaches are obtained for the following parameters
\begin{equation}
    \{ L \,[\mathrm{fm}]= 1, 2 \}
    \times \left\{ \epsilon \,[\mathrm{fm}] = \frac{1}{4}, \frac{1}{5}, \frac{1}{10}, \frac{1}{20}, \frac{1}{40}, \frac{1}{50} \right\}
    \times \{ n_s = 1, 2, 3, 4, \infty \}
    \, ,
\end{equation}
as long as $N = L / \epsilon \leq 50$.

\subsection{Continuum extrapolation before infinite volume limit}

\begin{figure}[!htb]
\scalebox{1.05}{\input{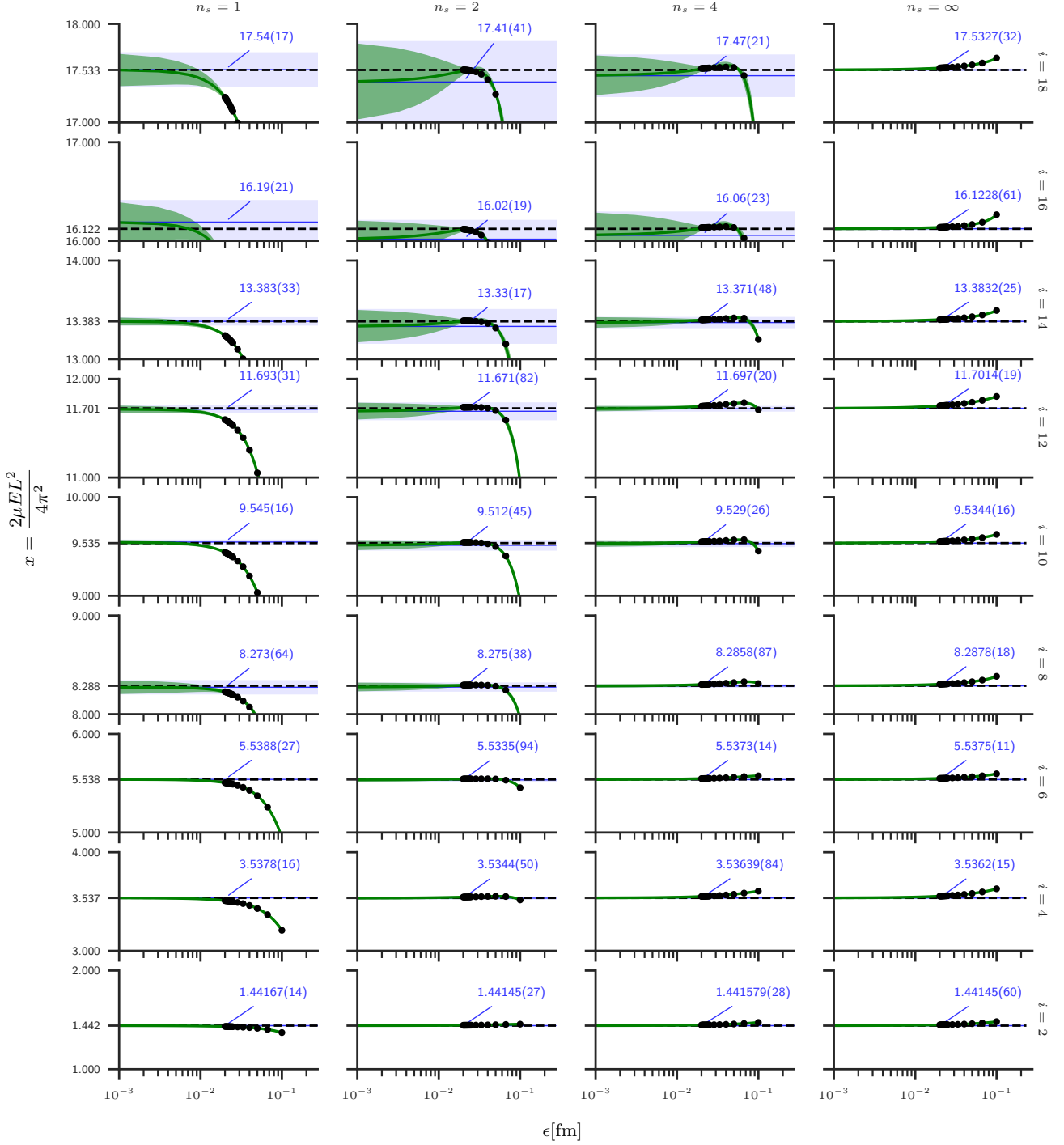}}
\caption{
    \label{fig:continuum-extrapolation-spectrum}
    Continuum extrapolation of the discrete finite volume spectrum with $L=1\,[\mathrm{fm}]$.
    Each column represents a different implementation of the kinetic operator, rows correspond to eigenvalues of the hamiltonian sorted by value.
    For visualization purposes we present each second eigenvalue starting at $E_2$ ($E_0$ was used to tune the interaction and is thus constant by construction).
    Black dots are the eigenvalues at different lattice spacings, the green band is the model averaged fit function for best parameters and the blue band parallel to the x-axis is the continuum-extrapolated energy.
    The uncertainty is dominated by the fluctuations over models; the propagated numerical uncertainty is negligible in comparison.
    The dashed line corresponds to the expected result obtained by computing the intersection of the zeta function $S^{\spherical}_3$ with the phase shifts.
    The boundary of each frame corresponds to the poles of the zeta function.
    Different energy extrapolations in the continuum agree with zeros of the \Luscher zeta within uncertainty.
    For finite discretization implementations ($n_s < \infty$), the uncertainty drastically increases with the number of excited states ($\sim 3$ orders of magnitude from $E_2^{(n_s)}$ to $E_{20}^{(n_s)}$).
}
\end{figure}

After tuning the contact interaction to the first zero of the spherical zeta function, we compute the spectrum of the hamiltonian.
Next, we extrapolate the obtained energy eigenvalues to the continuum $\epsilon \to 0$ using a polynomial fit
\begin{equation}
    E^{(n_s)}_i(\epsilon) = E_i^{(n_s)} + \sum\limits_{n=1}^{n_\mathrm{max}} e_{i,n}^{(n_s)} \epsilon^n \, .
\end{equation}
Because the contact interaction is expected to scale linear with the momentum cutoff and thus linear in $1/\epsilon$ (see \eqref{three-d-counterterm}), one cannot generally expect the fit coefficients $e_{i,n}^{(n_s)}$ to be zero for odd $n$ or $n < n_s$, despite the kinetic improvement~\eqref{gamma determination}.
Nevertheless, we would expect the small $n$ coefficient for larger $n_s$ to be relatively smaller then small $n$ coefficients for smaller $n_s$: $e_{i,n}^{(n_{s_1})} < e_{i,n}^{(n_{s_2})}$ on average for $n_{s_1} > n_{s_2}$.

We individually fit each discretization implementation to extract the continuum energies $E_i^{(n_s)}$ using the software provided by \Ref{peter_lepage_2016_60221}.
Because our numerical uncertainties have an estimated relative error at the order $~10^{-13}$, we must in principle fit the energy for relatively high values of $n_\mathrm{max}$ which would require having many data points over different scales of $\epsilon$.
For this reason we add further lattice spacings
\begin{equation}
	\left\{
		\epsilon \, [\mathrm{fm}] =
		\frac{1}{4}, \frac{1}{5}, \frac{1}{10},
		\frac{1}{15}, \frac{1}{20}, \frac{1}{25},
		\frac{1}{ 30}, \frac{1}{ 35}, \frac{1}{ 40},
		\frac{1}{ 41}, \frac{1}{ 42}, \frac{1}{ 43},
		\frac{1}{ 44}, \frac{1}{45}, \frac{1}{ 46},
		\frac{1}{ 47}, \frac{1}{ 48}, \frac{1}{ 49}, \frac{1}{ 50}
	\right\}
	\, .
\end{equation}
However, we still obtain $\chi^2_{\mathrm{d.o.f}} \gg 1$ up to the point where it is computationally not feasible to add new data points for even smaller lattice spacings as the dimension of the hamiltonian scales with $(L/\epsilon)^3$.

For this reason, we have decided to fit multiple fit models over the span of $n_\mathrm{max} = \{2, 3, 4, 5\}$ and compare their results to estimate a systematic extrapolation uncertainty (unweighted average and standard deviation of results over models).
We repeat this procedure for each discretization and compare different continuum energies to decide wether the fits are consistent.
These values are compared to the spectrum predicted by L\"uscher's formalism.

\begin{figure}[H]
	\centering
    \scalebox{0.8}{\input{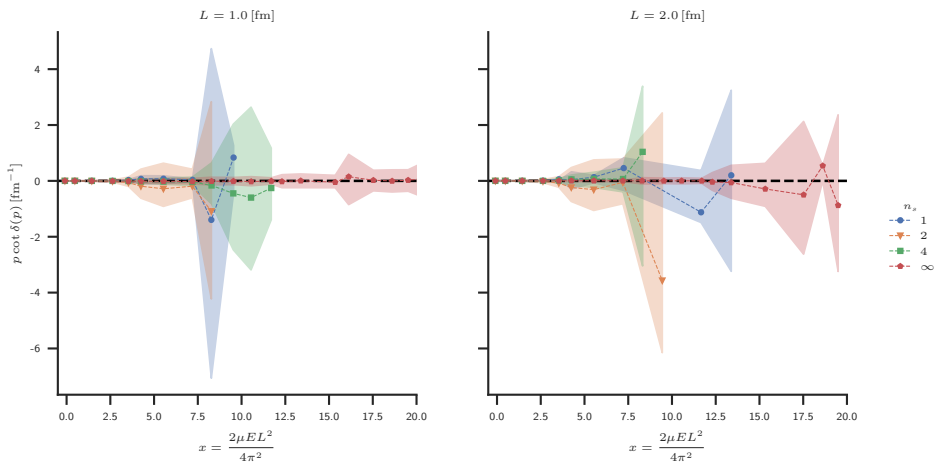}}
    \caption{
        \label{fig:continuum-extrapolation-ere}
        Phase shifts computed by inserting the continuum-extrapolated spectrum for different discretization implementations $n_s$ and finite volumes $L$ in the zeta function $S^{\spherical}_3$.
        Data points indicate locations of the eigenvalue.
        We show the propagated error associated with continuum extrapolation as an uncertainty band.
        The black dashed line represents physical phase shifts.
        Bands stop at different $x$ values because we stop presenting results after uncertainties become too large (but are still consistent with the physical phase shifts).
        }
\end{figure}

We present the model average over best fits of the spectrum in \figref{continuum-extrapolation-spectrum}.
Also, we provide access to the raw data and fitting scripts online at \cite{luescher-nd_201}.
We observe that the model average for polynomials of degree 2 up to 5 is consistent over different discretization and agrees with the expected continuum results.
We noted that including higher polynomials with $n_\mathrm{max} > 6$ resulted in overfitting of higher energy levels visible in oscillating fit functions which were generally were more favorable in model selection criteria\footnote{
    A potential cure for overfitting of higher polynomials would have been the marginalization of higher contributions which would cast the contributions of higher neglected epsilon terms into the uncertainty of the data.
    We eventually settled for an unweighted model average over smaller $n_\mathrm{max}$ because the continuum-extrapolated spectrum was more consistent over different $n_s$.
}.
As expected, the continuum limit becomes more uncertain for excited states.
Furthermore, the $n_s = \infty$ implementation provides the most precise results.
Surprisingly a few energy levels in the $n_s = 1$ implementation have a more precise continuum limit on average than some improved implementations -- even though non-extrapolated energy values are further apart from the continuum as in the improved cases.
This effect is related to the continuum convergence pattern.
While the $n_s = 1$ (and $n_s = \infty$) energy values seem to converge against the continuum result from below (and respectively from above) for all excited states, the improved derivative eigenvalues change their convergence pattern.
The slope of the extrapolation function changes it sign from $E_2 \to E_4$ for $n_s = 2$ and from $E_6 \to E_8$ for  $n_s = 4$.
This suggests that the importance of fit model coefficients $e_{i,n}^{(n_s)}$ changes and thus makes it more difficult to perform the continuum limit.

In the next step, we use the continuum-extrapolated spectrum to convert it to phase shifts using the spherical zeta function.
We present the phase shifts in \figref{continuum-extrapolation-ere}.
Independent of discretization scheme, we observe that the continuum-extrapolated results agree with the constant input phase shifts.
Because the zeta function is relatively steep, uncertainties in the continuum limit get drastically enhanced when converting to phase shifts (on average more than an order of magnitude).
We observe that for $x > 5$ all discretizations besides the exact-$p^2$ discretization come with significant uncertainties.

We emphasize that these findings are not unique to the unitary case, we obtain similar results for a non-zero scattering length.
We present data for an example non-unitarity scenario with $a_{30} = - 5$~fm in our repository \cite{luescher-nd_201}.

\subsection{Using \Luscher's formula before continuum extrapolation}
\begin{figure}[H]
	\centering
    \scalebox{0.8}{\input{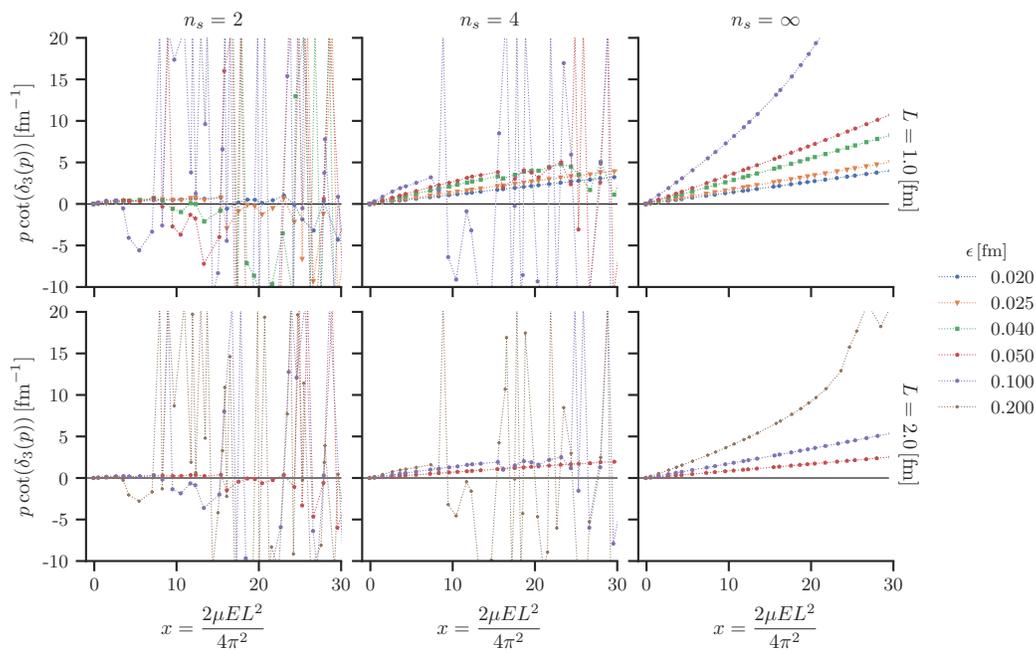}}
    \caption{
        \label{fig:unimproved spherical}
        We present energy eigenvalues presented in \figref{continuum-extrapolation-spectrum} directly inserted $S^{\spherical}_3$--without a continuum limit.
        In the top row we show results for $L=1.0$~fm, in bottom we show $L=2.0$~fm, while in different columns we show different discretization schemes.
        Even though results for $n_s = 2$ seem to be close to the continuum limit result, they start to drastically oscillate for higher energies.
        While more improved discretization schemes seem to oscillate less, they do not lay on top of the continuum result where the difference is related to the lattice spacing.
    }
\end{figure}

Next we want to discuss what effects finite discretization artifacts have when applying L\"uscher's formalism to a spectrum for finite lattice spacings.
We insert the energy levels presented in \figref{continuum-extrapolation-spectrum} before taking the continuum limit and present results in figure \figref{unimproved spherical}.

\begin{figure}[htb]
    \scalebox{1.0}{\input{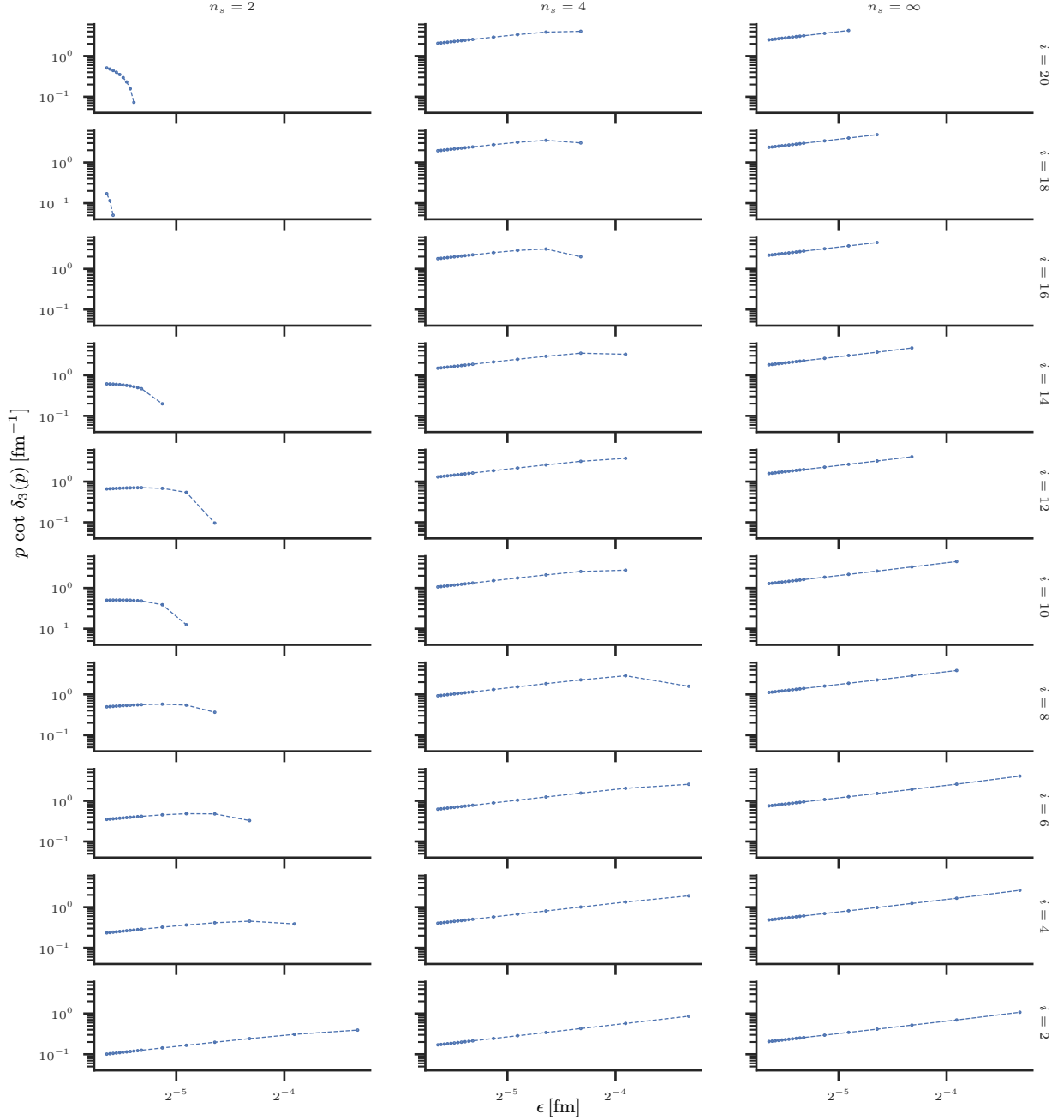}}
    \caption{
        \label{fig:iv-continuum}
        Continuum limit of different phase shift points computed by inserting finite lattice spacing eigenstates in $S^{\spherical}_3$ (see \figref{unimproved spherical}).
        Each column represents a different kinetic operator and each row tracks a different eigenvalue of the discrete finite-volume hamiltonian.
        Note that both axis have a log scale and thus on these scales, a linear trend for the phase shifts suggests that they extrapolate to zero.
    }
\end{figure}

We note that the phase shifts for $x > 10$ start to oscillate wildly.
This is the case because energy values are close to the poles of the zeta function (close to the frame boundaries in \figref{continuum-extrapolation-spectrum}).  With an imperfect kinetic operator, the lattice artifacts in the energy can push energy levels past a pole in the continuum zeta.  This leads to multiple interacting energy levels on a single segment of the zeta function.

Furthermore it seems like the small $x$ results for $n_s = 2$ seem to be closer to the expected flat result than other discretization schemes.
This behavior can be explained by \figref{continuum-extrapolation-spectrum}.
While other discretization schemes for $x < 8$ monotonically converge against the continuum limit, $n_s = 2$ data points converge non-monotonically and are therefore closer to the continuum by accident.
In this sense it is possible to select a discretization scheme which in principle converges slower against the continuum, but has an accidental good agreement with the continuum even though it is discrete.

For small energies, better discretization schemes or small lattice spacings, we observe that the phase shifts do not oscillate and monotonically increase in $x$ with no or small curvature.
This non-flat $x$-dependence seems to depend less on the employed discretization scheme but certainly on the value of the lattice spacing.
This suggests that artifacts of the imperfect kinetic operator are negligible compared to cutoff effects of the lattice spacing itself.
The non-zero lattice spacing induces effective-range-like effects.
As we will show in the next section, this effect arises from using the continuum $S^{\spherical}$ rather than the lattice-aware $S^{\dispersion}$.

We visualize the continuum limit of phase shift points in \figref{iv-continuum}.
Similar to the case where we first extrapolated the spectrum to the continuum and computed phase shifts afterwards, the best discretization allows to also extrapolate higher excited states to zero--visible by the linear log-log dependence of the phase shifts on epsilon.
We note that similar to the case where we first extrapolated the spectrum to the continuum, the extrapolation of the phase shifts seems to work best for the same discretization schemes in the same energy range.
For example, while we find a linear log-log scaling region in \figref{iv-continuum} for $n_s = 2$ and $x<6$, uncertainties of the $n_s =2$ extrapolation also start to increase in \figref{continuum-extrapolation-spectrum} after $x>6$.
However the $n_s = 4$ implementation seems to be stable longer in \figref{iv-continuum} which is related to the $x>9$ state having a relatively larger continuum extrapolation uncertainty while also being close to the continuum value.


\subsection{Results of the dispersion method in three dimensions}

\begin{figure}[!htb]
	\centering
    \scalebox{0.8}{\input{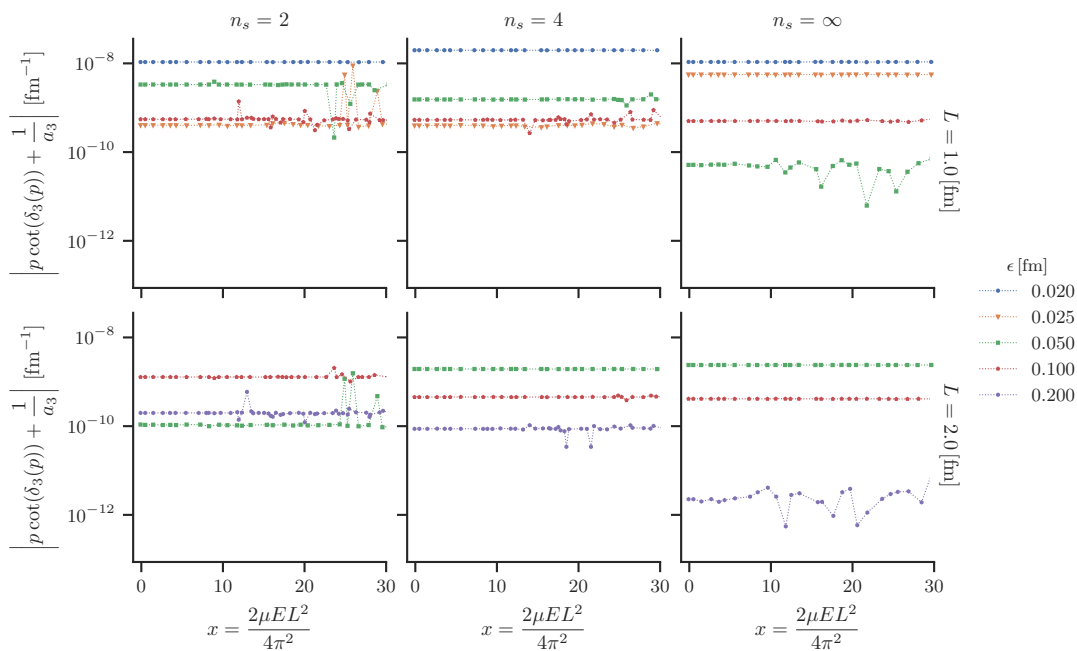}}
    \caption{The same as \Figref{unimproved spherical}, but tuned and subsequently analyzed using the appropriate latticized \Luscher function, matching the cutoff on the sum to the lattice scale and accounting for the dispersion relation.
    We emphasize the results are on a log scale, and the tuning was to $-1/a_3 = 0$.
    }
    \label{fig:unimproved dispersion}
\end{figure}

In this section, we again attempt to tune our contact interaction to unitarity by matching the first zero of the zeta function.
However, the difference is that at each lattice spacing we tune to that spacing's respective $S^{\dispersion}_3$, leveraging the dispersion relation for that derivative.
Then, when we extract finite-volume and finite-spacing energy levels, we put them through the dispersion equation \eqref{dispersion-zeta-contact} using the same $S^{\dispersion}$ function.
The numerical results of said procedure are shown in \Figref{unimproved dispersion}.
Note that the results for $p\cot\delta$ are now flat across the spectrum, matching the known result for a contact interaction.
Moreover, comparing the scale to that in, for example, \Figref{unimproved spherical}, there the deviations were of order~1, while here the results remain within $10^{-8}$ of zero, with the value entirely reflecting how well the contact interaction was tuned.
Put another way, we have verified that the dispersion zeta function provides exact finite-spacing energy levels for our contact interaction \eqref{p space hamiltonian}, just as one would hope for a contact interaction in the continuum.

\begin{figure}[!hbt]
	\centering
    \scalebox{0.7}{\input{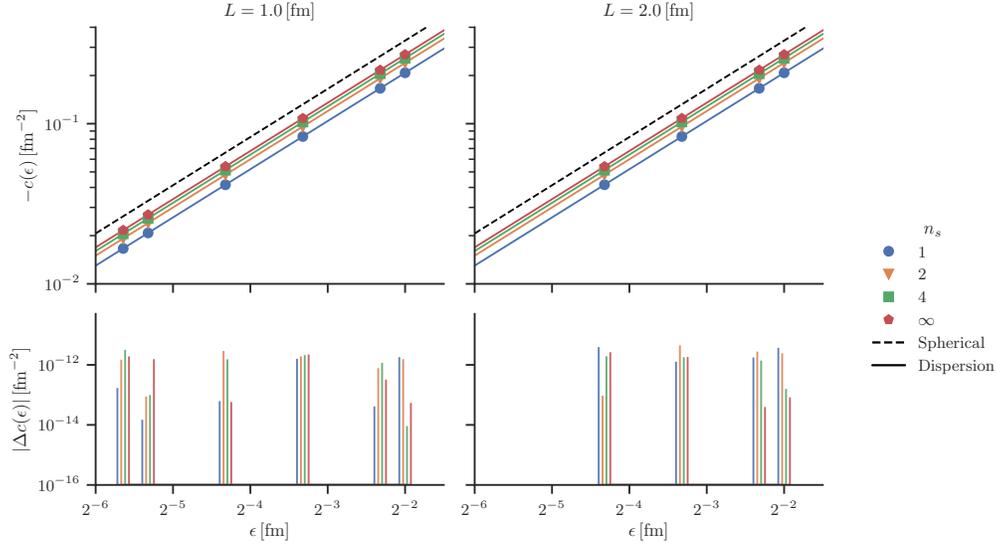}}
    \caption{
        Scaling of the contact interaction strength $C_R^{(n_s)}(\epsilon)$ fitted using the dispersion method at unitarity.
        Data points are values of the contact interaction fitted to the first intersection of the phase shifts with the dispersion zeta function.
        The solid lines are analytic scaling predictions following \eqref{dispersion-renormalization} and the dashed line corresponds to the spherical counter term $ \mathcal{L}^{\spherical}_3 = 2 \pi$.
        Bar diagrams below present the absolute error between prediction and extracted value.
    }
    \label{fig:dispersion running of strength}
\end{figure}

In \Figref{dispersion running of strength} we show how the strength of the contact interaction runs with the lattice scale according to the analytic expectation \eqref{dispersion-renormalization}.
Note that the lines are not fits to the data; though the difference is down at $10^{-12}$ or better.
Again, this difference depends on the accuracy of the tuning.

We note that
\begin{enumerate}
	\item when matching the contact interaction parameter using spherical \Luscher data and finite spacing eigenvalues, the data points did not exactly match the analytic spherical contact scaling.
	The error at the smallest lattice spacing had a relative error on the percent scale and it got worse for larger lattice spacings.
	\item even in the limit of $n_s \to \infty$ the dispersion counter term $\mathcal{L}^\dispersion_3$ will not match the spherical counter term $\mathcal{L}^{\spherical}_3$.
At any finite $N$ the spherical integral and cartesian integrals differ---if the radius of the sphere is $N/2$, the corners of the lattice's Brillouin zone are absent; the cartesian integral matches the Brillouin zone correctly, critical for any finite-$N$ result.
\end{enumerate}

\subsection{Momentum-induced terms of  \texorpdfstring{$S^\bigcirc_3(x^\dispersion)$}{S3-spherical} due to discretization\label{sec:3d induced momenta}}
The zeta function in L\"uscher's formula, $S^\bigcirc_3(x)$, is derived in the continuum.
As such, it requires continuum energies $x$ for its argument.
If one instead feeds discretized energies $x^\dispersion$ through $S^\bigcirc_3(x)$ then momentum-dependent terms are subsequently induced.

\begin{figure}[htb]
    \scalebox{0.8}{\input{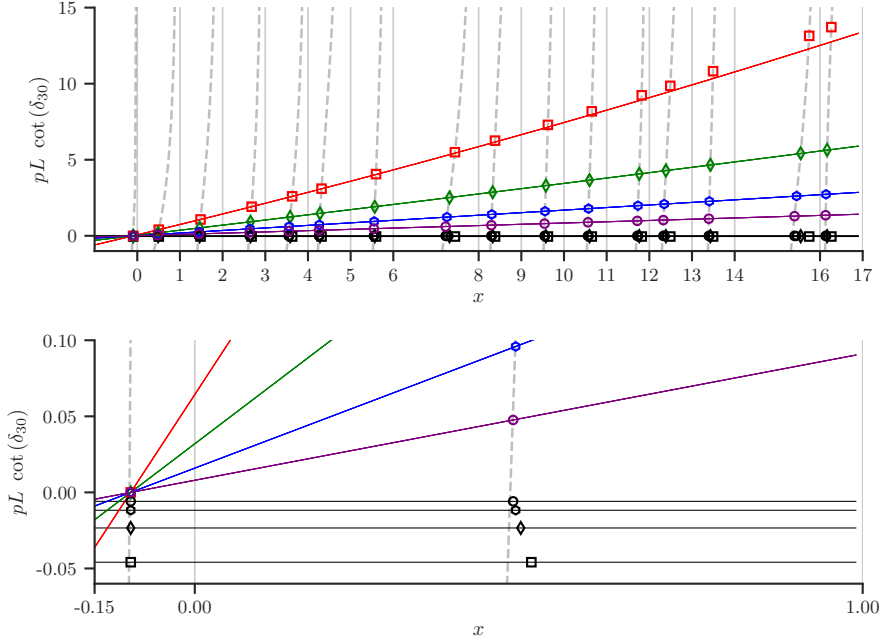}}
    \caption{
        Here we show a contact interaction in three dimensions with the ground state tuned to the first zero of the spherical zeta function $S^{\spherical}_3$ on cubic lattices with $N=10,$ 20, 40, 80 (squares, diamonds, hexagons, and circles, respectively), with the resulting spectrum analyzed with $S^{\spherical}_3$ (colored points) and the $N$-appropriate $S^{\dispersion}_3$ (black points).
        The gray dashed line is $S^\spherical_3$ and the thin vertical lines are at the non-interacting $x$s where it diverges.
        The colored lines are the second-order analytic prediction for the difference between the dispersion and spherical analysis as a function of $x$.
        For clarity of the continuum limit we show, in the bottom panel, a limited range in $x$ and $pL\cot\delta_{30}$, where it is clear that each $N$ hits the zero of $S^\spherical_3$ but that the flat behavior at any finite $N$ is away from an infinite scattering length when analyzed with $S^\dispersion_3$.
    }
    \label{fig:3d-corrections}
\end{figure}

This is particularly evident for the contact interaction as was observed, for example, in \Ref{Endres:2012cw}.
To understand the source of these terms, consider
\begin{multline}
\frac{1}{\pi L}S^\bigcirc_3(x^\dispersion)=\frac{1}{\pi L}\left(S^{\dispersion}_3(x^\dispersion)+\left(S^\bigcirc_3(x^\dispersion)-S^{\dispersion}_3(x^\dispersion)\right)\right)=\frac{-1}{a_3}+\frac{1}{\pi L}\left(S^\bigcirc_3(x^\dispersion)-S^{\dispersion}_3(x^\dispersion)\right)\\
=\frac{-1}{a_3}+\lim_{\eta \to\infty}\frac{1}{\pi L}\left(\sum_{\bm n\notin \mathrm{B.Z.}}^{|\bm n|<\eta / 2} \frac{1}{\bm n^{2}-x^\dispersion}-\mathcal{L}^\bigcirc_3\frac{\eta}{2}+\mathcal{L}^\dispersion_3\frac{N}{2}\right)\ .
\end{multline}
In the first line we added and subtracted $S^\dispersion_3$ and used the dispersion results~\eqref{dispersion-zeta-form} and~\eqref{dispersion-zeta-contact} to introduce the scattering length in the case of a contact interaction.
For convenience we assume $n_s=\infty$\footnote{
The logic of the following derivation remains them same also for $n_s < \infty$, but, in this case, the expressions $\bm n^2$ must be replaced with the proper dispersion $\tilde K^{(n_s)}_{\bm n \bm n}$ \eqref{normalized-kinetic-hamitlonian}, which makes it difficult to obtained closed expressions.  Also, within the Brillouin zone the different dispersion relations cause the two sums differ by \order{x/N^2} term-by-term.}.
In the second line, since $\bm n$ is now restricted to be \emph{outside} the Brillouin zone, we can assume that $\bm n^2\gg x^\dispersion$ and expand in small $x^\dispersion$ under the summation,
\begin{align}
	\label{eqn:observable}
	S^\bigcirc_3(x^\dispersion)
	&=
	\frac{-\pi L}{a_3}
	+\mathcal{L}^\dispersion_3\frac{N}{2}
	+\lim_{\eta \to\infty}\left(\sum_{\bm n\notin \mathrm{B.Z.}}^{|\bm n|<\eta / 2} \frac{1}{\bm n^{2}}
	-\mathcal{L}^\bigcirc_3\frac{\eta}{2}\right)
	+x^\dispersion\lim_{\eta \to\infty}\sum_{\bm n\notin \mathrm{B.Z.}}^{|\bm n|<\eta / 2} \frac{1}{\bm n^{4}}
	+(x^\dispersion)^2\lim_{\eta \to\infty}\sum_{\bm n\notin \mathrm{B.Z.}}^{|\bm n|<\eta / 2} \frac{1}{\bm n^{6}}
	+ \ldots
	\\
	\label{eq:small x}
	&\equiv \frac{-\pi L}{a_3}+\alpha_1(N)+\alpha_2(N)x^\dispersion+\alpha_3(N)(x^\dispersion)^2
	+\ldots
\end{align}
The last line above shows explicitly the induced momentum-dependence in $x^\dispersion$ and defines the coefficients $\alpha_i(N)$ in terms of particular lattice summations similar to those of the three-dimensional zeta function.  The dependence of these coefficients on $N$ comes from the \emph{exclusion} of momentum modes within the Brillouin zone in the summation.   The fact that these coefficients do \emph{not} depend on $L$ is a unique feature of the contact interaction.  The numerical values of the coefficients $\alpha_i(N)$ can be determined using standard acceleration techniques (see, for example, Appendix B of~\Ref{Luu:2011ep}).  We provide values for select cases of $N$ in \Tabref{slopes}.
\begin{table}
\caption{Coefficients $\alpha_i(N)$ as a function of $N$ in 3-D.\label{tab:slopes}}
\center
\begin{tabular}{S[table-format=1.0]S[table-format=-1.15,table-auto-round=false]S[table-format=-1.15,table-auto-round=false]S[table-format=-1.15,table-auto-round=false]}
{$N$} & {$\alpha_1$} & {$\alpha_2$} & {$\alpha_3$} \\ \midrule
10 & 0.34622847019345 &2.1088361299026 &0.02096728133239\\
20 & 0.17384029798483 &1.0470052482673 &0.00253774588732\\
40 & 0.08701147975728 &0.5225652776531 &0.00031456311910\\
50 & 0.06961796407968 &0.4179620004936 &0.00016089237674\\
80 & 0.04351717442702 &0.2611651268184 &0.00003923695720\\
100 &0.03481483765136 &0.2089208128674 &0.00002008418957
\end{tabular}
\end{table}

In \Figref{3d-corrections} we show the result of tuning a finite-spacing contact interaction to the first zero of the continuum zeta function $S^\spherical_3$.
At each spacing the spectrum is fed through the continuum zeta for analysis, resulting in an apparent spacing-dependent momentum dependence that matches the small-$x$ expansion~\eqref{small x} discussed in the next section.
The same spectrum is also fed through the spacing-appropriate dispersion zeta $S^\dispersion$, resulting in the flat black lines.
Shown in detail in the bottom panel, it's clear that the continuum limit taken this way results in any finite spacing having a nonzero scattering length that vanishes with the continuum limit.
In contrast, tuning to the dispersion function directly, as in \Figref{unimproved dispersion}, is flat and nearly zero at each individual lattice spacing.
We expect that this difference explains the induced momentum dependence of, for example, \Refs{Endres:2011er,Endres:2012cw}.

\section{One Dimension}\label{sec:1D}

Here we consider L\"uscher's formula in one dimension with a contact interaction.
Since the sum in the quantization condition \eqref{general luscher} or the one-loop finite-volume sum \eqref{I0 FV} with $D=1$ is convergent we have 
\begin{align}\label{eq:tuning-1d}
    C(\Lambda)
    &=
            -\frac{1}{\mu a_{1}}
    &
    \frac{a_{1}}{L}
        &=
            \frac{1}{2 \pi^{2}}
            \sum_{n=-\infty}^{\infty} \frac{1}{n^{2}-x}
        \equiv
            \frac{1}{2 \pi^{2}}
            S^\bigcirc_{1}\left(x\right)
    &
    \Bigg(x
        &=
            \frac{2\mu E L^2}{4\pi^2}\Bigg)
    \, ,
\end{align}
where the contact strength $C(\Lambda)$ does not run.
The energy $E$ is a finite-volume energy on a torus of circumference $L$.
In one dimension the sum in the zeta function is well behaved and has a compact form,
\begin{equation}\label{eq:1d luscher}
S^\bigcirc_{1}(x) \equiv \sum_{n=-\infty}^{\infty} \frac{1}{n^{2}-x}=-\pi \frac{\cot (\pi \sqrt{x})}{\sqrt{x}}\ ,
\end{equation}
which gives a closed form expression for L\"uscher's formula,
\begin{equation}\label{eq:1d luscher constant}
\frac{a_{1}}{L} =-\frac{1}{pL}\cot\left(\frac{pL}{2}\right),
\end{equation}
consistent with those found in \Refs{Luscher:1990ck,Zhu:2019dho}.

Since there is no counterterm in one dimension, the dispersion form of L\"uscher's formula is straightforward to obtain.  If one identifies the lattice spacing as the cutoff, then the sum in the zeta function is restricted to the Brillouin zone and one has
\begin{equation}
    \frac{a_{1}}{L}
    =
    \frac{1}{2 \pi^{2}} \sum_{n=-\frac{N}{2}}^{\frac{N}{2}-1} \frac{1}{\tilde{K}^N_{nn}-x}
    =
    \frac{1}{2 \pi^{2}} \sum_{n=-\frac{N}{2}}^{\frac{N}{2}-1} \frac{1}{\frac{N^2}{4\pi^2}\left(\sum_{s}\gamma_s^{(\nstep)}\cos\frac{2\pi n s}{N}\right)-x}
    \equiv
    \frac{1}{2 \pi^{2}} S^{\dispersion}_{1}\left(x\right),\label{eq:1d dispersion luscher}
\end{equation}
where we explicitly show that the dispersion function $S^{\dispersion}$ depends on $\nstep$ and $N$ but not on $L$ or $\epsilon$ explicitly.

As stressed in the previous section, only continuum-extrapolated energies should be used in the quantization condition \eqref{1d luscher}, or induced momentum dependence terms will result.
For example, in \Figref{luescher1d} we show the induced momentum dependence terms when non-continuum eigenvalues $x^\dispersion$ are inserted into $S^{\spherical}_1$ (colored points) for lattice sizes of $N=4$, 10, 12, and 14 and $\nstep=\infty$.
However, we also show the scattering data determined through $S^{\dispersion}_1(x^\dispersion)$ (black points), which lie on a flat line, as expected.

\begin{figure}
\center
    \center
    \input{figure/1d.pgf}
    \caption{
    	Phase shifts and zeta functions in the spherical and dispersion scenario in one dimension.
		\\
		\textit{(top)}
        	Finite-spacing eigenvalues $x^\dispersion=2\mu E^\dispersion L^2/4\pi^2$ of the Schr\"odinger equation are inserted into respective zeta functions to obtain phase shifts.
			The eigenvalues are obtained for a contact interactions fixed to $ a_{1}/L=1/10$ (closed symbols) and $a_{1}/L=0$ (open symbols) using the analytic result for the interaction strength $C(\Lambda)$ \eqref{tuning-1d} and not by tuning to a zero of any zeta.
			Different markers correspond to different discretizations: $N=4$ (triangles), 10 (squares), 12 (diamonds), and 14 (hexagons).
        	For analysis, the colored points are obtained using $S^\spherical_1(x^\dispersion)$; corresponding to $N=4$ (red), 10 (green), 12 (blue), and 14 (purple).
        	The thin colored lines are the derived induced momentum-dependent terms for each $N$ as given in \Tabref{induced terms in 1 d}.
        	The black points are obtained using the $N$-appropriate $S^{\dispersion}_1(x^\dispersion)$ and exhibit the correct flat-line behavior.
        	The dashed gray line is $S_1^\spherical$, as in the bottom panel.
        \\
        \textit{(bottom)}
        	Two one-dimensional zeta functions, the spherical function $S_1^{\spherical}$ (light gray) given in \eqref{1d luscher} and $S_1^{\dispersion}$ (red) given in \eqref{1d dispersion luscher} with $N=4$ and $\nstep=\infty$.
        	The difference between the dispersion and spherical curves is responsible for moving the red triangles to the black triangles in the top panel.
        }
        \label{fig:luescher1d}
\end{figure}

We can derive the functional form of these induced momentum-dependent terms following the exact steps taken in \Secref{3d induced momenta}, again assuming $\nstep=\infty$,
\begin{align*}
S^\bigcirc_{1}\left(x^\dispersion\right)&=S^{\dispersion}_{1}\left(x^\dispersion\right)+\left(S^\bigcirc_{1}\left(x^\dispersion\right)  -S^{\dispersion}_{1}\left(x^\dispersion\right)\right) 
=\frac{2\pi^2 a_{1}}{L}+ \sum_{n\notin \operatorname{B.Z.}} \frac{1}{n^{2}-x^\dispersion} \\
&=\frac{2\pi^2 a_{1}}{L}+\alpha_1(N)+ \alpha_2(N)x^\dispersion+\alpha_3(N)(x^\dispersion)^2+\ldots\ .
\end{align*}
where again we assume $\nstep=\infty$ energy.
In the second term we assumed $n^2\gg x^\dispersion$ since the sum is restricted to modes outside the Brillouin zone.
The coefficients $\alpha_i(N)$ can be determined to arbitrary precision.
\Tabref{induced terms in 1 d} shows these terms for $N=4$, 10, 12, and 14.
The thin colored lines in \Figref{luescher1d} correspond to the functions given in this table.
In the limit $N\goesto\infty$ all states are included in the Brillouin zone so all terms vanish and one recovers the flat, momentum-independent behavior.

\begin{table}
    \caption{
    The coefficients $\alpha_i(N)$ of the induced momentum-dependent terms to order $(x^\dispersion)^2$ due to a contact interaction using $S^\bigcirc_1(x^\dispersion)$ as a function of discretization $N$.
    Here $x^\dispersion$ is determined by a finite-spacing finite-volume $\nstep=\infty$ eigenenergy.}
    \label{tab:induced terms in 1 d}
    \begin{tabular}{S[table-format=2.1]S[table-format=2.6]S[table-format=2.7]S[table-format=2.8]}
{$N$} & {$\alpha_1$} & {$\alpha_2$} & {$\alpha_3$} \\ \midrule
      4 &   1.03987  & 0.102146 & 0.0190611 \\
    10  &   0.40265  & 0.005543 & 0.0001404 \\
    12  &   0.33487  & 0.003171 & 0.0000549 \\
    14  &   0.28668  & 0.001983 & 0.0000250 \\
    \end{tabular}
\end{table}

Analyzing the finite-spacing dispersion zeta function $S^{\dispersion}_1$ with $\nstep=\infty$ produces the flat behavior all the way through.
This demonstrates that in the one-dimensional case, it was not the contact operator that caused the momentum dependence, but that the dependence was induced by leveraging the continuum finite-volume formalism itself.

Finally, we draw the reader's attention to the structure of $S^\dispersion$ at any finite $N$ in the bottom panel of \Figref{luescher1d}, where the $N=4$ dispersion zeta is shown.
Note that any flat function of $x$ can only ever intercept the zeta function three times---which makes sense, as there are only three states in the parity-even sector of a one-dimensional $N=4$ lattice: $\ket{0}$, $(\ket{-1}+\ket{+1})/\sqrt{2}$ and $\ket{2}$ (which, being on the edge of the Brillouin zone, is the same state as $\ket{-2}$).
If one tunes a contact interaction so that the scattering amplitude vanishes, one will see one state with $0<x<1$, one with $1<x<4$, and one state with $\abs{x}$ very large and a sign depending on whether one is slightly above or below zero numerically.
The finiteness of the parity-even sector puts constraints on the interactions that can be faithfully put onto such a small lattice: one cannot create any interaction where the scattering amplitude intersects the $N=4$ $S^{\dispersion}$ four times, because that would entail too many finite-volume states.
Of course, this is a generic feature in any number of dimensions, and the constraint ultimately vanishes in the continuum limit $N\goesto\infty$; as $N$ increases the number of accessible $n^2$ shells grows in a dimension-dependent way.
       \FloatBarrier
\section{Two-dimensions}\label{sec:2D}

In two dimensions, assuming a contact interaction, it is convenient to write the effective range expansion \eqref{ere} in terms of a \emph{reduced} scattering length,
\begin{equation}\label{eq:2d contact phase shift}
    \cot \left(\delta_{2}(p)\right)
    =
    \frac{2}{\pi} \ln \left(p \tilde a_{2}\right),
    \hspace{0.5in}
    \text{where}
    \hspace{0.5in}
    \tilde a_{2}
    =
    R_{20}\exp\left\{-\frac{\pi}{2 a_{20}}\right\}
\end{equation}
with no additional shape parameters.
Using the infinite-volume relation \eqref{IV pole} with $p=0$, $D=2$, and a finite cutoff $\Lambda$ gives
\begin{equation}\label{eq:C2}
    C(\Lambda)
    =
    -\frac{\pi}{\mu \log \left(\tilde a_{2} \Lambda\right)}.
\end{equation}
Using this in the finite-volume relation \eqref{FV pole} and quantization condition \eqref{general luscher} yields
\begin{equation}\label{eq:first 2d}
    \frac{2}{\pi} \log \left(p\tilde a_{2}\right)
    =
    \lim_{\Lambda\to\infty}
    \left(
        -\frac{2}{\mu L^{2}} \sum_{\vec{q}}^{\Lambda} \frac{1}{E-\frac{\vec{q}^{2}}{2\mu}}
        -\frac{2}{\pi} \log (\Lambda / p)
    \right).
\end{equation}
The logarithmic dependence on $p$ makes this relation difficult for analysis, particularly for small $p$.
Furthermore, for $\tilde a_{2}$ sufficiently small (but positive)\footnote{
    Our definition of the scattering length in 2-d requires $\tilde a_{2}\ge 0$~\cite{Pupyshev:2014}.
} a bound state can occur, with imaginary momentum $p\to i\gamma$.
Then both sides become complex, further complicating the analysis.
Also, the momentum-\emph{independent} logarithmic counterterm needed to regulate the infinite sum is not manifest in the above expression.
To make the counterterm manifest, and to address the issue of small $p$ states and bound states, we subtract $\frac{2}{\pi}\log\left(\frac{pL}{2\pi}\right)$ on both sides,
\begin{align}
    \frac{2}{\pi} \log \left(\frac{2\pi \tilde a_{2}}{L}\right)
    &=
    \lim_{\Lambda\to\infty}
    \left(
        -\frac{2}{\mu L^{2}} \sum_{\vec{q}}^{\Lambda} \frac{1}{E-\frac{\vec{q}^{2}}{2\mu}}
        -\frac{2}{\pi} \log \left(\frac{\Lambda L}{2\pi}\right)
    \right)
    \nonumber\\
    &=
    \lim_{\Lambda\to\infty}
    \left(
        \frac{1}{\pi^2} \sum_{\vec{q}}^{\Lambda} \frac{1}{\left(\frac{\vec{q}L}{2\pi}\right)^2-x}-\frac{2}{\pi} \log \left(\frac{\Lambda L}{2\pi}\right)
    \right)
    \label{eq:second 2d}
\end{align}
where $x=2\mu EL^2/4\pi^2$ as always.
Setting $N=\Lambda L/\pi$ and $\left(\frac{\vec{q}L}{2\pi}\right)^2=\vec{n}^2$ gives
\begin{equation}
    \frac{2}{\pi} \log \left(\frac{2\pi \tilde a_{2}}{L}\right)
    =
    \frac{1}{\pi^2}
    \lim_{N\to\infty}
    \left(
        \sum_{|\vec{n}|\le \frac{N}{2}} \frac{1}{\vec{n}^2-x}
        -2\pi \log \left(\frac{N}{2}\right)
    \right)
    \equiv
    \frac{1}{\pi^2}S^{\spherical}_2\left(x\right)\ ,\label{eq:2d luscher}
\end{equation}
which defines the two-dimensional zeta function $S^{\spherical}_2$.
This matches the general result \eqref{spherical S} as long as we allow the limit
\begin{equation}
    \lim_{D\to2}\counterterm_D^\spherical \left(\frac{N}{2}\right)^{D-2}=2\pi \log \left(\frac{N}{2}\right).
\end{equation}

This two-dimensional \Luscher function \eqref{2d luscher} encompasses both bound and scattering states for the contact interaction.
Note the logarithmic dependence of the scattering length $\tilde a_{2}$ which requires an accompanying scale to render the argument of the logarithm dimensionless---we choose the infrared scale $L$, the linear size of the finite volume.
Finally, we note that for general finite-range \emph{S-wave} interactions, the L\"uscher formula in 2-D is
\begin{equation}\label{eq:full 2d luescher}
\cot(\delta_{2}(p))-\frac{2}{\pi}\log\left(\frac{pL}{2\pi}\right) = \frac{1}{\pi^2}S^\bigcirc_2\left(\left(\frac{pL}{2\pi}\right)^2\right)\ .
\end{equation}
This form was originally derived in \Ref{Beane:2010ny}, and is also consistent with \Ref{Zhu:2019dho} once the subtraction of the logarithm and the difference in definition of our zeta functions are taken into account. For higher partial waves we refer the reader to Ref.~\cite{Fiebig:1994qi}.

\subsection{Dispersion L\"uscher in 2 dimensions}

The discussion above is valid only in the continuum.
For a discretized lattice, an additional length scale is introduced that must be accounted for.
As is the case in both 3-D and 1-D, there exists a dispersion L\"uscher equation that is valid for the contact interaction and accounts for the discretization.
In \Appref{3D dispersion} we derive this dispersion L\"uscher formula for 2D and only state the result here.

Identifying the lattice spacing $\epsilon=N/L$, we have
\begin{align}
    \frac{2}{\pi} \log \left(\frac{2\pi \tilde a_{2}}{L}\right)
    &=
    \frac{1}{\pi^2}
    \left(
        \sum_{n_x,n_y=-\frac{N}{2}}^{\frac{N}{2}-1}\frac{1}{\tilde{K}^N_{nn}-x^\dispersion}
        -2\pi \log \left(\mathcal{L}^\dispersion_2\frac{N}{2}\right)
    \right)
    \nonumber\\
    &\equiv\frac{1}{\pi^2}S^{\dispersion}_2\left(x^\dispersion\right)\, ,\label{eq:2d dispersion luscher}
	\qquad
    \mathcal{L}^\dispersion_{2}
    =
    \exp \left(\log (2)-G \frac{2}{\pi}\right)
    =
    1.116306393581637659468497 \ldots
\end{align}
where $G$ is Catalan's constant. 

To demonstrate the success of this formula we tuned lattices with $N=10$, 20, and 40 to $\tilde a_{2}/L = {1}/{10}$, which allows for a bound state, using $S^{\dispersion}$.
In \Figref{luescher2d} the black points were analyzed through $S^\dispersion_2$, and lie on a flat line, indicating that our dispersion \Luscher formula has correctly accounted for discretization effects.
On the other hand, if we use the same energies but analyze them with the usual continuum \Luscher function $S^{\spherical}_2$, shown as colored points, we see induced momentum-dependence and the flat line behavior is lost.

\begin{figure}
    \center
    \scalebox{0.8}{\input{figure/2d.pgf}}
    \caption{
    	Phase shifts and zeta functions in the spherical and dispersion scenario in two dimensions.
		\\
		\textit{(top)}
        	Finite-spacing eigenvalues $x^\dispersion=2\mu E^\dispersion L^2/4\pi^2$ of the Schr\"odinger equation are inserted into respective zeta functions to obtain phase shifts.
			The eigenvalues are obtained for a contact interaction analytically determined for the spherical case~\eqref{C2} and the dispersion case~\eqref{C2-dispersion}.
			Both tunings are fixed to $\tilde a_{2}/L=1/10$ (closed symbols) and $\tilde a_{2}/L=0$ (open symbols).
			Different markers correspond to different discretizations: $N=4$ (triangles), 10 (squares), 12 (diamonds), and 14 (hexagons).
        	For analysis, the colored points are obtained using $S^\spherical_2(x^\dispersion)$; corresponding to $N=4$ (red), 10 (green), 20 (blue), and 40 (purple).
        	The thin colored lines are the derived induced momentum-dependent terms for each $N$ as given in \Tabref{induced terms in 2 d}.
        	The black points are obtained using the $N$-appropriate $S^{\dispersion}_2(x^\dispersion)$ and exhibit the correct flat-line behavior.
        	The dashed gray line is $S_2^\spherical$, as in the bottom panel.
        \\
        \textit{(bottom)}
        	Two one-dimensional zeta functions, the spherical function $S_2^{\spherical}$ (light gray) given in \eqref{2d luscher} and $S_2^{\dispersion}$ (red) given in \eqref{2d dispersion luscher} with $N=4$ and $\nstep=\infty$.
        	The difference between the dispersion and spherical curves is responsible for moving the red triangles to the black triangles in the top panel.
	}
    \label{fig:luescher2d}
\end{figure}

As was done in the three- and one-dimensional cases, we can derive the functional form of the induced momentum-dependent terms.
The derivation is identical to those cases; for concision we show only the end result.
Expanded around small $x^\dispersion$ one finds
\begin{equation}
    \label{eq:2D corrections}
    S^\bigcirc_{2}\left(x^\dispersion\right)
    =
    2\pi\log\left(2\pi \frac{\tilde a_{2}}{L}\right)
    + \alpha_1(N)
    + \alpha_2(N) x^\dispersion
    + \alpha_3(N) (x^\dispersion)^2
    + \ldots
\end{equation}
The coefficients $\alpha_i(N)$ have an implicit dependence on $N$ since the sums are restricted outside of the Brillouin zone.  Further, in 2-D the sums involved in $\alpha_i$ converge sufficiently fast and there exist various techniques for evaluating these sums (see Appendices A of \Ref{Fiebig:1994qi,Beane:2010ny}).
We provide the numerical values of $\alpha_i(N)$ in \Tabref{induced terms in 2 d} for the discretizations shown in \Figref{luescher2d}.
These functions were also used to calculate the thin colored lines in \Figref{luescher2d}, where we see the small-$x$ expansion lose accuracy quickly for $N=4$ (consider the bound state, for example) but hold deeper into the spectrum for larger $N$.

\begin{table}
    \caption{The coefficients $\alpha_i$ of the induced momentum-dependent terms in \eqref{2D corrections} due to a contact interaction using $S^\bigcirc_2(x^\dispersion)$ as a function of discretization $N$, assuming $\nstep=\infty$.
    }
    \label{tab:induced terms in 2 d}
    \begin{tabular}{S[table-format=2.1]S[table-format=2.6]S[table-format=2.7]S[table-format=2.8]}
{$N$} & {{$\alpha_1$}} & {$\alpha_2$} & {$\alpha_3$} \\ \midrule
    4   &   0.20642   &   0.726184 & 0.0937105           \\
    10  &   0.03402   &   0.105117 & 0.0018607           \\
    20  &   0.00086   &   0.025850 & 0.0001108           \\
    40  &   0.00851   &   0.006433 & 0.0000068    \\
\end{tabular}
\end{table}

      \FloatBarrier
\section{Conclusion}\label{sec:conclusion}

We presented a tuning prescription for a two-particle lattice system interacting through a contact interaction in 1-, 2- and 3-dimensions.
For this interaction, the tuning prescription allows us to compute infinite volume continuum scattering observables 
from data computed in the finite volume and discrete space.
Furthermore we derived a \Luscher-like formalism which directly converts the associated finite-volume finite-spacing spectra to infinite-volume continuum phase shifts for the contact interaction.

In 3-dimensions, we analyzed three different approaches in detail:
\begin{enumerate}
	\item we tuned the interaction parameter in a finite volume with a finite lattice spacing to the intersections of the \Luscher zeta function and the phase shifts, extracted the continuum-extrapolated spectrum, and used the same \Luscher zeta function to re-obtain the phase shifts,
	\item we repeated the same procedure without extrapolating the spectrum to the continuum and found phase shifts with induced energy-dependence,
	\item we derived a dispersion-aware zeta function which removed the energy dependence in the phase shifts,
	\item we perturbatively computed the discretization dependent coefficients which describe the difference in the effective range expansion between continuum extrapolated results and results obtained at a finite spacing.
\end{enumerate}

The first approach follows the logic of \Luscher's original work and reproduces the expected phase shifts.
Even though we had full control over numerical errors, the continuum extrapolation of the spectrum suffered from systematic artifacts and induced significant uncertainties (on a relative scale) when put through the zeta function.
In general the best discretization allows the best extrapolation and for smaller energy values, continuum results are more precise.
It is possible to find discretizations in which the finite spacing spectrum is close to its continuum result but the extrapolation uncertainties can be larger because of non-monotonic behavior of individual energy levels in dependence of the lattice spacing.

The second approach, applying the infinite-volume map to finite-spacing energy levels---the approach of most recent lattice QCD work---suffers in the case of the analyzed interaction from notable discretization artifacts.
These artifacts induce an energy dependence in the phase shifts at any finite spacing.
For example, we found induced effective range (and higher order) effects which we analytically estimated.
These induced terms can be extrapolated to zero in a stable manner in the continuum if one only considers energy values in the scaling region.
We provide tables of coefficients which estimate the size of errors in the phase shifts caused by the discretization.

The third approach allows a direct conversion from finite-spacing finite-volume energy levels to continuum infinite-volume phase shifts without any extrapolation.
Thus it was possible to consistently tune the interaction parameter to high precision.
Further, this tuning allows one to distinguish between kinetic discretization effects and discretization effects affecting the regulator of the theory and thus allows one to determine the interaction consistently.

Finally, we repeated our three-dimensional analysis above to both one- and two-dimensional systems.
The latter is further complicated by logarithmic singularities as opposed to power law divergences, and so here we proposed a slightly modified \Luscher equation in two dimensions to account for the logarithmic singularity near $p\sim 0$.
In both cases our results are consistent with those found in the literature.

We expect our discretization-specific tuning for the contact interaction parameter can be carried beyond the two-body sector and used in many-body computations, so that calculations of the Bertsch parameter should benefit from having a systematically correct tuned interaction, which we plan to investigate in future work.
We note, however, that while it would be desirable to find a similar dispersion formalism and tuning prescription for any interaction (for example, finite-range interactions), the derivation of this prescription in this case would depend on an explicit knowledge of the short-distance parts of the interaction.
We do not rule out, however, that our dispersion formalism might be applicable to other specific interactions, or maybe even generalizes in a perturbative manner for general interactions.  


\section*{Acknowledgements}\label{sec:acknowledgements}

The authors thank
Tom Cohen,
Ben H\"{o}rz,
Ken McElvain,
Colin Morningstar,
Andr\'{e} Walker-Loud,
and
Jan-Lukas Wynen
for stimulating discussion, feedback, technical help and computational resources during the course of this work.
C.K. gratefully acknowledges funding through the Alexander von Humboldt Foundation through a Feodor Lynen Research Fellowship.
This work was done in part through financial support from the Deutsche Forschungsgemeinschaft (Sino-German CRC 110).
E.B. thanks Stefan Krieg for the invitation to speak at Lattice Practices 2018\cite{lattice-practices}; the exercises\cite{lattice-practices-exercises} prepared for that school in part grew into this paper.
E.B. is supported by the U.S. Department of Energy under Contract No. DE-FG02-93ER-40762.

\appendix

\section{Dispersion Relation Coefficients}\label{sec:coefficients}

In \eqref{laplacian} and \eqref{gamma determination} we give the definition and how to determine the $\gamma_s^{(\nstep)}$ coefficients that give us finite difference formulas.
This is done by matching the expansion of the cosine to the continuum dispersion
\begin{align}
	p^2
	&\mapsto
	- \frac{1}{\epsilon^2}\sum_{s=0}^{n_s} \gamma_s^{(n_s)} \cos(s p \epsilon)
	=
	\frac{1}{\epsilon^2}\sum_{s=0}^{n_s} \sum_{m=0}^\infty \gamma_s^{(n_s)}\frac{(-)^m}{(2m)!} (s p \epsilon)^{2m}
	\overset{!}{=}
	p^2 \left[ 1 + \mathcal O \left( (\epsilon p)^{2 n_s} \right) \right]
	\, .
\end{align}
Matching this expression order by order in $\epsilon p$ or equivalently $m \leq n_s$ effectively results in a matrix equation for the coefficients $\gamma_s^{(n_s)}$
\begin{align}
	A_{ms} &\equiv \frac{(-)^m}{(2m)!} s^{2m}
	\, , &
	\sum_{s=0}^{n_s} A_{m s} \gamma_s^{(n_s)}
	\overset{!}{=}
	v_m
	&=
	\begin{cases}
		1 & m=1 \\ 0 & \text{otherwise}
	\end{cases}
	\, , &
	\vec \gamma^{(n_s)}
	&= A^{-1} \vec v
\end{align}
Results for order $n_s \leq 4$ are displayed in \tabref{dispersion coefficients}.

\begin{table}[ht]
    \caption{Values for $\gamma_s^{(\nstep)}$ for a variety of different $\nstep$s that give the optimal approximation $\omega^{(\nstep)}(p,\epsilon) = (\epsilon p)^2\left[1+ \order{(\epsilon p)^{2 \nstep}}\right]$.}
    \label{tab:dispersion coefficients}
    \begin{tabular}{c | ccccc}
        $\gamma_s^{(\nstep)}$   &   $s=0$   &   $s=1$   &   $s=2$   &   $s=3$       &   $s=4$   \\ \hline
        $\nstep=1$              &   $2$     &   $-2$    &           &               &           \\
        $\nstep=2$              &   $5/2$   &   $-8/3$  &   $1/6$   &               &           \\
        $\nstep=3$              &   $49/18$ &   $-3$    &   $3/10$  &   $-1/45$     &           \\
        $\nstep=4$              &   $205/72$&   $-16/5$ &   $2/5$   &   $-16/315$   &   $1/280$
    \end{tabular}
\end{table}

\section{The Usual Counterterm}\label{sec:counterterm/spherical}

The spherical integrals $I_D$ are cut off by a radius of $N/2$.
In \eqref{spherical quantization} a variety of dimensionful parameters are separated from the integral itself.
For convenience, the integrals that appear in $S_D^\spherical$ for $x>0$ itself are given by
\begin{equation}
    \label{eq:spherical S improved regulator}
     \Omega_D \int_0^{N/2}  \mathcal P \frac{ n^{D-1} \mathrm{d}n}{n^2-x} = \begin{cases}
        4\pi\left( \frac{N}{2} - \sqrt{x} \tanh\inverse \frac{\sqrt{x}}{N/2}\right)
            &   (D=3)\\[0.25cm]
        \pi \log\left( \frac{(N/2)^2}{x} - 1\right)
            &   (D=2)\\[0.25cm]
        -\frac{2}{\sqrt{x}} \tanh\inverse \frac{\sqrt{x}}{N/2}                     &   (D=1)
    \end{cases}.
\end{equation}

The regulating behavior needed to cancel the divergence in the sum in $S_D^\spherical$ can be found by expanding these integrals around the large-$N$ behavior,
\begin{equation}
     \Omega_D\int_0^{N/2} \mathcal P \frac{ n^{D-1} \mathrm{d}n}{n^2-x} \goesto \begin{cases}
        4\pi \left(\frac{N}{2}\right) - \frac{4 \pi x}{N/2} + \order{\left(\frac{N}{2}\right)^{-2}}
            &   (D=3)\\[0.25cm]
        \pi \log \left(\frac{(N/2)^2}{x}\right) - \frac{\pi x}{(N/2)^2} + \order{\left(\frac{N}{2}\right)^{-4}}
            &   (D=2)\\[0.25cm]
        -\frac{2}{(N/2)} - \frac{2 x}{3 (N/2)^3} + \order{\left(\frac{N}{2}\right)^{-4}}
            &   (D=1)
    \end{cases}
\end{equation}
However, rather than including only the leading divergent behavior, one can use the exact integral values in \eqref{spherical S improved regulator}, accelerating the convergence to the large-$N$ limit.

In two dimensions, we can rewrite separate the large-$N$ behavior of the infinite-volume integral
\begin{align}
    \Omega_D\int_0^{N/2} \mathcal P \frac{ n^{D-1} \mathrm{d}n}{n^2-x}
    &=
    \pi \log \left(\frac{(N/2)^2}{x}\right) + \pi \log\left(1-\frac{x}{(N/2)^2}\right)
    &
    (D&=2)
\end{align}
but cannot separate out an $x$-independent counterterm.

\section{The dispersion method in three and two dimensions}\label{sec:3D dispersion}
In this section we explicitly derive the dispersion formalism in both three and two dimensions by renormalizing the contact interaction on a lattice.
This non-perturbative renormalization allows to extract regularization independent observables (see also \Refs{Seki:2005ns, Epelbaum:2018zli}).
We show that it is possible to tune the contact strength parameter in a finite volume for a given discretization scheme such that one directly obtains continuum infinite volume results when using the dispersion formalism---without any further extrapolation.

\subsection{Three dimensions}

According to \eqref{T matrix}, \eqref{I0} and \eqref{spherical FD}, we find that the phase shifts are related to the contact interaction by
\begin{equation}\label{eq:blah blah}
	p \cot \delta_3(p)
	= \lim\limits_{\Lambda \to \infty}\frac{2 \pi}{\mu}\frac{1}{T(p, \Lambda)} + i p
	= \lim\limits_{\Lambda \to \infty}
		\frac{2 \pi}{\mu} \left[
			\frac{1}{C(\Lambda)} - I_3(p, \Lambda)
		\right]
	\, ,
\end{equation}
with
\begin{equation}
	I_3(p, \Lambda)
	=
	-\frac{\mu}{2 \pi}
	\left[
	i p + \frac{2 \Lambda}{\pi} + \frac{2  p}{\pi} \log \left( \frac{\Lambda - p}{\Lambda + p}\right)
	\right]
\end{equation}
The contact interaction cannot depend on any dynamic momenta; it is only possible to absorb momentum independent regulator terms when renormalizing the contact interaction.
It is still possible to renormalize the interaction such that the phase shifts, in the limit of $\Lambda \to \infty$, are independent of the cutoff by choosing the renormalized strength $C_R$ according to
\begin{equation}\label{eq:three-d-counterterm}
	\frac{2 \pi}{\mu} \frac{1}{C_R(\Lambda)} + \frac{2 \Lambda}{\pi}
	\equiv
	- \frac{1}{a_3}
	=
	p \cot \delta_3(p)
	\, .
\end{equation}

In particular, because the limit of $\Lambda \to \infty$ is well defined for this choice of the contact interaction parameter $C_R(\Lambda)$, one is able to evaluate both sides for a given momentum, such as $p=0$
\begin{equation}
	- \frac{1}{a_3}
	=
	\lim\limits_{p \to 0}\lim\limits_{\Lambda \to \infty}
		\left[
			\frac{2 \pi}{\mu}\frac{1}{T(p, \Lambda)} \bigg|_{C=C_R} + i p
	\right]
	=
	\lim\limits_{\Lambda \to \infty}
	\frac{2 \pi}{\mu}
		\left[
		\frac{1}{C_R(\Lambda)} - I_3(0, \Lambda)
		\right]
	\, .
\end{equation}

We now want to find an equivalent expression to the finite-volume zeta functions in presence of a discretization scheme.
In particular, the discretization scheme depends on the implementation of the kinetic operator $K^{(n_s)}$ and thus depends on the $n_s$ parameter.
The lattice spacing can be identified with the hard momentum cutoff $\Lambda = \pi / \epsilon$.
That is, the expectation value of the dispersion scales as $\hat K^{(n_s)}(\epsilon) \ket{p} = p^2 [1 + \mathcal O(\epsilon p)^{2 n_s}]\ket{p}$.

If one replaces the continuum momentum dispersion $q^2$ in $I_3$ with the kinetic operator for a given lattice spacing and discretization, one defines a sequence in $n_s$ which converges against $I_3$ in the limit of $n_s \to \infty$
\begin{equation}
	\lim\limits_{n_s \to \infty} I^{(n_s)}_3(p, \Lambda) = I_3(p, \Lambda)
	\, , \qquad
	I^{(n_s)}_3\left(p, \Lambda=\frac{\pi}{\epsilon} \right)
	=
	    \int\limits_{-\pi/\epsilon}^{+\pi/\epsilon}
        \mathrm{d}^3 \vec{q}
        \left[
            \PV \left(
                \frac{1}{
                    E - \frac{1}{2\mu} K_{qq}^{(n_s)} }
                \right)
            -i \pi \delta\left(E - \frac{1}{2\mu}K_{qq}^{(n_s)}\right)
        \right]
        \, .
\end{equation}
We furthermore define a sequence for the contact strength parameter depending on the cutoff and the employed discretization scheme which equivalently converges against the continuum result.
This sequence is determined by matching against the dispersion integral for each value of the cutoff and for each discretization scheme
\begin{equation}\label{eq:dispersion-renormalization}
	\lim\limits_{n_s \to \infty} C^{(n_s)}_R(\Lambda) = C_R(\Lambda) \, ,
	\qquad
	- \frac{1}{a_3}
	\equiv
	\frac{2 \pi}{\mu}
		\left[
		\frac{1}{C_R^{(n_s)}(\Lambda)} - I_3^{(n_s)}(0, \Lambda)
		\right]
	\, .
\end{equation}
It is possible to make this choice since both terms do not depend on any external momentum $p$.
This is specific for the contact interaction.
One can view this choice as the renormalization equation for contact interaction in presence of lattice discretization which, by definition, trivially satisfies
\begin{equation}
	- \frac{1}{a_3}
	=
	\lim\limits_{\Lambda \to \infty} \lim\limits_{n_s \to \infty}
	\frac{2 \pi}{\mu}
		\left[
		\frac{1}{C_R^{(n_s)}(\Lambda)} - I_3^{(n_s)}(0, \Lambda)
		\right]
	\, .
\end{equation}
In fact, it satisfies this equation even without the limits.

Next we address how this renormalization choice relates to the dispersion zeta function.
For any lattice implementation of a contact interaction with strength $c^\dispersion$ in finite volume, the Schr\"odinger equation can be rewritten as
\begin{equation}\label{eq:schroe}
	\hat G(E) \hat V \ket{\psi} = E\ket{\psi}
	\quad \Rightarrow \quad
	0 = 1 - c^\dispersion I_{3, \FV}^{(n_s)}\left(\sqrt{2 \mu E^\dispersion}, \Lambda = \frac{\pi}{\epsilon}\right) \, ,
\end{equation}
where $E^\dispersion$ are the finite volume energy levels, which depend on the employed discretization scheme and on the contact interaction of strength $c^\dispersion$.
The finite volume sum $I_{3, \FV}^{(n_s)}(p, \pi/\epsilon)$ is obtained by replacing the integral $d^3 \vec q$ in $I_3^{(n_s)}(p, \Lambda)$ with a sum  over vectors $\vec q = 2 \pi \vec n / L$.
Because the above equation is true for any value of $c^\dispersion$ and its corresponding spectrum, it is especially true for $c^\dispersion = C_R^{(n_s)}(\Lambda)$.
This means that
\begin{equation}
	- \frac{1}{a_3}
	=
	\frac{2 \pi}{\mu}
		\left[
		I_{3, \FV}^{(n_s)}\left(\sqrt{2 \mu E_i}, \frac{\pi}{\epsilon}\right)
		- I_3^{(n_s)}\left(0, \frac{\pi}{\epsilon}\right)
		\right]
	\, ,
\end{equation}
which defines the dispersion zeta function
\begin{align}\label{eq:dispersion-zeta-form}
	- \frac{1}{a_3}
	=
	\frac{1}{\pi L}
	S^{\dispersion}_3(x^\dispersion)
	&=\frac{1}{\pi L}\left(\sum\limits_{n \in \BZ}\frac{1}{K_{nn}^{(n_s)} - x^\dispersion} - \mathcal{L}_3^{\dispersion} \frac{N}{2}\right)
	\, ,
	\\ \label{eq:dispersion-zeta-contact}
	\mathcal{L}_3^\dispersion
	&=
	\frac{2 \pi^2 L}{\mu}
	I_3^{(n_s)}\left(0, \Lambda = \frac{\pi}{\epsilon}\right)
	\overset{n_s\to\infty}{\longrightarrow} 15.348
	\, .
\end{align}
See \Secref{dispersion-counterterm} for the computation of this coefficient.
Equation \eqref{dispersion-zeta-form} explains why results directly match the continuum infinite volume phase shifts when computed with this modified zeta function.
Note that this result does not hold for general finite-range interactions if it is not possible to make an equivalent choice as in \eqref{dispersion-renormalization}.
We stress that this derivation uses the analytic  expression for the $T$-matrix and simplifies drastically because the phase shifts for a renormalized contact interaction are momentum independent.
This momentum independence had the consequence that the counter term in \eqref{dispersion-zeta-contact} is momentum independent as well.

\subsection{Two dimensions}

In two dimensions the analog of~\eqref{blah blah} is
\begin{equation}
\cot \delta_2(p) - i =\lim_{\Lambda\to\infty}\frac{2}{\mu}\left(\frac{1}{C(\Lambda)}- I_2(p, \Lambda)\right)\ , 
\end{equation}
with
\begin{equation}
I_2(p, \Lambda)
=
-\frac{\mu}{\pi } \log \left(\frac{p}{\sqrt{\Lambda ^2-p^2}}\right)
+ i\frac{\mu }{2}
\ .
\end{equation}
Our renormalized coefficient is defined by using the phase shift condition for a contact interaction in 2-D~\eqref{2d contact phase shift} in the $\Lambda\to\infty$ limit,
\begin{equation}\label{eq:log stuff}
	\frac{2}{\mu}\frac{1}{C_R(\Lambda)} + \frac{2}{\pi } \log \left(\frac{p}{\Lambda}\right)
	=
	\frac { 2 } { \pi } \log \left( p \tilde a _ { 2 } \right)\ ,
\end{equation}
which ensures the renormalized contact strength $C_R(\Lambda)$ is momentum independent.
With the kinetic operator for a given lattice spacing and discretization, we again define a sequence in $n_s$ which converges against $I_2$ in the limit of $n_s \to \infty$,
\begin{equation}
	\lim\limits_{n_s \to \infty} I^{(n_s)}_2(p, \Lambda) = I_2(p, \Lambda)
	\, , \qquad
	I^{(n_s)}_2\left(p, \Lambda=\frac{\pi}{\epsilon} \right)
	=
	    \int\limits_{-\pi/\epsilon}^{+\pi/\epsilon}
        \mathrm{d}^2 \vec{q}
        \left[
            \PV \left(
                \frac{1}{
                    E - \frac{1}{2\mu} K_{qq}^{(n_s)} }
                \right)
            -i \pi \delta\left(E - \frac{1}{2\mu}K_{qq}^{(n_s)}\right)
        \right]
        \, .
\end{equation}
As was done prior to~\eqref{dispersion-renormalization}, we also define a sequence for the discrete coefficient $C^{(n_s)}_R(\Lambda)$ that is determined by matching against the dispersion integral for each value of the cutoff and for each discretization scheme.  However, in this case, due to the presence of logarithms in~\eqref{log stuff}, we first subtract the expression $\frac{2}{\pi}\log(pL/2\pi)$ prior to setting $p=0$,
\begin{equation}
	\lim\limits_{p\to 0}
	\left[
		\frac { 2 } { \pi } \log \left( p \tilde a _ { 2 } \right)-\frac{2}{\pi } \log \left(\frac{pL}{2\pi}\right)
	\right]
	=
	\frac { 2 } { \pi } \log \left(2\pi \frac{\tilde a _ { 2 }}{L} \right)
	\equiv
	\frac{2 \pi}{\mu}
		\left[
		\frac{1}{C_R^{(n_s)}(\Lambda)} - \left.\left(I_2^{(n_s)}(p, \Lambda) - i\frac{\mu}{2}+\frac{\mu}{\pi^2} \log \left(\frac{pL}{2\pi}\right)\right)\right|_{p=0}
		\right]
	\, .
\end{equation}
Our sequence $\lim_{n_s\to\infty}C^{(n_s)}_R(\Lambda) = C_R(\Lambda)$ is well defined but implicitly depends on an external length scale $L$ due to the presence of the logarithm.  To arrive at the dispersion equation in two dimensions one repeats the steps from~\eqref{schroe} leading up to~\eqref{dispersion-zeta-form}, but now~\eqref{dispersion-zeta-form} becomes
\begin{equation}
    \frac{2}{\pi} \log \left(\frac{2\pi \tilde a_{2}}{L}\right)=\frac{1}{\pi^2}S^{\dispersion}_2\left(x^\dispersion\right)
    =
    \frac{1}{\pi^2}
    \left(
        \sum_{n\in\operatorname{B.Z.}}\frac{1}{\tilde{K}^N_{nn}-x^\dispersion}
        -2\pi \log \left(\mathcal{L}^\dispersion_2\frac{N}{2}\right)
    \right)\ .
\end{equation}
Here
\begin{equation}
    \mathcal{L}^\dispersion_{2}
    =
    \exp \left(\log (2)-G \frac{2}{\pi}\right)
    =
    1.116306393581637659468497 \ldots
\end{equation}
and $G$ is Catalan's constant.  We derive this counterterm in \ref{sec:dispersion-counterterm}.  The renormalized coefficient in this case is
\begin{equation}\label{eq:C2-dispersion}
C^{(n_s)}_R(\Lambda)=-\frac{ \pi}{\mu \log \left(\tilde a_{2} \counterterm^\dispersion_2\Lambda\right)}\ ,
\end{equation}
where now the coefficient $\counterterm_2^\dispersion$ carries a $n_s$ dependence.

\section{The Dispersion Counterterm}
\label{sec:dispersion-counterterm}

To evaluate the infinite-volume integral in \eqref{dispersion-counter-integral}, we rescale the $n$ integration to extract $N$ out of the integral and rescale  $x \to \xtilde = 4 \pi^2  x/(N/2)^2$
\begin{align}\label{eq:rescaled-counterterm-integral}
    \int_{-N/2}^{+N/2} \mathrm{d}^Dn\; \PV \frac{1}{\tilde K_{nn}^N - x}
    =
    4\pi^2 \left(\frac{N}{2}\right)^{D-2} \int_{-1}^{+1} \mathrm{d}^D\nu\; \PV \frac{1}{4\sum_{ds} \gamma^{(\nstep)}_s \cos \pi \nu s - \xtilde}
    \, ,
\end{align}
which is well defined for any dimension if $\xtilde\neq0$ and for $\xtilde=0$ if $D>2$.
For $\xtilde\leq0$ the sum over dimensions can be isolated by introducing another integral
\begin{equation}
    \label{eq:dispersion counterterm}
    \eqref{rescaled-counterterm-integral}
    =
    4 \pi^2 \left(\frac{N}{2}\right)^{D-2}\int_{0}^{\infty} 2y\; \mathrm{d}y\; e^{\xtilde y^2}\left(\int_{-1}^{+1} \mathrm{d}\nu\; e^{-4y^2 \sum_s \gamma_s^{(\nstep)} \cos \pi \nu s}\right)^D
\end{equation}
which can be numerically evaluated.
This trick relies on the Laplacian stencil not coupling momenta in different directions \eqref{kinetic}.

\begin{figure}[htb]
    \scalebox{0.9}{\input{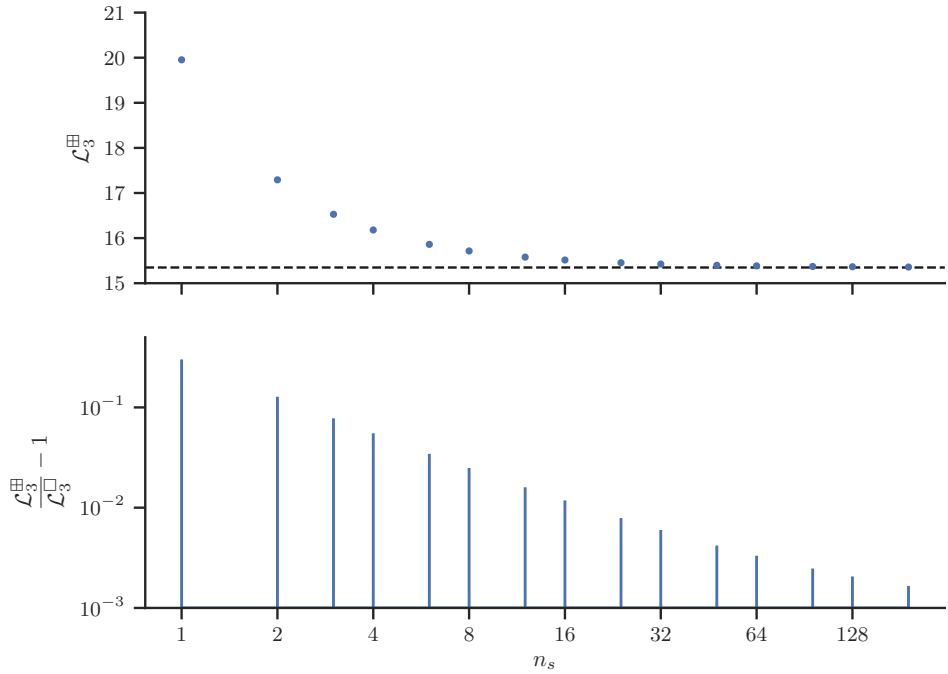}}
    \caption{
    	In the top panel we show the dispersion counterterm $\counterterm^{\dispersion}_{3}$ in \eqref{dispersion-zeta-contact} as a function of $\nstep$, and the $\nstep=\infty$ result, as a dashed line.
	In the bottom panel we show an alternate view into how the counterterm converges to the $\nstep=\infty$ value.
    }
    \label{fig:nstep counterterm}
\end{figure}

The counterterm for the leading divergence $\counterterm_D^{\dispersion}$ is the $\xtilde=0$ value.
For three dimensions, we show this counterterm and how it differs from the $\nstep \to \infty$ counterterm in \Figref{nstep counterterm} and provide precise values in table \tabref{diserpersion-zeta-3d-counterterm-counterterm}.
\begin{table}[htb]
    \begin{tabular}[t]{S[table-format=1.0]S[table-format=-1.15,table-auto-round=false]}
\toprule
  {$n_{s}$} &  {$\mathcal L^{\dispersion}_3$} \\ \midrule
        1 &                   19.95484069754250 \\
        2 &                   17.29373815124490 \\
        3 &                   16.52937382320310 \\
        4 &                   16.18180866760400 \\
        5 &                   15.98674421926657 \\
        6 &                   15.86306583941247 \\
        7 &                   15.77814195390823 \\
        8 &                   15.71645955032004 \\
        9 &                   15.66975024312220 \\
       10 &                   15.63322294937334 \\
       11 &                   15.60391847090050 \\
       12 &                   15.57991454331338 \\
       13 &                   15.55991041141987 \\
       14 &                   15.54299563540736 \\
       15 &                   15.52851460484342 \\
       16 &                   15.51598363500783 \\
       17 &                   15.50503834281211 \\
\bottomrule
\end{tabular}
\hspace{2em}
\begin{tabular}[t]{S[table-format=1.0]S[table-format=-1.15,table-auto-round=false]}
\toprule
  {$n_{s}$} &  {$\mathcal L^{\dispersion}_3$} \\ \midrule
       18 &                   15.49539917228799 \\
       19 &                   15.48684818183953 \\
       20 &                   15.47921303345776 \\
       21 &                   15.47235571159063 \\
       22 &                   15.46616442211375 \\
       23 &                   15.46054767500336 \\
       24 &                   15.45542989514885 \\
       25 &                   15.45074812098955 \\
       26 &                   15.44644948965527 \\
       27 &                   15.44248929886918 \\
       28 &                   15.43882949733264 \\
       29 &                   15.43543749725523 \\
       30 &                   15.43228523176677 \\
       31 &                   15.42934840038550 \\
       32 &                   15.42660586027747 \\
       33 &                   15.42403913154090 \\
       34 &                   15.42163199240652 \\
\bottomrule
\end{tabular}
\hspace{2em}
\begin{tabular}[t]{S[table-format=1.0]S[table-format=-1.15,table-auto-round=false]}
\toprule
  {$n_{s}$} &  {$\mathcal L^{\dispersion}_3$} \\ \midrule
       35 &                   15.41937014589034 \\
       36 &                   15.41724094363697 \\
       37 &                   15.41523315584890 \\
       38 &                   15.41333677859132 \\
       39 &                   15.41154287159014 \\
       40 &                   15.40984342105034 \\
       41 &                   15.40823122311402 \\
       42 &                   15.40669978443130 \\
       43 &                   15.40524323698864 \\
       44 &                   15.40385626486986 \\
       45 &                   15.40253404104781 \\
       46 &                   15.40127217264322 \\
       47 &                   15.40006665335855 \\
       48 &                   15.39891382201564 \\
       49 &                   15.39781032630417 \\
       50 &                   15.39675309099405 \\
 $\infty$ &                   15.34824844606382 \\
\bottomrule
\end{tabular}

    \caption{
    	\label{tab:diserpersion-zeta-3d-counterterm-counterterm}
		Counter term coefficients for the three-dimensional dispersion zeta function defined in \eqref{dispersion-zeta-contact}.
    }
\end{table}

If we assume $n_{s}=\infty$ we can obtain analytic solutions when $\tilde x=0$.  For $D=3$ we find
\begin{multline}
\counterterm_3^{\dispersion}=-8 G-4 i \left\{2 \text{Li}_2\left(1-\sqrt[4]{-1}\right)-2 \text{Li}_2\left(1+(-1)^{3/4}\right)+\text{Li}_2\left(3 i-2 i
   \sqrt{2}\right)-2 \text{Li}_2\left(\frac{1}{2} \left((-1-i)+\sqrt{2}\right)\right)\right.\\
   \left.+2 \text{Li}_2\left(\frac{1}{2}
   \left((-1+i)+\sqrt{2}\right)\right)-2 \text{Li}_2\left(\frac{2}{(1-i)+\sqrt{2}}\right)+2
   \text{Li}_2\left(\frac{2}{(1+i)+\sqrt{2}}\right)+2 \text{Li}_2\left(\frac{2 i}{(1+3 i)+(1+2 i)
   \sqrt{2}}\right)\right.\\
   \left.-\text{Li}_2\left(i \left(-3+2 \sqrt{2}\right)\right)-2 \text{Li}_2\left(\frac{2}{(3+i)+(2+i)
   \sqrt{2}}\right)\right\}+\pi  \log \left(7880+5572 \sqrt{2}\right)\ ,
   \end{multline}
where $G$ is Catalan's constant and $\text{Li}_2$ is a polylogarithm of order 2.  For $D=2$ the dominant $N$ part of~\eqref{rescaled-counterterm-integral}, after subtracting off the logarithmic singularity in $\sqrt{x}$, is logarithmic,
\begin{equation}
 \int_{-N/2}^{+N/2} \mathrm{d}^2n\; \PV \frac{1}{\bm n^2 - x}-2\pi\log\left(\sqrt{x}\right)= 2\pi\log\left(\frac{N}{2}\right)-4\left(G-\frac{\pi }{2}\log(2)\right)+\mathcal{O}(N^{-1})=2\pi\log\left(\counterterm_2^{\dispersion}\frac{N}{2}\right)+\mathcal{O}(N^{-1})\ ,
\end{equation}
with
\begin{equation}
\counterterm_2^\dispersion=\exp\left(\log(2)-G\frac{2}{\pi}\right)\ .
\end{equation}

\bibliography{master}

\begin{thebibliography}{66}%
\makeatletter
\providecommand \@ifxundefined [1]{%
 \@ifx{#1\undefined}
}%
\providecommand \@ifnum [1]{%
 \ifnum #1\expandafter \@firstoftwo
 \else \expandafter \@secondoftwo
 \fi
}%
\providecommand \@ifx [1]{%
 \ifx #1\expandafter \@firstoftwo
 \else \expandafter \@secondoftwo
 \fi
}%
\providecommand \natexlab [1]{#1}%
\providecommand \enquote  [1]{``#1''}%
\providecommand \bibnamefont  [1]{#1}%
\providecommand \bibfnamefont [1]{#1}%
\providecommand \citenamefont [1]{#1}%
\providecommand \href@noop [0]{\@secondoftwo}%
\providecommand \href [0]{\begingroup \@sanitize@url \@href}%
\providecommand \@href[1]{\@@startlink{#1}\@@href}%
\providecommand \@@href[1]{\endgroup#1\@@endlink}%
\providecommand \@sanitize@url [0]{\catcode `\\12\catcode `\$12\catcode
  `\&12\catcode `\#12\catcode `\^12\catcode `\_12\catcode `\%12\relax}%
\providecommand \@@startlink[1]{}%
\providecommand \@@endlink[0]{}%
\providecommand \url  [0]{\begingroup\@sanitize@url \@url }%
\providecommand \@url [1]{\endgroup\@href {#1}{\urlprefix }}%
\providecommand \urlprefix  [0]{URL }%
\providecommand \Eprint [0]{\href }%
\providecommand \doibase [0]{http://dx.doi.org/}%
\providecommand \selectlanguage [0]{\@gobble}%
\providecommand \bibinfo  [0]{\@secondoftwo}%
\providecommand \bibfield  [0]{\@secondoftwo}%
\providecommand \translation [1]{[#1]}%
\providecommand \BibitemOpen [0]{}%
\providecommand \bibitemStop [0]{}%
\providecommand \bibitemNoStop [0]{.\EOS\space}%
\providecommand \EOS [0]{\spacefactor3000\relax}%
\providecommand \BibitemShut  [1]{\csname bibitem#1\endcsname}%
\let\auto@bib@innerbib\@empty
\bibitem [{\citenamefont {Baker}(1999)}]{PhysRevC.60.054311}%
  \BibitemOpen
  \bibfield  {author} {\bibinfo {author} {\bibfnamefont {George~A.}\
  \bibnamefont {Baker}},\ }\bibfield  {title} {\enquote {\bibinfo {title}
  {Neutron matter model},}\ }\href {\doibase 10.1103/PhysRevC.60.054311}
  {\bibfield  {journal} {\bibinfo  {journal} {Phys. Rev. C}\ }\textbf {\bibinfo
  {volume} {60}},\ \bibinfo {pages} {054311} (\bibinfo {year}
  {1999})}\BibitemShut {NoStop}%
\bibitem [{\citenamefont {Wigner}(1955)}]{Wigner:1955zz}%
  \BibitemOpen
  \bibfield  {author} {\bibinfo {author} {\bibfnamefont {Eugene~P.}\
  \bibnamefont {Wigner}},\ }\bibfield  {title} {\enquote {\bibinfo {title}
  {{Lower Limit for the Energy Derivative of the Scattering Phase Shift}},}\
  }\href {\doibase 10.1103/PhysRev.98.145} {\bibfield  {journal} {\bibinfo
  {journal} {Phys. Rev.}\ }\textbf {\bibinfo {volume} {98}},\ \bibinfo {pages}
  {145--147} (\bibinfo {year} {1955})}\BibitemShut {NoStop}%
\bibitem [{\citenamefont {Phillips}\ and\ \citenamefont
  {Cohen}(1997)}]{Phillips:1996ae}%
  \BibitemOpen
  \bibfield  {author} {\bibinfo {author} {\bibfnamefont {Daniel~R.}\
  \bibnamefont {Phillips}}\ and\ \bibinfo {author} {\bibfnamefont {Thomas~D.}\
  \bibnamefont {Cohen}},\ }\bibfield  {title} {\enquote {\bibinfo {title} {{How
  short is too short? Constraining contact interactions in nucleon-nucleon
  scattering}},}\ }\href {\doibase 10.1016/S0370-2693(96)01411-6} {\bibfield
  {journal} {\bibinfo  {journal} {Phys. Lett.}\ }\textbf {\bibinfo {volume}
  {B390}},\ \bibinfo {pages} {7--12} (\bibinfo {year} {1997})},\ \Eprint
  {http://arxiv.org/abs/nucl-th/9607048} {arXiv:nucl-th/9607048 [nucl-th]}
  \BibitemShut {NoStop}%
\bibitem [{\citenamefont {Hammer}\ and\ \citenamefont
  {Lee}(2010)}]{Hammer:2010fw}%
  \BibitemOpen
  \bibfield  {author} {\bibinfo {author} {\bibfnamefont {H.~W.}\ \bibnamefont
  {Hammer}}\ and\ \bibinfo {author} {\bibfnamefont {Dean}\ \bibnamefont
  {Lee}},\ }\bibfield  {title} {\enquote {\bibinfo {title} {{Causality and the
  effective range expansion}},}\ }\href {\doibase 10.1016/j.aop.2010.06.006}
  {\bibfield  {journal} {\bibinfo  {journal} {Annals Phys.}\ }\textbf {\bibinfo
  {volume} {325}},\ \bibinfo {pages} {2212--2233} (\bibinfo {year} {2010})},\
  \Eprint {http://arxiv.org/abs/1002.4603} {arXiv:1002.4603 [nucl-th]}
  \BibitemShut {NoStop}%
\bibitem [{\citenamefont {Hamber}\ \emph {et~al.}(1983)\citenamefont {Hamber},
  \citenamefont {Marinari}, \citenamefont {Parisi},\ and\ \citenamefont
  {Rebbi}}]{Hamber198399}%
  \BibitemOpen
  \bibfield  {author} {\bibinfo {author} {\bibfnamefont {H.W.}\ \bibnamefont
  {Hamber}}, \bibinfo {author} {\bibfnamefont {E.}~\bibnamefont {Marinari}},
  \bibinfo {author} {\bibfnamefont {G.}~\bibnamefont {Parisi}}, \ and\ \bibinfo
  {author} {\bibfnamefont {C.}~\bibnamefont {Rebbi}},\ }\bibfield  {title}
  {\enquote {\bibinfo {title} {Numerical simulations of quantum
  chromodynamics},}\ }\href {\doibase 10.1016/0370-2693(83)91412-0} {\bibfield
  {journal} {\bibinfo  {journal} {Physics Letters B}\ }\textbf {\bibinfo
  {volume} {124}},\ \bibinfo {pages} {99 -- 104} (\bibinfo {year}
  {1983})}\BibitemShut {NoStop}%
\bibitem [{\citenamefont {L\"uscher}(1986{\natexlab{a}})}]{luscher:1986I}%
  \BibitemOpen
  \bibfield  {author} {\bibinfo {author} {\bibfnamefont {M.}~\bibnamefont
  {L\"uscher}},\ }\bibfield  {title} {\enquote {\bibinfo {title} {Volume
  dependence of the energy spectrum in massive quantum field theories i},}\
  }\href@noop {} {\bibfield  {journal} {\bibinfo  {journal} {Communications in
  Mathematical Physics}\ }\textbf {\bibinfo {volume} {104}},\ \bibinfo {pages}
  {177--206} (\bibinfo {year} {1986}{\natexlab{a}})},\ \bibinfo {note}
  {10.1007/BF01211589}\BibitemShut {NoStop}%
\bibitem [{\citenamefont {L\"uscher}(1986{\natexlab{b}})}]{luscher:1986II}%
  \BibitemOpen
  \bibfield  {author} {\bibinfo {author} {\bibfnamefont {M.}~\bibnamefont
  {L\"uscher}},\ }\bibfield  {title} {\enquote {\bibinfo {title} {Volume
  dependence of the energy spectrum in massive quantum field theories ii},}\
  }\href@noop {} {\bibfield  {journal} {\bibinfo  {journal} {Communications in
  Mathematical Physics}\ }\textbf {\bibinfo {volume} {105}},\ \bibinfo {pages}
  {153--188} (\bibinfo {year} {1986}{\natexlab{b}})},\ \bibinfo {note}
  {10.1007/BF01211097}\BibitemShut {NoStop}%
\bibitem [{\citenamefont {Wiese}(1989)}]{wiese1989}%
  \BibitemOpen
  \bibfield  {author} {\bibinfo {author} {\bibfnamefont {U.-J.}\ \bibnamefont
  {Wiese}},\ }\bibfield  {title} {\enquote {\bibinfo {title} {Identification of
  resonance parameters from the finite volume energy spectrum},}\ }\href
  {\doibase 10.1016/0920-5632(89)90171-0} {\bibfield  {journal} {\bibinfo
  {journal} {Nuclear Physics B - Proceedings Supplements}\ }\textbf {\bibinfo
  {volume} {9}},\ \bibinfo {pages} {609 -- 613} (\bibinfo {year}
  {1989})}\BibitemShut {NoStop}%
\bibitem [{\citenamefont {L\"uscher}(1991{\natexlab{a}})}]{Luscher1991}%
  \BibitemOpen
  \bibfield  {author} {\bibinfo {author} {\bibfnamefont {Martin}\ \bibnamefont
  {L\"uscher}},\ }\bibfield  {title} {\enquote {\bibinfo {title} {Two-particle
  states on a torus and their relation to the scattering matrix},}\ }\href
  {\doibase 10.1016/0550-3213(91)90366-6} {\bibfield  {journal} {\bibinfo
  {journal} {Nuclear Physics B}\ }\textbf {\bibinfo {volume} {354}},\ \bibinfo
  {pages} {531 -- 578} (\bibinfo {year} {1991}{\natexlab{a}})}\BibitemShut
  {NoStop}%
\bibitem [{\citenamefont {L\"uscher}(1991{\natexlab{b}})}]{Luscher1991237}%
  \BibitemOpen
  \bibfield  {author} {\bibinfo {author} {\bibfnamefont {Martin}\ \bibnamefont
  {L\"uscher}},\ }\bibfield  {title} {\enquote {\bibinfo {title} {Signatures of
  unstable particles in finite volume},}\ }\href {\doibase
  10.1016/0550-3213(91)90584-K} {\bibfield  {journal} {\bibinfo  {journal}
  {Nuclear Physics B}\ }\textbf {\bibinfo {volume} {364}},\ \bibinfo {pages}
  {237 -- 251} (\bibinfo {year} {1991}{\natexlab{b}})}\BibitemShut {NoStop}%
\bibitem [{\citenamefont {Valiente}\ and\ \citenamefont
  {Zinner}(2016)}]{Valiente:2015oya}%
  \BibitemOpen
  \bibfield  {author} {\bibinfo {author} {\bibfnamefont {Manuel}\ \bibnamefont
  {Valiente}}\ and\ \bibinfo {author} {\bibfnamefont {Nikolaj~Thomas}\
  \bibnamefont {Zinner}},\ }\bibfield  {title} {\enquote {\bibinfo {title}
  {{Unitary fermions and L\"uscher's formula on a crystal}},}\ }\href {\doibase
  10.1007/s11433-016-0205-x} {\bibfield  {journal} {\bibinfo  {journal} {Sci.
  China Phys. Mech. Astron.}\ }\textbf {\bibinfo {volume} {59}},\ \bibinfo
  {pages} {114211} (\bibinfo {year} {2016})},\ \Eprint
  {http://arxiv.org/abs/1506.05458} {arXiv:1506.05458 [cond-mat.quant-gas]}
  \BibitemShut {NoStop}%
\bibitem [{\citenamefont {Seki}\ and\ \citenamefont {van
  Kolck}(2006)}]{Seki:2005ns}%
  \BibitemOpen
  \bibfield  {author} {\bibinfo {author} {\bibfnamefont {Ryoichi}\ \bibnamefont
  {Seki}}\ and\ \bibinfo {author} {\bibfnamefont {U.}~\bibnamefont {van
  Kolck}},\ }\bibfield  {title} {\enquote {\bibinfo {title} {{Effective field
  theory of nucleon-nucleon scattering on large discrete lattices}},}\ }\href
  {\doibase 10.1103/PhysRevC.73.044006} {\bibfield  {journal} {\bibinfo
  {journal} {Phys. Rev.}\ }\textbf {\bibinfo {volume} {C73}},\ \bibinfo {pages}
  {044006} (\bibinfo {year} {2006})},\ \Eprint
  {http://arxiv.org/abs/nucl-th/0509094} {arXiv:nucl-th/0509094 [nucl-th]}
  \BibitemShut {NoStop}%
\bibitem [{\citenamefont {Ishii}\ \emph {et~al.}(2007)\citenamefont {Ishii},
  \citenamefont {Aoki},\ and\ \citenamefont {Hatsuda}}]{Ishii:2006ec}%
  \BibitemOpen
  \bibfield  {author} {\bibinfo {author} {\bibfnamefont {N.}~\bibnamefont
  {Ishii}}, \bibinfo {author} {\bibfnamefont {S.}~\bibnamefont {Aoki}}, \ and\
  \bibinfo {author} {\bibfnamefont {T.}~\bibnamefont {Hatsuda}},\ }\bibfield
  {title} {\enquote {\bibinfo {title} {{The Nuclear Force from Lattice QCD}},}\
  }\href {\doibase 10.1103/PhysRevLett.99.022001} {\bibfield  {journal}
  {\bibinfo  {journal} {Phys. Rev. Lett.}\ }\textbf {\bibinfo {volume} {99}},\
  \bibinfo {pages} {022001} (\bibinfo {year} {2007})},\ \Eprint
  {http://arxiv.org/abs/nucl-th/0611096} {arXiv:nucl-th/0611096 [nucl-th]}
  \BibitemShut {NoStop}%
\bibitem [{\citenamefont {Nemura}\ \emph {et~al.}(2009)\citenamefont {Nemura},
  \citenamefont {Ishii}, \citenamefont {Aoki},\ and\ \citenamefont
  {Hatsuda}}]{Nemura:2008sp}%
  \BibitemOpen
  \bibfield  {author} {\bibinfo {author} {\bibfnamefont {Hidekatsu}\
  \bibnamefont {Nemura}}, \bibinfo {author} {\bibfnamefont {Noriyoshi}\
  \bibnamefont {Ishii}}, \bibinfo {author} {\bibfnamefont {Sinya}\ \bibnamefont
  {Aoki}}, \ and\ \bibinfo {author} {\bibfnamefont {Tetsuo}\ \bibnamefont
  {Hatsuda}},\ }\bibfield  {title} {\enquote {\bibinfo {title}
  {{Hyperon-nucleon force from lattice QCD}},}\ }\href {\doibase
  10.1016/j.physletb.2009.02.003} {\bibfield  {journal} {\bibinfo  {journal}
  {Phys. Lett.}\ }\textbf {\bibinfo {volume} {B673}},\ \bibinfo {pages}
  {136--141} (\bibinfo {year} {2009})},\ \Eprint
  {http://arxiv.org/abs/0806.1094} {arXiv:0806.1094 [nucl-th]} \BibitemShut
  {NoStop}%
\bibitem [{\citenamefont {Aoki}\ \emph {et~al.}(2010)\citenamefont {Aoki},
  \citenamefont {Hatsuda},\ and\ \citenamefont {Ishii}}]{Aoki:2009ji}%
  \BibitemOpen
  \bibfield  {author} {\bibinfo {author} {\bibfnamefont {Sinya}\ \bibnamefont
  {Aoki}}, \bibinfo {author} {\bibfnamefont {Tetsuo}\ \bibnamefont {Hatsuda}},
  \ and\ \bibinfo {author} {\bibfnamefont {Noriyoshi}\ \bibnamefont {Ishii}},\
  }\bibfield  {title} {\enquote {\bibinfo {title} {{Theoretical Foundation of
  the Nuclear Force in QCD and its applications to Central and Tensor Forces in
  Quenched Lattice QCD Simulations}},}\ }\href {\doibase 10.1143/PTP.123.89}
  {\bibfield  {journal} {\bibinfo  {journal} {Prog. Theor. Phys.}\ }\textbf
  {\bibinfo {volume} {123}},\ \bibinfo {pages} {89--128} (\bibinfo {year}
  {2010})},\ \Eprint {http://arxiv.org/abs/0909.5585} {arXiv:0909.5585
  [hep-lat]} \BibitemShut {NoStop}%
\bibitem [{\citenamefont {Murano}\ \emph {et~al.}(2011)\citenamefont {Murano},
  \citenamefont {Ishii}, \citenamefont {Aoki},\ and\ \citenamefont
  {Hatsuda}}]{Murano:2011nz}%
  \BibitemOpen
  \bibfield  {author} {\bibinfo {author} {\bibfnamefont {Keiko}\ \bibnamefont
  {Murano}}, \bibinfo {author} {\bibfnamefont {Noriyoshi}\ \bibnamefont
  {Ishii}}, \bibinfo {author} {\bibfnamefont {Sinya}\ \bibnamefont {Aoki}}, \
  and\ \bibinfo {author} {\bibfnamefont {Tetsuo}\ \bibnamefont {Hatsuda}},\
  }\bibfield  {title} {\enquote {\bibinfo {title} {{Nucleon-Nucleon Potential
  and its Non-locality in Lattice QCD}},}\ }\href {\doibase
  10.1143/PTP.125.1225} {\bibfield  {journal} {\bibinfo  {journal} {Prog.
  Theor. Phys.}\ }\textbf {\bibinfo {volume} {125}},\ \bibinfo {pages}
  {1225--1240} (\bibinfo {year} {2011})},\ \Eprint
  {http://arxiv.org/abs/1103.0619} {arXiv:1103.0619 [hep-lat]} \BibitemShut
  {NoStop}%
\bibitem [{\citenamefont {Aoki}\ \emph {et~al.}(2013)\citenamefont {Aoki},
  \citenamefont {Charron}, \citenamefont {Doi}, \citenamefont {Hatsuda},
  \citenamefont {Inoue},\ and\ \citenamefont {Ishii}}]{Aoki:2012bb}%
  \BibitemOpen
  \bibfield  {author} {\bibinfo {author} {\bibfnamefont {Sinya}\ \bibnamefont
  {Aoki}}, \bibinfo {author} {\bibfnamefont {Bruno}\ \bibnamefont {Charron}},
  \bibinfo {author} {\bibfnamefont {Takumi}\ \bibnamefont {Doi}}, \bibinfo
  {author} {\bibfnamefont {Tetsuo}\ \bibnamefont {Hatsuda}}, \bibinfo {author}
  {\bibfnamefont {Takashi}\ \bibnamefont {Inoue}}, \ and\ \bibinfo {author}
  {\bibfnamefont {Noriyoshi}\ \bibnamefont {Ishii}},\ }\bibfield  {title}
  {\enquote {\bibinfo {title} {{Construction of energy-independent potentials
  above inelastic thresholds in quantum field theories}},}\ }\href {\doibase
  10.1103/PhysRevD.87.034512} {\bibfield  {journal} {\bibinfo  {journal} {Phys.
  Rev.}\ }\textbf {\bibinfo {volume} {D87}},\ \bibinfo {pages} {034512}
  (\bibinfo {year} {2013})},\ \Eprint {http://arxiv.org/abs/1212.4896}
  {arXiv:1212.4896 [hep-lat]} \BibitemShut {NoStop}%
\bibitem [{\citenamefont {Kurth}\ \emph {et~al.}(2013)\citenamefont {Kurth},
  \citenamefont {Ishii}, \citenamefont {Doi}, \citenamefont {Aoki},\ and\
  \citenamefont {Hatsuda}}]{Kurth:2013tua}%
  \BibitemOpen
  \bibfield  {author} {\bibinfo {author} {\bibfnamefont {T.}~\bibnamefont
  {Kurth}}, \bibinfo {author} {\bibfnamefont {N.}~\bibnamefont {Ishii}},
  \bibinfo {author} {\bibfnamefont {T.}~\bibnamefont {Doi}}, \bibinfo {author}
  {\bibfnamefont {S.}~\bibnamefont {Aoki}}, \ and\ \bibinfo {author}
  {\bibfnamefont {T.}~\bibnamefont {Hatsuda}},\ }\bibfield  {title} {\enquote
  {\bibinfo {title} {{Phase shifts in $I=2 {\pi}{\pi}$-scattering from two
  lattice approaches}},}\ }\href {\doibase 10.1007/JHEP12(2013)015} {\bibfield
  {journal} {\bibinfo  {journal} {JHEP}\ }\textbf {\bibinfo {volume} {12}},\
  \bibinfo {pages} {015} (\bibinfo {year} {2013})},\ \Eprint
  {http://arxiv.org/abs/1305.4462} {arXiv:1305.4462 [hep-lat]} \BibitemShut
  {NoStop}%
\bibitem [{\citenamefont {Sugiura}\ \emph {et~al.}(2017)\citenamefont
  {Sugiura}, \citenamefont {Ishii},\ and\ \citenamefont
  {Oka}}]{Sugiura:2017vwo}%
  \BibitemOpen
  \bibfield  {author} {\bibinfo {author} {\bibfnamefont {Takuya}\ \bibnamefont
  {Sugiura}}, \bibinfo {author} {\bibfnamefont {Noriyoshi}\ \bibnamefont
  {Ishii}}, \ and\ \bibinfo {author} {\bibfnamefont {Makoto}\ \bibnamefont
  {Oka}},\ }\bibfield  {title} {\enquote {\bibinfo {title} {{Derivative
  Expansion of Wave Function Equivalent Potentials}},}\ }\href {\doibase
  10.1103/PhysRevD.95.074514} {\bibfield  {journal} {\bibinfo  {journal} {Phys.
  Rev.}\ }\textbf {\bibinfo {volume} {D95}},\ \bibinfo {pages} {074514}
  (\bibinfo {year} {2017})},\ \Eprint {http://arxiv.org/abs/1701.08268}
  {arXiv:1701.08268 [nucl-th]} \BibitemShut {NoStop}%
\bibitem [{\citenamefont {Yamazaki}\ and\ \citenamefont
  {Kuramashi}(2018{\natexlab{a}})}]{Yamazaki:2019vid}%
  \BibitemOpen
  \bibfield  {author} {\bibinfo {author} {\bibfnamefont {Takeshi}\ \bibnamefont
  {Yamazaki}}\ and\ \bibinfo {author} {\bibfnamefont {Yoshinobu}\ \bibnamefont
  {Kuramashi}},\ }\bibfield  {title} {\enquote {\bibinfo {title} {{Relation
  between scattering amplitude and Bethe-Salpeter wave function in quantum
  field theory}},}\ }\bibfield  {booktitle} {\emph {\bibinfo {booktitle}
  {{Proceedings, 36th International Symposium on Lattice Field Theory (Lattice
  2018): East Lansing, MI, United States, July 22-28, 2018}}},\ }\href
  {\doibase 10.22323/1.334.0077} {\bibfield  {journal} {\bibinfo  {journal}
  {PoS}\ }\textbf {\bibinfo {volume} {LATTICE2018}},\ \bibinfo {pages} {077}
  (\bibinfo {year} {2018}{\natexlab{a}})},\ \Eprint
  {http://arxiv.org/abs/1901.01670} {arXiv:1901.01670 [hep-lat]} \BibitemShut
  {NoStop}%
\bibitem [{\citenamefont {Aoki}\ \emph {et~al.}(2018)\citenamefont {Aoki},
  \citenamefont {Doi}, \citenamefont {Hatsuda},\ and\ \citenamefont
  {Ishii}}]{Aoki:2017yru}%
  \BibitemOpen
  \bibfield  {author} {\bibinfo {author} {\bibfnamefont {Sinya}\ \bibnamefont
  {Aoki}}, \bibinfo {author} {\bibfnamefont {Takumi}\ \bibnamefont {Doi}},
  \bibinfo {author} {\bibfnamefont {Tetsuo}\ \bibnamefont {Hatsuda}}, \ and\
  \bibinfo {author} {\bibfnamefont {Noriyoshi}\ \bibnamefont {Ishii}},\
  }\bibfield  {title} {\enquote {\bibinfo {title} {{Comment on “Relation
  between scattering amplitude and Bethe-Salpeter wave function in quantum
  field theory”}},}\ }\href {\doibase 10.1103/PhysRevD.98.038501} {\bibfield
  {journal} {\bibinfo  {journal} {Phys. Rev.}\ }\textbf {\bibinfo {volume}
  {D98}},\ \bibinfo {pages} {038501} (\bibinfo {year} {2018})},\ \Eprint
  {http://arxiv.org/abs/1711.09344} {arXiv:1711.09344 [hep-lat]} \BibitemShut
  {NoStop}%
\bibitem [{\citenamefont {Yamazaki}\ and\ \citenamefont
  {Kuramashi}(2018{\natexlab{b}})}]{Yamazaki:2018qut}%
  \BibitemOpen
  \bibfield  {author} {\bibinfo {author} {\bibfnamefont {Takeshi}\ \bibnamefont
  {Yamazaki}}\ and\ \bibinfo {author} {\bibfnamefont {Yoshinobu}\ \bibnamefont
  {Kuramashi}},\ }\bibfield  {title} {\enquote {\bibinfo {title} {{Reply to
  “Comment on ‘Relation between scattering amplitude and Bethe-Salpeter
  wave function in quantum field theory”’}},}\ }\href {\doibase
  10.1103/PhysRevD.98.038502} {\bibfield  {journal} {\bibinfo  {journal} {Phys.
  Rev.}\ }\textbf {\bibinfo {volume} {D98}},\ \bibinfo {pages} {038502}
  (\bibinfo {year} {2018}{\natexlab{b}})},\ \Eprint
  {http://arxiv.org/abs/1808.06299} {arXiv:1808.06299 [hep-lat]} \BibitemShut
  {NoStop}%
\bibitem [{\citenamefont {Iritani}\ \emph {et~al.}(2017)\citenamefont
  {Iritani}, \citenamefont {Aoki}, \citenamefont {Doi}, \citenamefont
  {Hatsuda}, \citenamefont {Ikeda}, \citenamefont {Inoue}, \citenamefont
  {Ishii}, \citenamefont {Nemura},\ and\ \citenamefont
  {Sasaki}}]{Iritani:2017rlk}%
  \BibitemOpen
  \bibfield  {author} {\bibinfo {author} {\bibfnamefont {Takumi}\ \bibnamefont
  {Iritani}}, \bibinfo {author} {\bibfnamefont {Sinya}\ \bibnamefont {Aoki}},
  \bibinfo {author} {\bibfnamefont {Takumi}\ \bibnamefont {Doi}}, \bibinfo
  {author} {\bibfnamefont {Testuo}\ \bibnamefont {Hatsuda}}, \bibinfo {author}
  {\bibfnamefont {Yoichi}\ \bibnamefont {Ikeda}}, \bibinfo {author}
  {\bibfnamefont {Takashi}\ \bibnamefont {Inoue}}, \bibinfo {author}
  {\bibfnamefont {Noriyoshi}\ \bibnamefont {Ishii}}, \bibinfo {author}
  {\bibfnamefont {Hidekatsu}\ \bibnamefont {Nemura}}, \ and\ \bibinfo {author}
  {\bibfnamefont {Kenji}\ \bibnamefont {Sasaki}},\ }\bibfield  {title}
  {\enquote {\bibinfo {title} {{Are two nucleons bound in lattice QCD for heavy
  quark masses? Consistency check with L\"{u}scher's finite volume formula}},}\
  }\href {\doibase 10.1103/PhysRevD.96.034521} {\bibfield  {journal} {\bibinfo
  {journal} {Phys. Rev.}\ }\textbf {\bibinfo {volume} {D96}},\ \bibinfo {pages}
  {034521} (\bibinfo {year} {2017})},\ \Eprint
  {http://arxiv.org/abs/1703.07210} {arXiv:1703.07210 [hep-lat]} \BibitemShut
  {NoStop}%
\bibitem [{\citenamefont {Iritani}\ \emph {et~al.}(2019)\citenamefont
  {Iritani}, \citenamefont {Aoki}, \citenamefont {Doi}, \citenamefont {Gongyo},
  \citenamefont {Hatsuda}, \citenamefont {Ikeda}, \citenamefont {Inoue},
  \citenamefont {Ishii}, \citenamefont {Nemura},\ and\ \citenamefont
  {Sasaki}}]{Iritani:2018zbt}%
  \BibitemOpen
  \bibfield  {author} {\bibinfo {author} {\bibfnamefont {Takumi}\ \bibnamefont
  {Iritani}}, \bibinfo {author} {\bibfnamefont {Sinya}\ \bibnamefont {Aoki}},
  \bibinfo {author} {\bibfnamefont {Takumi}\ \bibnamefont {Doi}}, \bibinfo
  {author} {\bibfnamefont {Shinya}\ \bibnamefont {Gongyo}}, \bibinfo {author}
  {\bibfnamefont {Tetsuo}\ \bibnamefont {Hatsuda}}, \bibinfo {author}
  {\bibfnamefont {Yoichi}\ \bibnamefont {Ikeda}}, \bibinfo {author}
  {\bibfnamefont {Takashi}\ \bibnamefont {Inoue}}, \bibinfo {author}
  {\bibfnamefont {Noriyoshi}\ \bibnamefont {Ishii}}, \bibinfo {author}
  {\bibfnamefont {Hidekatsu}\ \bibnamefont {Nemura}}, \ and\ \bibinfo {author}
  {\bibfnamefont {Kenji}\ \bibnamefont {Sasaki}} (\bibinfo {collaboration} {HAL
  QCD}),\ }\bibfield  {title} {\enquote {\bibinfo {title} {{Systematics of the
  HAL QCD Potential at Low Energies in Lattice QCD}},}\ }\href {\doibase
  10.1103/PhysRevD.99.014514} {\bibfield  {journal} {\bibinfo  {journal} {Phys.
  Rev.}\ }\textbf {\bibinfo {volume} {D99}},\ \bibinfo {pages} {014514}
  (\bibinfo {year} {2019})},\ \Eprint {http://arxiv.org/abs/1805.02365}
  {arXiv:1805.02365 [hep-lat]} \BibitemShut {NoStop}%
\bibitem [{\citenamefont {Gongyo}\ and\ \citenamefont
  {Aoki}(2018)}]{Gongyo:2018gou}%
  \BibitemOpen
  \bibfield  {author} {\bibinfo {author} {\bibfnamefont {Shinya}\ \bibnamefont
  {Gongyo}}\ and\ \bibinfo {author} {\bibfnamefont {Sinya}\ \bibnamefont
  {Aoki}},\ }\bibfield  {title} {\enquote {\bibinfo {title} {{Asymptotic
  behavior of Nambu–Bethe–Salpeter wave functions for scalar systems with a
  bound state}},}\ }\href {\doibase 10.1093/ptep/pty097} {\bibfield  {journal}
  {\bibinfo  {journal} {PTEP}\ }\textbf {\bibinfo {volume} {2018}},\ \bibinfo
  {pages} {093B03} (\bibinfo {year} {2018})},\ \Eprint
  {http://arxiv.org/abs/1807.02967} {arXiv:1807.02967 [hep-lat]} \BibitemShut
  {NoStop}%
\bibitem [{\citenamefont {Akahoshi}\ \emph {et~al.}(2019)\citenamefont
  {Akahoshi}, \citenamefont {Aoki}, \citenamefont {Aoyama}, \citenamefont
  {Doi}, \citenamefont {Miyamoto},\ and\ \citenamefont
  {Sasaki}}]{Akahoshi:2019klc}%
  \BibitemOpen
  \bibfield  {author} {\bibinfo {author} {\bibfnamefont {Yutaro}\ \bibnamefont
  {Akahoshi}}, \bibinfo {author} {\bibfnamefont {Sinya}\ \bibnamefont {Aoki}},
  \bibinfo {author} {\bibfnamefont {Tatsumi}\ \bibnamefont {Aoyama}}, \bibinfo
  {author} {\bibfnamefont {Takumi}\ \bibnamefont {Doi}}, \bibinfo {author}
  {\bibfnamefont {Takaya}\ \bibnamefont {Miyamoto}}, \ and\ \bibinfo {author}
  {\bibfnamefont {Kenji}\ \bibnamefont {Sasaki}},\ }\bibfield  {title}
  {\enquote {\bibinfo {title} {{$I=2$ $\pi\pi$ potential in the HAL QCD method
  with all-to-all propagators}},}\ }\href@noop {} {\  (\bibinfo {year}
  {2019})},\ \Eprint {http://arxiv.org/abs/1904.09549} {arXiv:1904.09549
  [hep-lat]} \BibitemShut {NoStop}%
\bibitem [{\citenamefont {Namekawa}\ and\ \citenamefont
  {Yamazaki}(2019)}]{Namekawa:2019xiy}%
  \BibitemOpen
  \bibfield  {author} {\bibinfo {author} {\bibfnamefont {Yusuke}\ \bibnamefont
  {Namekawa}}\ and\ \bibinfo {author} {\bibfnamefont {Takeshi}\ \bibnamefont
  {Yamazaki}},\ }\bibfield  {title} {\enquote {\bibinfo {title} {{Quark mass
  dependence of on-shell and half off-shell scattering amplitudes from
  Bethe-Salpeter wave function inside the interaction range}},}\ }\href
  {\doibase 10.1103/PhysRevD.99.114508} {\bibfield  {journal} {\bibinfo
  {journal} {Phys. Rev.}\ }\textbf {\bibinfo {volume} {D99}},\ \bibinfo {pages}
  {114508} (\bibinfo {year} {2019})},\ \Eprint
  {http://arxiv.org/abs/1904.00387} {arXiv:1904.00387 [hep-lat]} \BibitemShut
  {NoStop}%
\bibitem [{\citenamefont {McElvain}\ and\ \citenamefont
  {Haxton}(2019)}]{McElvain:2019ltw}%
  \BibitemOpen
  \bibfield  {author} {\bibinfo {author} {\bibfnamefont {K.~S.}\ \bibnamefont
  {McElvain}}\ and\ \bibinfo {author} {\bibfnamefont {W.~C.}\ \bibnamefont
  {Haxton}},\ }\bibfield  {title} {\enquote {\bibinfo {title} {{Nuclear physics
  without high-momentum potentials: Constructing the nuclear effective
  interaction directly from scattering observables}},}\ }\href {\doibase
  10.1016/j.physletb.2019.134880} {\bibfield  {journal} {\bibinfo  {journal}
  {Phys. Lett.}\ }\textbf {\bibinfo {volume} {B797}},\ \bibinfo {pages}
  {134880} (\bibinfo {year} {2019})},\ \Eprint
  {http://arxiv.org/abs/1902.03543} {arXiv:1902.03543 [nucl-th]} \BibitemShut
  {NoStop}%
\bibitem [{\citenamefont {Borasoy}\ \emph {et~al.}(2007)\citenamefont
  {Borasoy}, \citenamefont {Epelbaum}, \citenamefont {Krebs}, \citenamefont
  {Lee},\ and\ \citenamefont {Mei{\ss}ner}}]{Borasoy:2007vy}%
  \BibitemOpen
  \bibfield  {author} {\bibinfo {author} {\bibfnamefont {Bugra}\ \bibnamefont
  {Borasoy}}, \bibinfo {author} {\bibfnamefont {Evgeny}\ \bibnamefont
  {Epelbaum}}, \bibinfo {author} {\bibfnamefont {Hermann}\ \bibnamefont
  {Krebs}}, \bibinfo {author} {\bibfnamefont {Dean}\ \bibnamefont {Lee}}, \
  and\ \bibinfo {author} {\bibfnamefont {Ulf-G.}\ \bibnamefont {Mei{\ss}ner}},\
  }\bibfield  {title} {\enquote {\bibinfo {title} {{Two-particle scattering on
  the lattice: Phase shifts, spin-orbit coupling, and mixing angles}},}\ }\href
  {\doibase 10.1140/epja/i2007-10500-9} {\bibfield  {journal} {\bibinfo
  {journal} {Eur. Phys. J.}\ }\textbf {\bibinfo {volume} {A34}},\ \bibinfo
  {pages} {185--196} (\bibinfo {year} {2007})},\ \Eprint
  {http://arxiv.org/abs/0708.1780} {arXiv:0708.1780 [nucl-th]} \BibitemShut
  {NoStop}%
\bibitem [{\citenamefont {Borasoy}\ \emph {et~al.}(2008)\citenamefont
  {Borasoy}, \citenamefont {Epelbaum}, \citenamefont {Krebs}, \citenamefont
  {Lee},\ and\ \citenamefont {Mei{\ss}ner}}]{Borasoy:2007vi}%
  \BibitemOpen
  \bibfield  {author} {\bibinfo {author} {\bibfnamefont {Bugra}\ \bibnamefont
  {Borasoy}}, \bibinfo {author} {\bibfnamefont {Evgeny}\ \bibnamefont
  {Epelbaum}}, \bibinfo {author} {\bibfnamefont {Hermann}\ \bibnamefont
  {Krebs}}, \bibinfo {author} {\bibfnamefont {Dean}\ \bibnamefont {Lee}}, \
  and\ \bibinfo {author} {\bibfnamefont {Ulf-G.}\ \bibnamefont {Mei{\ss}ner}},\
  }\bibfield  {title} {\enquote {\bibinfo {title} {{Chiral effective field
  theory on the lattice at next-to-leading order}},}\ }\href {\doibase
  10.1140/epja/i2008-10544-3} {\bibfield  {journal} {\bibinfo  {journal} {Eur.
  Phys. J.}\ }\textbf {\bibinfo {volume} {A35}},\ \bibinfo {pages} {343--355}
  (\bibinfo {year} {2008})},\ \Eprint {http://arxiv.org/abs/0712.2990}
  {arXiv:0712.2990 [nucl-th]} \BibitemShut {NoStop}%
\bibitem [{\citenamefont {Lee}(2009)}]{Lee:2008fa}%
  \BibitemOpen
  \bibfield  {author} {\bibinfo {author} {\bibfnamefont {Dean}\ \bibnamefont
  {Lee}},\ }\bibfield  {title} {\enquote {\bibinfo {title} {{Lattice
  simulations for few- and many-body systems}},}\ }\href {\doibase
  10.1016/j.ppnp.2008.12.001} {\bibfield  {journal} {\bibinfo  {journal} {Prog.
  Part. Nucl. Phys.}\ }\textbf {\bibinfo {volume} {63}},\ \bibinfo {pages}
  {117--154} (\bibinfo {year} {2009})},\ \Eprint
  {http://arxiv.org/abs/0804.3501} {arXiv:0804.3501 [nucl-th]} \BibitemShut
  {NoStop}%
\bibitem [{\citenamefont {Epelbaum}\ \emph {et~al.}(2009)\citenamefont
  {Epelbaum}, \citenamefont {Krebs}, \citenamefont {Lee},\ and\ \citenamefont
  {Mei{\ss}ner}}]{Epelbaum:2008vj}%
  \BibitemOpen
  \bibfield  {author} {\bibinfo {author} {\bibfnamefont {Evgeny}\ \bibnamefont
  {Epelbaum}}, \bibinfo {author} {\bibfnamefont {Hermann}\ \bibnamefont
  {Krebs}}, \bibinfo {author} {\bibfnamefont {Dean}\ \bibnamefont {Lee}}, \
  and\ \bibinfo {author} {\bibfnamefont {Ulf-G.}\ \bibnamefont {Mei{\ss}ner}},\
  }\bibfield  {title} {\enquote {\bibinfo {title} {{Ground state energy of
  dilute neutron matter at next-to-leading order in lattice chiral effective
  field theory}},}\ }\href {\doibase 10.1140/epja/i2009-10755-0} {\bibfield
  {journal} {\bibinfo  {journal} {Eur. Phys. J.}\ }\textbf {\bibinfo {volume}
  {A40}},\ \bibinfo {pages} {199--213} (\bibinfo {year} {2009})},\ \Eprint
  {http://arxiv.org/abs/0812.3653} {arXiv:0812.3653 [nucl-th]} \BibitemShut
  {NoStop}%
\bibitem [{\citenamefont {Epelbaum}\ \emph {et~al.}(2010)\citenamefont
  {Epelbaum}, \citenamefont {Krebs}, \citenamefont {Lee},\ and\ \citenamefont
  {Mei{\ss}ner}}]{Epelbaum:2010xt}%
  \BibitemOpen
  \bibfield  {author} {\bibinfo {author} {\bibfnamefont {Evgeny}\ \bibnamefont
  {Epelbaum}}, \bibinfo {author} {\bibfnamefont {Hermann}\ \bibnamefont
  {Krebs}}, \bibinfo {author} {\bibfnamefont {Dean}\ \bibnamefont {Lee}}, \
  and\ \bibinfo {author} {\bibfnamefont {Ulf-G.}\ \bibnamefont {Mei{\ss}ner}},\
  }\bibfield  {title} {\enquote {\bibinfo {title} {{Lattice calculations for
  A=3,4,6,12 nuclei using chiral effective field theory}},}\ }\href {\doibase
  10.1140/epja/i2010-11009-x} {\bibfield  {journal} {\bibinfo  {journal} {Eur.
  Phys. J.}\ }\textbf {\bibinfo {volume} {A45}},\ \bibinfo {pages} {335--352}
  (\bibinfo {year} {2010})},\ \Eprint {http://arxiv.org/abs/1003.5697}
  {arXiv:1003.5697 [nucl-th]} \BibitemShut {NoStop}%
\bibitem [{\citenamefont {{Lu, Bing-Nan and L\"ahde, Timo A. and Lee, Dean and
  Mei{\ss}ner, Ulf-G.}}(2016)}]{Lu:2015riz}%
  \BibitemOpen
  \bibfield  {author} {\bibinfo {author} {\bibnamefont {{Lu, Bing-Nan and
  L\"ahde, Timo A. and Lee, Dean and Mei{\ss}ner, Ulf-G.}}},\ }\bibfield
  {title} {\enquote {\bibinfo {title} {{Precise determination of lattice phase
  shifts and mixing angles}},}\ }\href {\doibase
  10.1016/j.physletb.2016.06.081} {\bibfield  {journal} {\bibinfo  {journal}
  {Phys. Lett.}\ }\textbf {\bibinfo {volume} {B760}},\ \bibinfo {pages}
  {309--313} (\bibinfo {year} {2016})},\ \Eprint
  {http://arxiv.org/abs/1506.05652} {arXiv:1506.05652 [nucl-th]} \BibitemShut
  {NoStop}%
\bibitem [{\citenamefont {{Elhatisari, Serdar and Lee, Dean and Rupak, Gautam
  and Epelbaum, Evgeny and Krebs, Hermann and L\"ahde, Timo A. and Luu, Thomas
  and Mei{\ss}ner, Ulf-G.}}(2015)}]{Elhatisari:2015iga}%
  \BibitemOpen
  \bibfield  {author} {\bibinfo {author} {\bibnamefont {{Elhatisari, Serdar and
  Lee, Dean and Rupak, Gautam and Epelbaum, Evgeny and Krebs, Hermann and
  L\"ahde, Timo A. and Luu, Thomas and Mei{\ss}ner, Ulf-G.}}},\ }\bibfield
  {title} {\enquote {\bibinfo {title} {{Ab initio alpha-alpha scattering}},}\
  }\href {\doibase 10.1038/nature16067} {\bibfield  {journal} {\bibinfo
  {journal} {Nature}\ }\textbf {\bibinfo {volume} {528}},\ \bibinfo {pages}
  {111} (\bibinfo {year} {2015})},\ \Eprint {http://arxiv.org/abs/1506.03513}
  {arXiv:1506.03513 [nucl-th]} \BibitemShut {NoStop}%
\bibitem [{\citenamefont {Elhatisari}\ \emph
  {et~al.}(2016{\natexlab{a}})\citenamefont {Elhatisari} \emph
  {et~al.}}]{Elhatisari:2016owd}%
  \BibitemOpen
  \bibfield  {author} {\bibinfo {author} {\bibfnamefont {Serdar}\ \bibnamefont
  {Elhatisari}} \emph {et~al.},\ }\bibfield  {title} {\enquote {\bibinfo
  {title} {{Nuclear binding near a quantum phase transition}},}\ }\href
  {\doibase 10.1103/PhysRevLett.117.132501} {\bibfield  {journal} {\bibinfo
  {journal} {Phys. Rev. Lett.}\ }\textbf {\bibinfo {volume} {117}},\ \bibinfo
  {pages} {132501} (\bibinfo {year} {2016}{\natexlab{a}})},\ \Eprint
  {http://arxiv.org/abs/1602.04539} {arXiv:1602.04539 [nucl-th]} \BibitemShut
  {NoStop}%
\bibitem [{\citenamefont {Elhatisari}\ \emph
  {et~al.}(2016{\natexlab{b}})\citenamefont {Elhatisari}, \citenamefont {Lee},
  \citenamefont {Mei{\ss}ner},\ and\ \citenamefont
  {Rupak}}]{Elhatisari:2016hby}%
  \BibitemOpen
  \bibfield  {author} {\bibinfo {author} {\bibfnamefont {Serdar}\ \bibnamefont
  {Elhatisari}}, \bibinfo {author} {\bibfnamefont {Dean}\ \bibnamefont {Lee}},
  \bibinfo {author} {\bibfnamefont {Ulf-G.}\ \bibnamefont {Mei{\ss}ner}}, \
  and\ \bibinfo {author} {\bibfnamefont {Gautam}\ \bibnamefont {Rupak}},\
  }\bibfield  {title} {\enquote {\bibinfo {title} {{Nucleon-deuteron scattering
  using the adiabatic projection method}},}\ }\href {\doibase
  10.1140/epja/i2016-16174-2} {\bibfield  {journal} {\bibinfo  {journal} {Eur.
  Phys. J.}\ }\textbf {\bibinfo {volume} {A52}},\ \bibinfo {pages} {174}
  (\bibinfo {year} {2016}{\natexlab{b}})},\ \Eprint
  {http://arxiv.org/abs/1603.02333} {arXiv:1603.02333 [nucl-th]} \BibitemShut
  {NoStop}%
\bibitem [{\citenamefont {{Klein, Nico and Elhatisari, Serdar and L\"ahde, Timo
  A. and Lee, Dean and Mei{\ss}ner, Ulf-G.}}(2018)}]{Klein:2018lqz}%
  \BibitemOpen
  \bibfield  {author} {\bibinfo {author} {\bibnamefont {{Klein, Nico and
  Elhatisari, Serdar and L\"ahde, Timo A. and Lee, Dean and Mei{\ss}ner,
  Ulf-G.}}},\ }\bibfield  {title} {\enquote {\bibinfo {title} {{The Tjon Band
  in Nuclear Lattice Effective Field Theory}},}\ }\href {\doibase
  10.1140/epja/i2018-12553-y} {\bibfield  {journal} {\bibinfo  {journal} {Eur.
  Phys. J.}\ }\textbf {\bibinfo {volume} {A54}},\ \bibinfo {pages} {121}
  (\bibinfo {year} {2018})},\ \Eprint {http://arxiv.org/abs/1803.04231}
  {arXiv:1803.04231 [nucl-th]} \BibitemShut {NoStop}%
\bibitem [{\citenamefont {Li}\ \emph {et~al.}(2019{\natexlab{a}})\citenamefont
  {Li}, \citenamefont {Elhatisari}, \citenamefont {Epelbaum}, \citenamefont
  {Lee}, \citenamefont {Lu},\ and\ \citenamefont {Mei{\ss}ner}}]{Li:2019ldq}%
  \BibitemOpen
  \bibfield  {author} {\bibinfo {author} {\bibfnamefont {Ning}\ \bibnamefont
  {Li}}, \bibinfo {author} {\bibfnamefont {Serdar}\ \bibnamefont {Elhatisari}},
  \bibinfo {author} {\bibfnamefont {Evgeny}\ \bibnamefont {Epelbaum}}, \bibinfo
  {author} {\bibfnamefont {Dean}\ \bibnamefont {Lee}}, \bibinfo {author}
  {\bibfnamefont {Bingnan}\ \bibnamefont {Lu}}, \ and\ \bibinfo {author}
  {\bibfnamefont {Ulf-G}\ \bibnamefont {Mei{\ss}ner}},\ }\bibfield  {title}
  {\enquote {\bibinfo {title} {{Galilean invariance restoration on the
  lattice}},}\ }\href {\doibase 10.1103/PhysRevC.99.064001} {\bibfield
  {journal} {\bibinfo  {journal} {Phys. Rev.}\ }\textbf {\bibinfo {volume}
  {C99}},\ \bibinfo {pages} {064001} (\bibinfo {year} {2019}{\natexlab{a}})},\
  \Eprint {http://arxiv.org/abs/1902.01295} {arXiv:1902.01295 [nucl-th]}
  \BibitemShut {NoStop}%
\bibitem [{\citenamefont {Bovermann}\ \emph {et~al.}(2019)\citenamefont
  {Bovermann}, \citenamefont {Epelbaum}, \citenamefont {Krebs},\ and\
  \citenamefont {Lee}}]{Bovermann:2019jbt}%
  \BibitemOpen
  \bibfield  {author} {\bibinfo {author} {\bibfnamefont {Lukas}\ \bibnamefont
  {Bovermann}}, \bibinfo {author} {\bibfnamefont {Evgeny}\ \bibnamefont
  {Epelbaum}}, \bibinfo {author} {\bibfnamefont {Hermann}\ \bibnamefont
  {Krebs}}, \ and\ \bibinfo {author} {\bibfnamefont {Dean}\ \bibnamefont
  {Lee}},\ }\bibfield  {title} {\enquote {\bibinfo {title} {{Scattering phase
  shifts and mixing angles for an arbitrary number of coupled channels on the
  lattice}},}\ }\href@noop {} {\  (\bibinfo {year} {2019})},\ \Eprint
  {http://arxiv.org/abs/1905.02492} {arXiv:1905.02492 [nucl-th]} \BibitemShut
  {NoStop}%
\bibitem [{\citenamefont {{L\"ahde, Timo A. and Mei{\ss}ner,
  Ulf-G.}}(2019)}]{Lahde:2019npb}%
  \BibitemOpen
  \bibfield  {author} {\bibinfo {author} {\bibnamefont {{L\"ahde, Timo A. and
  Mei{\ss}ner, Ulf-G.}}},\ }\bibfield  {title} {\enquote {\bibinfo {title}
  {{Nuclear Lattice Effective Field Theory}},}\ }\href {\doibase
  10.1007/978-3-030-14189-9} {\bibfield  {journal} {\bibinfo  {journal} {Lect.
  Notes Phys.}\ }\textbf {\bibinfo {volume} {957}},\ \bibinfo {pages} {1--396}
  (\bibinfo {year} {2019})}\BibitemShut {NoStop}%
\bibitem [{\citenamefont {Endres}\ \emph {et~al.}(2011)\citenamefont {Endres},
  \citenamefont {Kaplan}, \citenamefont {Lee},\ and\ \citenamefont
  {Nicholson}}]{Endres:2011er}%
  \BibitemOpen
  \bibfield  {author} {\bibinfo {author} {\bibfnamefont {Michael~G.}\
  \bibnamefont {Endres}}, \bibinfo {author} {\bibfnamefont {David~B.}\
  \bibnamefont {Kaplan}}, \bibinfo {author} {\bibfnamefont {Jong-Wan}\
  \bibnamefont {Lee}}, \ and\ \bibinfo {author} {\bibfnamefont {Amy~N.}\
  \bibnamefont {Nicholson}},\ }\bibfield  {title} {\enquote {\bibinfo {title}
  {{Lattice Monte Carlo calculations for unitary fermions in a harmonic
  trap}},}\ }\href {\doibase 10.1103/PhysRevA.84.043644} {\bibfield  {journal}
  {\bibinfo  {journal} {Phys. Rev.}\ }\textbf {\bibinfo {volume} {A84}},\
  \bibinfo {pages} {043644} (\bibinfo {year} {2011})},\ \Eprint
  {http://arxiv.org/abs/1106.5725} {arXiv:1106.5725 [hep-lat]} \BibitemShut
  {NoStop}%
\bibitem [{\citenamefont {Lee}(2008)}]{Lee:2007ae}%
  \BibitemOpen
  \bibfield  {author} {\bibinfo {author} {\bibfnamefont {Dean}\ \bibnamefont
  {Lee}},\ }\bibfield  {title} {\enquote {\bibinfo {title} {{The Symmetric
  heavy-light ansatz}},}\ }\href {\doibase 10.1140/epja/i2008-10537-2}
  {\bibfield  {journal} {\bibinfo  {journal} {Eur. Phys. J.}\ }\textbf
  {\bibinfo {volume} {A35}},\ \bibinfo {pages} {171--187} (\bibinfo {year}
  {2008})},\ \Eprint {http://arxiv.org/abs/0704.3439} {arXiv:0704.3439
  [cond-mat.supr-con]} \BibitemShut {NoStop}%
\bibitem [{\citenamefont {Endres}\ \emph {et~al.}(2013)\citenamefont {Endres},
  \citenamefont {Kaplan}, \citenamefont {Lee},\ and\ \citenamefont
  {Nicholson}}]{Endres:2012cw}%
  \BibitemOpen
  \bibfield  {author} {\bibinfo {author} {\bibfnamefont {Michael~G.}\
  \bibnamefont {Endres}}, \bibinfo {author} {\bibfnamefont {David~B.}\
  \bibnamefont {Kaplan}}, \bibinfo {author} {\bibfnamefont {Jong-Wan}\
  \bibnamefont {Lee}}, \ and\ \bibinfo {author} {\bibfnamefont {Amy~N.}\
  \bibnamefont {Nicholson}},\ }\bibfield  {title} {\enquote {\bibinfo {title}
  {{Lattice Monte Carlo calculations for unitary fermions in a finite box}},}\
  }\href {\doibase 10.1103/PhysRevA.87.023615} {\bibfield  {journal} {\bibinfo
  {journal} {Phys. Rev.}\ }\textbf {\bibinfo {volume} {A87}},\ \bibinfo {pages}
  {023615} (\bibinfo {year} {2013})},\ \Eprint {http://arxiv.org/abs/1203.3169}
  {arXiv:1203.3169 [hep-lat]} \BibitemShut {NoStop}%
\bibitem [{\citenamefont {He}\ \emph {et~al.}(2019)\citenamefont {He},
  \citenamefont {Li}, \citenamefont {Lu},\ and\ \citenamefont
  {Lee}}]{He:2019ipt}%
  \BibitemOpen
  \bibfield  {author} {\bibinfo {author} {\bibfnamefont {Rongzheng}\
  \bibnamefont {He}}, \bibinfo {author} {\bibfnamefont {Ning}\ \bibnamefont
  {Li}}, \bibinfo {author} {\bibfnamefont {Bing-Nan}\ \bibnamefont {Lu}}, \
  and\ \bibinfo {author} {\bibfnamefont {Dean}\ \bibnamefont {Lee}},\
  }\bibfield  {title} {\enquote {\bibinfo {title} {{Superfluid Condensate
  Fraction and Pairing Wave Function of the Unitary Fermi Gas}},}\ }\href@noop
  {} {\  (\bibinfo {year} {2019})},\ \Eprint {http://arxiv.org/abs/1910.01257}
  {arXiv:1910.01257 [cond-mat.quant-gas]} \BibitemShut {NoStop}%
\bibitem [{\citenamefont {Fiebig}\ \emph {et~al.}(1994)\citenamefont {Fiebig},
  \citenamefont {Dominguez},\ and\ \citenamefont {Woloshyn}}]{Fiebig:1994qi}%
  \BibitemOpen
  \bibfield  {author} {\bibinfo {author} {\bibfnamefont {H.~R.}\ \bibnamefont
  {Fiebig}}, \bibinfo {author} {\bibfnamefont {A.}~\bibnamefont {Dominguez}}, \
  and\ \bibinfo {author} {\bibfnamefont {R.~M.}\ \bibnamefont {Woloshyn}},\
  }\bibfield  {title} {\enquote {\bibinfo {title} {{Meson meson scattering
  phase shifts in (2+1)-dimensional lattice QED}},}\ }\href {\doibase
  10.1016/0550-3213(94)90534-7} {\bibfield  {journal} {\bibinfo  {journal}
  {Nucl. Phys.}\ }\textbf {\bibinfo {volume} {B418}},\ \bibinfo {pages}
  {649--685} (\bibinfo {year} {1994})}\BibitemShut {NoStop}%
\bibitem [{\citenamefont {Beane}(2010)}]{Beane:2010ny}%
  \BibitemOpen
  \bibfield  {author} {\bibinfo {author} {\bibfnamefont {Silas~R.}\
  \bibnamefont {Beane}},\ }\bibfield  {title} {\enquote {\bibinfo {title}
  {{Ground state energy of the interacting Bose gas in two dimensions: An
  Explicit construction}},}\ }\href {\doibase 10.1103/PhysRevA.82.063610}
  {\bibfield  {journal} {\bibinfo  {journal} {Phys. Rev.}\ }\textbf {\bibinfo
  {volume} {A82}},\ \bibinfo {pages} {063610} (\bibinfo {year} {2010})},\
  \Eprint {http://arxiv.org/abs/1002.3815} {arXiv:1002.3815
  [cond-mat.quant-gas]} \BibitemShut {NoStop}%
\bibitem [{\citenamefont {K\"orber}\ \emph {et~al.}(2019)\citenamefont
  {K\"orber}, \citenamefont {Berkowitz},\ and\ \citenamefont
  {Luu}}]{luescher-nd_201}%
  \BibitemOpen
  \bibfield  {author} {\bibinfo {author} {\bibfnamefont {Christopher}\
  \bibnamefont {K\"orber}}, \bibinfo {author} {\bibfnamefont {Evan}\
  \bibnamefont {Berkowitz}}, \ and\ \bibinfo {author} {\bibfnamefont {Thomas}\
  \bibnamefont {Luu}},\ }\href@noop {} {\enquote {\bibinfo {title}
  {\texttt{luescher-nd}},}\ }\bibinfo {howpublished} {GitHub Repository
  \url{https://github.com/ckoerber/luescher-nd/releases/tag/v1.0.0}} (\bibinfo
  {year} {2019})\BibitemShut {NoStop}%
\bibitem [{\citenamefont {Li}\ \emph {et~al.}(2019{\natexlab{b}})\citenamefont
  {Li}, \citenamefont {Wu}, \citenamefont {Abell}, \citenamefont {Leinweber},\
  and\ \citenamefont {Thomas}}]{Li:2019qvh}%
  \BibitemOpen
  \bibfield  {author} {\bibinfo {author} {\bibfnamefont {Yan}\ \bibnamefont
  {Li}}, \bibinfo {author} {\bibfnamefont {Jia-Jun}\ \bibnamefont {Wu}},
  \bibinfo {author} {\bibfnamefont {Curtis~D.}\ \bibnamefont {Abell}}, \bibinfo
  {author} {\bibfnamefont {Derek~B.}\ \bibnamefont {Leinweber}}, \ and\
  \bibinfo {author} {\bibfnamefont {Anthony~W.}\ \bibnamefont {Thomas}},\
  }\bibfield  {title} {\enquote {\bibinfo {title} {{Partial Wave Mixing in
  Hamiltonian Effective Field Theory}},}\ }\href@noop {} {\  (\bibinfo {year}
  {2019}{\natexlab{b}})},\ \Eprint {http://arxiv.org/abs/1910.04973}
  {arXiv:1910.04973 [hep-lat]} \BibitemShut {NoStop}%
\bibitem [{\citenamefont {Beane}\ \emph {et~al.}(2004)\citenamefont {Beane},
  \citenamefont {Bedaque}, \citenamefont {Parreno},\ and\ \citenamefont
  {Savage}}]{Beane:2003da}%
  \BibitemOpen
  \bibfield  {author} {\bibinfo {author} {\bibfnamefont {S.~R.}\ \bibnamefont
  {Beane}}, \bibinfo {author} {\bibfnamefont {P.~F.}\ \bibnamefont {Bedaque}},
  \bibinfo {author} {\bibfnamefont {A.}~\bibnamefont {Parreno}}, \ and\
  \bibinfo {author} {\bibfnamefont {M.~J.}\ \bibnamefont {Savage}},\ }\bibfield
   {title} {\enquote {\bibinfo {title} {{Two nucleons on a lattice}},}\ }\href
  {\doibase 10.1016/j.physletb.2004.02.007} {\bibfield  {journal} {\bibinfo
  {journal} {Phys. Lett.}\ }\textbf {\bibinfo {volume} {B585}},\ \bibinfo
  {pages} {106--114} (\bibinfo {year} {2004})},\ \Eprint
  {http://arxiv.org/abs/hep-lat/0312004} {arXiv:hep-lat/0312004 [hep-lat]}
  \BibitemShut {NoStop}%
\bibitem [{\citenamefont {Ozaki}\ and\ \citenamefont
  {Sasaki}(2013)}]{Ozaki:2012ce}%
  \BibitemOpen
  \bibfield  {author} {\bibinfo {author} {\bibfnamefont {Sho}\ \bibnamefont
  {Ozaki}}\ and\ \bibinfo {author} {\bibfnamefont {Shoichi}\ \bibnamefont
  {Sasaki}},\ }\bibfield  {title} {\enquote {\bibinfo {title} {{L\"uscher's
  finite size method with twisted boundary conditions: An application to the
  $J/\psi$-$\phi$ system to search for a narrow resonance}},}\ }\href {\doibase
  10.1103/PhysRevD.87.014506} {\bibfield  {journal} {\bibinfo  {journal} {Phys.
  Rev.}\ }\textbf {\bibinfo {volume} {D87}},\ \bibinfo {pages} {014506}
  (\bibinfo {year} {2013})},\ \Eprint {http://arxiv.org/abs/1211.5512}
  {arXiv:1211.5512 [hep-lat]} \BibitemShut {NoStop}%
\bibitem [{\citenamefont {Hansen}\ and\ \citenamefont
  {Sharpe}(2012)}]{Hansen:2012tf}%
  \BibitemOpen
  \bibfield  {author} {\bibinfo {author} {\bibfnamefont {Maxwell~T.}\
  \bibnamefont {Hansen}}\ and\ \bibinfo {author} {\bibfnamefont {Stephen~R.}\
  \bibnamefont {Sharpe}},\ }\bibfield  {title} {\enquote {\bibinfo {title}
  {{Multiple-channel generalization of Lellouch-L\"uscher formula}},}\ }\href
  {\doibase 10.1103/PhysRevD.86.016007} {\bibfield  {journal} {\bibinfo
  {journal} {Phys. Rev.}\ }\textbf {\bibinfo {volume} {D86}},\ \bibinfo {pages}
  {016007} (\bibinfo {year} {2012})},\ \Eprint {http://arxiv.org/abs/1204.0826}
  {arXiv:1204.0826 [hep-lat]} \BibitemShut {NoStop}%
\bibitem [{\citenamefont {Briceno}\ \emph {et~al.}(2014)\citenamefont
  {Briceno}, \citenamefont {Davoudi}, \citenamefont {Luu},\ and\ \citenamefont
  {Savage}}]{Briceno:2013hya}%
  \BibitemOpen
  \bibfield  {author} {\bibinfo {author} {\bibfnamefont {Raul~A.}\ \bibnamefont
  {Briceno}}, \bibinfo {author} {\bibfnamefont {Zohreh}\ \bibnamefont
  {Davoudi}}, \bibinfo {author} {\bibfnamefont {Thomas~C.}\ \bibnamefont
  {Luu}}, \ and\ \bibinfo {author} {\bibfnamefont {Martin~J.}\ \bibnamefont
  {Savage}},\ }\bibfield  {title} {\enquote {\bibinfo {title} {{Two-Baryon
  Systems with Twisted Boundary Conditions}},}\ }\href {\doibase
  10.1103/PhysRevD.89.074509} {\bibfield  {journal} {\bibinfo  {journal} {Phys.
  Rev.}\ }\textbf {\bibinfo {volume} {D89}},\ \bibinfo {pages} {074509}
  (\bibinfo {year} {2014})},\ \Eprint {http://arxiv.org/abs/1311.7686}
  {arXiv:1311.7686 [hep-lat]} \BibitemShut {NoStop}%
\bibitem [{\citenamefont {Briceno}\ \emph {et~al.}(2013)\citenamefont
  {Briceno}, \citenamefont {Davoudi},\ and\ \citenamefont
  {Luu}}]{Briceno:2013lba}%
  \BibitemOpen
  \bibfield  {author} {\bibinfo {author} {\bibfnamefont {Raul~A.}\ \bibnamefont
  {Briceno}}, \bibinfo {author} {\bibfnamefont {Zohreh}\ \bibnamefont
  {Davoudi}}, \ and\ \bibinfo {author} {\bibfnamefont {Thomas~C.}\ \bibnamefont
  {Luu}},\ }\bibfield  {title} {\enquote {\bibinfo {title} {{Two-Nucleon
  Systems in a Finite Volume: (I) Quantization Conditions}},}\ }\href {\doibase
  10.1103/PhysRevD.88.034502} {\bibfield  {journal} {\bibinfo  {journal} {Phys.
  Rev.}\ }\textbf {\bibinfo {volume} {D88}},\ \bibinfo {pages} {034502}
  (\bibinfo {year} {2013})},\ \Eprint {http://arxiv.org/abs/1305.4903}
  {arXiv:1305.4903 [hep-lat]} \BibitemShut {NoStop}%
\bibitem [{\citenamefont {Li}\ \emph {et~al.}(2014)\citenamefont {Li},
  \citenamefont {Li},\ and\ \citenamefont {Liu}}]{Li:2014wga}%
  \BibitemOpen
  \bibfield  {author} {\bibinfo {author} {\bibfnamefont {Ning}\ \bibnamefont
  {Li}}, \bibinfo {author} {\bibfnamefont {Song-Yuan}\ \bibnamefont {Li}}, \
  and\ \bibinfo {author} {\bibfnamefont {Chuan}\ \bibnamefont {Liu}},\
  }\bibfield  {title} {\enquote {\bibinfo {title} {{Generalized L\"uscher's
  formula in multichannel baryon-baryon scattering}},}\ }\href {\doibase
  10.1103/PhysRevD.90.034509} {\bibfield  {journal} {\bibinfo  {journal} {Phys.
  Rev.}\ }\textbf {\bibinfo {volume} {D90}},\ \bibinfo {pages} {034509}
  (\bibinfo {year} {2014})},\ \Eprint {http://arxiv.org/abs/1401.5569}
  {arXiv:1401.5569 [hep-lat]} \BibitemShut {NoStop}%
\bibitem [{\citenamefont {Zhu}\ and\ \citenamefont {Tan}(2019)}]{Zhu:2019dho}%
  \BibitemOpen
  \bibfield  {author} {\bibinfo {author} {\bibfnamefont {Shangguo}\
  \bibnamefont {Zhu}}\ and\ \bibinfo {author} {\bibfnamefont {Shina}\
  \bibnamefont {Tan}},\ }\bibfield  {title} {\enquote {\bibinfo {title}
  {{$d$-dimensional L\"uscher's formula and the near-threshold three-body
  states in a finite volume}},}\ }\href@noop {} {\  (\bibinfo {year} {2019})},\
  \Eprint {http://arxiv.org/abs/1905.05117} {arXiv:1905.05117 [nucl-th]}
  \BibitemShut {NoStop}%
\bibitem [{\citenamefont {Kaplan}\ \emph {et~al.}(1998)\citenamefont {Kaplan},
  \citenamefont {Savage},\ and\ \citenamefont {Wise}}]{Kaplan:1998we}%
  \BibitemOpen
  \bibfield  {author} {\bibinfo {author} {\bibfnamefont {David~B.}\
  \bibnamefont {Kaplan}}, \bibinfo {author} {\bibfnamefont {Martin~J.}\
  \bibnamefont {Savage}}, \ and\ \bibinfo {author} {\bibfnamefont {Mark~B.}\
  \bibnamefont {Wise}},\ }\bibfield  {title} {\enquote {\bibinfo {title} {{Two
  nucleon systems from effective field theory}},}\ }\href {\doibase
  10.1016/S0550-3213(98)00440-4} {\bibfield  {journal} {\bibinfo  {journal}
  {Nucl. Phys.}\ }\textbf {\bibinfo {volume} {B534}},\ \bibinfo {pages}
  {329--355} (\bibinfo {year} {1998})},\ \Eprint
  {http://arxiv.org/abs/nucl-th/9802075} {arXiv:nucl-th/9802075 [nucl-th]}
  \BibitemShut {NoStop}%
\bibitem [{\citenamefont {Morningstar}\ \emph {et~al.}(2017)\citenamefont
  {Morningstar}, \citenamefont {Bulava}, \citenamefont {Singha}, \citenamefont
  {Brett}, \citenamefont {Fallica}, \citenamefont {Hanlon},\ and\ \citenamefont
  {{H\"{o}rz}}}]{Morningstar:2017spu}%
  \BibitemOpen
  \bibfield  {author} {\bibinfo {author} {\bibfnamefont {Colin}\ \bibnamefont
  {Morningstar}}, \bibinfo {author} {\bibfnamefont {John}\ \bibnamefont
  {Bulava}}, \bibinfo {author} {\bibfnamefont {Bijit}\ \bibnamefont {Singha}},
  \bibinfo {author} {\bibfnamefont {{Ruair\'{i}}}\ \bibnamefont {Brett}},
  \bibinfo {author} {\bibfnamefont {Jacob}\ \bibnamefont {Fallica}}, \bibinfo
  {author} {\bibfnamefont {Andrew}\ \bibnamefont {Hanlon}}, \ and\ \bibinfo
  {author} {\bibfnamefont {Ben}\ \bibnamefont {{H\"{o}rz}}},\ }\bibfield
  {title} {\enquote {\bibinfo {title} {{Estimating the two-particle $K$-matrix
  for multiple partial waves and decay channels from finite-volume
  energies}},}\ }\href {\doibase 10.1016/j.nuclphysb.2017.09.014} {\bibfield
  {journal} {\bibinfo  {journal} {Nucl. Phys.}\ }\textbf {\bibinfo {volume}
  {B924}},\ \bibinfo {pages} {477--507} (\bibinfo {year} {2017})},\ \Eprint
  {http://arxiv.org/abs/1707.05817} {arXiv:1707.05817 [hep-lat]} \BibitemShut
  {NoStop}%
\bibitem [{\citenamefont {Morningstar}()}]{Morningstar:2hib}%
  \BibitemOpen
  \bibfield  {author} {\bibinfo {author} {\bibfnamefont {Colin}\ \bibnamefont
  {Morningstar}},\ }\bibfield  {title} {\enquote {\bibinfo {title}
  {\texttt{TwoHadronsInBox}},}\ }\href
  {https://github.com/cjmorningstar10/TwoHadronsInBox} {\ }\bibinfo {note}
  {GitHub Repository
  \url{https://github.com/cjmorningstar10/TwoHadronsInBox}}\BibitemShut
  {NoStop}%
\bibitem [{\citenamefont {Lepage}\ and\ \citenamefont
  {Gohlke}(2016)}]{peter_lepage_2016_60221}%
  \BibitemOpen
  \bibfield  {author} {\bibinfo {author} {\bibfnamefont {Peter}\ \bibnamefont
  {Lepage}}\ and\ \bibinfo {author} {\bibfnamefont {Christoph}\ \bibnamefont
  {Gohlke}},\ }\href {\doibase 10.5281/zenodo.60221} {\enquote {\bibinfo
  {title} {lsqfit: lsqfit version 8.0.1},}\ } (\bibinfo {year}
  {2016})\BibitemShut {NoStop}%
\bibitem [{\citenamefont {Luu}\ and\ \citenamefont
  {Savage}(2011)}]{Luu:2011ep}%
  \BibitemOpen
  \bibfield  {author} {\bibinfo {author} {\bibfnamefont {Thomas}\ \bibnamefont
  {Luu}}\ and\ \bibinfo {author} {\bibfnamefont {Martin~J.}\ \bibnamefont
  {Savage}},\ }\bibfield  {title} {\enquote {\bibinfo {title} {{Extracting
  Scattering Phase-Shifts in Higher Partial-Waves from Lattice QCD
  Calculations}},}\ }\href {\doibase 10.1103/PhysRevD.83.114508} {\bibfield
  {journal} {\bibinfo  {journal} {Phys. Rev.}\ }\textbf {\bibinfo {volume}
  {D83}},\ \bibinfo {pages} {114508} (\bibinfo {year} {2011})},\ \Eprint
  {http://arxiv.org/abs/1101.3347} {arXiv:1101.3347 [hep-lat]} \BibitemShut
  {NoStop}%
\bibitem [{\citenamefont {L\"uscher}\ and\ \citenamefont
  {Wolff}(1990)}]{Luscher:1990ck}%
  \BibitemOpen
  \bibfield  {author} {\bibinfo {author} {\bibfnamefont {Martin}\ \bibnamefont
  {L\"uscher}}\ and\ \bibinfo {author} {\bibfnamefont {Ulli}\ \bibnamefont
  {Wolff}},\ }\bibfield  {title} {\enquote {\bibinfo {title} {{How to Calculate
  the Elastic Scattering Matrix in Two-dimensional Quantum Field Theories by
  Numerical Simulation}},}\ }\href {\doibase 10.1016/0550-3213(90)90540-T}
  {\bibfield  {journal} {\bibinfo  {journal} {Nucl. Phys.}\ }\textbf {\bibinfo
  {volume} {B339}},\ \bibinfo {pages} {222--252} (\bibinfo {year}
  {1990})}\BibitemShut {NoStop}%
\bibitem [{\citenamefont {Pupyshev}(2014)}]{Pupyshev:2014}%
  \BibitemOpen
  \bibfield  {author} {\bibinfo {author} {\bibfnamefont {V.~V.}\ \bibnamefont
  {Pupyshev}},\ }\bibfield  {title} {\enquote {\bibinfo {title} {{The Length
  and Effective Radius of Two-Dimensional Scattering of a Quantum Particle by a
  Centrally Symmetric Short-Range Potential}},}\ }\href {\doibase
  0040-5779/14/1803-1051} {\bibfield  {journal} {\bibinfo  {journal} {Th. Math.
  Phys.}\ }\textbf {\bibinfo {volume} {180(3)}},\ \bibinfo {pages}
  {1051–1072} (\bibinfo {year} {2014})}\BibitemShut {NoStop}%
\bibitem [{lat(2018)}]{lattice-practices}%
  \BibitemOpen
  \href
  {https://www.fz-juelich.de/ias/jsc/EN/Expertise/Workshops/Conferences/LAP18/_node.html}
  {\emph {\bibinfo {title} {Lattice Practices 2018}}}\ (\bibinfo {address}
  {J\"{u}lich Supercomputing Center, Forschungszentrum J\"{u}lich},\ \bibinfo
  {year} {2018})\BibitemShut {NoStop}%
\bibitem [{\citenamefont {Berkowitz}(2018)}]{lattice-practices-exercises}%
  \BibitemOpen
  \bibfield  {author} {\bibinfo {author} {\bibfnamefont {Evan}\ \bibnamefont
  {Berkowitz}},\ }\href@noop {} {\enquote {\bibinfo {title} {{1D Quantum
  Mechanics Example of the L\"uscher Formalism}},}\ }\bibinfo {howpublished}
  {\url{https://github.com/evanberkowitz/qm-luescher-example/}} (\bibinfo
  {year} {2018})\BibitemShut {NoStop}%
\bibitem [{\citenamefont {Epelbaum}\ \emph {et~al.}(2018)\citenamefont
  {Epelbaum}, \citenamefont {Gasparyan}, \citenamefont {Gegelia},\ and\
  \citenamefont {Mei{\ss}ner}}]{Epelbaum:2018zli}%
  \BibitemOpen
  \bibfield  {author} {\bibinfo {author} {\bibfnamefont {E.}~\bibnamefont
  {Epelbaum}}, \bibinfo {author} {\bibfnamefont {A.~M.}\ \bibnamefont
  {Gasparyan}}, \bibinfo {author} {\bibfnamefont {J.}~\bibnamefont {Gegelia}},
  \ and\ \bibinfo {author} {\bibfnamefont {Ulf-G.}\ \bibnamefont
  {Mei{\ss}ner}},\ }\bibfield  {title} {\enquote {\bibinfo {title} {{How (not)
  to renormalize integral equations with singular potentials in effective field
  theory}},}\ }\href {\doibase 10.1140/epja/i2018-12632-1} {\bibfield
  {journal} {\bibinfo  {journal} {Eur. Phys. J.}\ }\textbf {\bibinfo {volume}
  {A54}},\ \bibinfo {pages} {186} (\bibinfo {year} {2018})},\ \Eprint
  {http://arxiv.org/abs/1810.02646} {arXiv:1810.02646 [nucl-th]} \BibitemShut
  {NoStop}%
\end{thebibliography}%

\end{document}